\documentclass[12pt,nocenter,upper,crosshair,sfbold,chapterbib]{thema}

\usepackage{amsmath}
\usepackage{amsfonts}
\usepackage{amssymb}
\usepackage{amstext}
\usepackage{lscape}

\usepackage[dvips]{epsfig}

\captionheaderfont{\sl\bfseries}
\captionbodyfont{\sl}

\chapapp{\chaptername}

\begin{document}

%%%%%%%%%%%%%%%%%%%%%%%%%%%%%%%%%%%%%%%%%%%%%%%%%%%%%%%%%%%%%%%%%%%%%%%%%%%%%%

\frontmatter
\title{Combinatorics of boson normal ordering\\and some applications}
\translator{\textsf{PhD Dissertation}}
\author{\Large \textsc{Pawe\l{} B\l{}asiak}}
\institution{Institute of Nuclear Physics of Polish Academy of Sciences, Krak\'ow\\
and\\
Universit\'e Pierre et Marie Curie, Paris\\
\newpage}

\dedication{\textsf{\em --- To my parents ---}}
\date{2005}

\maketitle

\tableofcontents

%%%%%%%%%%%%%%%%%%%%%%%%%%%%%%%%%%%%%%%%%%%%%%%%%%%%%%%%%%%%%%%%%%%%%%%%%%%%%%%%%%%%%

\chapter{Preface}

The subject of this thesis is the investigation of the combinatorial
structures arising in the boson normal ordering problem. This
research project arose from the collaboration between the Henryk
Niewodnicza\'nski Institute of Nuclear of Polish Academy of
Sciences in Cracow and Laboratoire de Physique Th\'{e}orique de la
Mati\`{e}re Condens\'{e}e of the University of Pierre and Marie
Curie in Paris.
\footnote{This project was financially supported by a French Government Scholarship,
the H. Niewodnicza\'nski Institute of Nuclear Physics in Cracow
and the Polish Ministry of Scientific Research and Information
Technology Grant no: 1P03B 051 26.}
\\
The thesis was written under the common supervision of Prof. Karol
A. Penson (LPTMC, Paris VI) and Prof. Edward Kapu\'scik (IFJ PAN,
Cracow) under the program of \emph{co-tutelle}. 
I am deeply grateful to them for the help and
knowledge they have generously shared with me during my studies.
My appreciation goes also to our collaborators Dr Andrzej Horzela
(IFJ PAN, Cracow), Prof. Allan I. Solomon (Open University, UK)
and Prof. Gerard Duchamp (LIPN, Paris XIII).
\\
I have benefited a lot from discussions with Prof. Labib Haddad,
Prof. Pinaki Roy, Prof. Mark Yor, Prof. Giuseppe Dattoli, Prof.
Miguel A. M\'endez, Prof. Cestmir Burdik, Prof. Krzysztof Kowalski
and Dr. Jerzy Cislo.
\\
I would also like to thank Prof. Bertrand Guillot, Prof. Ryszard
Kerner from LPTMC (Paris VI) and Prof. Marek Kutschera, Prof.
Wojciech Broniowski, Prof. Krzysztof Golec-Biernat, Prof. Wojciech
Florkowski, Prof. Piotr \.Zenczykowski and Prof. Piotr Zieli\'nski 
from IFJ PAN (Cracow) for their interest and warm support.
\\
My utmost gratitude goes to my parents, to whom I dedicate this work.

\mainmatter

%%%%%%%%%%%%%%%%%%%%%%%%%%%%%%%%%%%%%%%%%%%%%%%%%%%%%%%%%%%%%%%%%%%%%%%%%%%%%%%%%%%%%%%%%%%%

\chapter{Introduction}
\label{Introduction}

In this work we are concerned with the one mode boson creation
$a^\dag$ and annihilation $a$ operators satisfying the commutation
relation $[a,a^\dag]=1$. We are interested in the combinatorial
structures arising in the problem of normal ordering of a wide
class of boson expressions. It shall provide us with the effective
tools for systematic treatment of these problems.
\\
By normal ordering of an operator expressed through the boson $a$
and $a^\dag$ operators we mean moving all annihilation operators
to the right of all creation operators with the use of commutation
relation. This procedure yields an operator which is equivalent
(in the operator sense) to the original one but has a different
functional representation. It is of both mathematical and physical
interest. From the mathematical point of view it allows one to
represent the operator by a function of two, in a sense
'commuting', variables. It is also connected with the physicist's
perspective of the problem. Commonly used, the so called coherent
state representation (see Appendix \ref{Coherent states})
implicitly requires the knowledge of the normally ordered form of
the operators in question. Put the other way, for the normally
ordered operator its coherent state matrix elements may be immediately
 read off. This representation is widely used {\it e.g.} in
quantum optics. We mention also that calculation of the vacuum
expectation values is much easier for operators in the
normally ordered form. For other applications see \cite{Klauder}.
\\
A standard approach to the normal ordering problem is through the
Wick theorem. It directly links the problem to combinatorics, {\it
i.e.} searching for all possible contractions in the boson
expression and then summing up the resulting terms. This may be
efficiently used for solving problems with finite number of boson
operators (especially when one uses the computer algebra
packages). Although this looks very simple in that form it is not
very constructive in more sophisticated cases. The main
disadvantage is that it does not give much help in solving
problems concerning operators defined through infinite series
expansions. To do this we would have to know the underlying
structure of the numbers involved. This still requires a lot of
careful analysis (not accessible to computers). In this work we
approach these problems using methods of advanced combinatorial
analysis \cite{Comtet}. It proves to be an efficient way for
obtaining compact formulas for normally ordered expansion
coefficients and then analyzing theirs properties.
\\
A great body of work was already put in the field. In his seminal
paper \cite{Katriel} Jacob Katriel pointed out that the numbers
which come up in the normal ordering problem for $(a^\dag a)^n$
are the Stirling numbers of the second kind. Later on, the connection
between the exponential generating function of the Bell
polynomials and the coherent state matrix elements of $e^{\lambda
a^\dag a}$ was provided \cite{Katriel2000}\cite{Katriel2002}. In
Chapter \ref{Generic} we give a modern review of these results
with special emphasis on the Dobi\'nski relations. We also make
use of a specific realization of the commutation relation
$[a,a^\dag]=1$ in terms of the multiplication $X$ and derivative
$D$ operators. It may be thought of as an introduction to the
methods used later on in this text. Chapter \ref{Boson
Expressions} is written in that spirit. We use this methodology
to investigate and obtain compact formulas for the coefficients
arising in normal ordering of a boson monomial (string of boson
creation and annihilation operators). Then we proceed to the normal
ordering of powers of a boson string and more generally
homogeneous boson polynomial ({\it i.e.} the combinations of the
of the boson strings with the same excess of creation over
annihilation operators). The numbers appearing in the
solution generalize Stirling and Bell numbers of the second kind.
For this problem we also supply the exponential generating
functions which are connected with the exponentials of the
operators in question. In each case we provide the coherent state
matrix elements of the boson expressions.
\\
Recalling the current state of the knowledge in the field we
should also mention the approach based on Lie-group
methodology \cite{Wilcox}. It proves useful for normal ordering
problems for the exponentials of expressions which are quadratic
in boson operators \cite{MehtaI}\cite{MehtaII}.
\\
In a series of papers by various authors
\cite{Witschel1975}\cite{Mikhailov1983}\cite{Mikhailov1985}\cite{Katriel1983}
some effort  in extending these results to operators having the
specific form $(a^\dag a+(a^\dag)^r)^n$ was made. In Chapter
\ref{Sheffer} we systematically extend this class to operators of the form $(q(a^\dag) a+v(a^\dag))^n$ and $e^{\lambda (q(a^\dag) a+v(a^\dag))}$
where $q(x)$ and $v(x)$ are arbitrary functions. This is done by
the use of umbral calculus methods \cite{Roman} in finding
representations of the monomiality principle ({\it i.e.}
representations of the Heisenberg-Weyl algebra in the space of
polynomials) and application of the coherent state methodology.
Moreover, we establish a one-to-one connection between this class of
normal ordering problems and the family of Sheffer-type
polynomials.
\\
These two classes of problems extend the current state of
knowledge on the subject in two different directions in a quite
systematic way. We believe this to be a considerable push forward in
the normal ordering problem. We also emphasize the use of
combinatorial methods which we give explicitly and which prove to be very
efficient in this kind of analysis.
\\
We illustrate this approach by examples taken both
from combinatorics and physics ({\it e.g.} for the hamiltonian of a
generalized Kerr medium). Moreover we also comment on applications
and extensions of the formalism. We give just a few of them in
Chapter \ref{Miscellany}. First we observe how to extend this
formalism to the case of deformed commutation relations. In
general it can be done at the expense of introducing 
operator-valued Stirling numbers. 
Next, we use the fact that the use of the
Dobi\'nski relations allow us to find the solutions to the Stieltjes
moment problem for the numbers arising in the normal ordering
procedure. We use these solutions to construct and analyze new families of
generalized coherent states \cite{Klauder}. We end by commenting
on the specific form of the formulas obtained in Chapter
\ref{Sheffer} which may be used to derive the explicit action of generalized
shift operators, called  the substitution theorem.  These
three examples already show that the methods we  use in solving normal
ordering problems lead to many diverse applications. At the end
this text we append complete list of publications
which indicate some other developments on the subject.

%%%%%%%%%%%%%%%%%%%%%%%%%%%%%%%%%%%%%%%%%%%%%%%%%%%%%%%%%%%%%%%%%%%%%%%%%%%%%%%%%%

\chapter{Preliminaries}
\label{Preliminaries}

\begin{chapterabstract}
\rule[1mm]{\textwidth}{.3pt}
In this chapter we give basic notions exploited later on in the
text. We start by fixing some conventions. Next we recall the
occupation number representation along with the creation $a^\dag$
and annihilation $a$ operator formalism. This serves to define and
comment on the normal ordering problem for boson operators. We end
by pointing out a particular representation of the above in terms
of the multiplication and derivative operators.
\\
\rule[1mm]{\textwidth}{.3pt}
\end{chapterabstract}

\section{Conventions}

In the mathematical literature there is always a certain freedom
in making basic definitions. Sometimes it is confusing, though.
For that reason it is reasonable to establish  explicitly some
conventions in the beginning.
\\
In the following by an indeterminate we primarily mean a formal
variable in the context of formal power series (see Section
\ref{Umbral}). It may be thought as a real or complex
number whenever the analytic properties are assured.
\\
We frequently make use of summation and the product operations. We
give the following conventions concerning their limits
\begin{eqnarray}\nonumber
\sum_{n=N_0}^N...=0\ \ \ \ \ \ \ \ \text{and} \ \ \ \ \ \ \ \ \prod_{n=N_0}^N...=1\ \ \ \ \ \ \ \ \ \ \ \  \text{for}\ \ N<N_0.
\end{eqnarray}
Also the convention $0^0=1$ is applied. 
\\
Moreover we define the so called \emph{falling factorial} symbol by
\begin{eqnarray}\nonumber
    x^{\underline {k}}=x\cdot(x-1)\cdot...\cdot(x-k+1),\ \ \ \ \ \ \ \ \ \ x^{\underline {0}}=1,
\end{eqnarray}
for nonnegative integer $k$ and indeterminate $x$.
\\
By the {\em ceiling} function $\lceil x\rceil$, for $x$ real, we
mean the nearest integer greater or equal to $x$.

\section{Occupation number representation: Boson operators}
\label{Boson operators}

We consider a pair of one mode boson annihilation $a$ and creation
$a^\dag$ operators satisfying the commutation relation
\begin{eqnarray}\label{HW}
[a,a^\dag]=1.
\end{eqnarray}
Together with the identity operator the generators $\{a,a^\dag,1\}$
constitute the
\emph{Heisenberg-Weyl} algebra.
\\
The \emph{occupation number representation} arises from the
interpretation of $a$ and $a^\dag$ as operators annihilating and
creating a particle (object) in a system. From this point of view
the Hilbert space $\mathcal{H}$ of states (sometimes called Fock
space) is generated by the \emph{number states} $|n\rangle$, where
$n=0,1,2,...$ count the number of particles (for bosons up to
infinity). We assume here the existence of a unique \emph{vacuum}
state $|0\rangle$ such that
\begin{eqnarray}\label{a0}
    a|0\rangle=0.
\end{eqnarray}
Then the number states $\{|n\rangle\}_{n=0}^\infty$ may be taken
as an orthonormal basis in $\mathcal{H}$, {\it i.e.}
\begin{eqnarray}
    \langle n|k\rangle=\delta_{n,k}
\end{eqnarray}
and
\begin{eqnarray}
    \sum_{n=0}^\infty |n\rangle\langle n|=1.
\end{eqnarray}
The last operator equality is called a \emph{resolution of unity}
which is equivalent to
the completeness property.\\
It may be deduced from Eqs.(\ref{HW}) and (\ref{a0}) that
operators $a$ and $a^\dag$ act on the number states as
\begin{eqnarray}\label{an}
\begin{array}{rcr}
a\ |n\rangle &=&\sqrt{n}\ |n-1\rangle,\\
a^\dag|n\rangle &=&\sqrt{n+1}\ |n+1\rangle.
\end{array}
\end{eqnarray}
(Indetermined phase factors may be incorporated into the states.)
\\
Then all states may be created from the vacuum through
\begin{eqnarray}\label{n0}
    |n\rangle=\frac{1}{\sqrt{n!}}(a^\dag)^n|0\rangle.
\end{eqnarray}
It  also follows that the \emph{number operator} $N$ counting the
number of particles in a system and  defined by
\begin{eqnarray}
    N|n\rangle=n|n\rangle
\end{eqnarray}
may be represented as $N=a^\dag a$ and satisfies the the following
commutation relations
\begin{eqnarray}\label{NN}
\begin{array}{l}
{[a,N]}=a,\\
{[a^\dag,N]}=-a^\dag.
\end{array}
\end{eqnarray}
This construction may be easily extended to the multi-boson case.
\\
The canonical form of the commutator Eq.(\ref{HW}) originates from
the study of a quantum particle in the harmonic oscillator
potential and quantization of the electromagnetic field. Any
standard textbook on Quantum Mechanics may serve to survey of
these topics.
\\
We note that the commutation relations of Eqs.(\ref{HW}) and
(\ref{NN}) may be easily extended to the deformed case
\cite{SolomonPLA} (see also Section \ref{Deformed
bosons}).

\section{Normal ordering}

The boson creation $a^\dag$ and annihilation $a$ operators
considered in previous section do not commute. This is the reason
for some ambiguities in the definitions of the operator functions
in Quantum Mechanics. To solve this problem one has to
additionally define the order of the operators involved.
\\
These difficulties with  operator ordering led to the definition
of the \emph{normally ordered} form of the boson operator in which
all the creation operators $a^\dag$ stand to the left of the
annihilation operators $a$.
\\
There are two well defined procedures on the boson expressions
yielding a normally ordered form. Namely, the \emph{normal
ordering} $\mathcal{N}$ and the \emph{double dot} $:\ :$
operations.
\\
By normal ordering of a general function $F(a,a^{\dag})$ we mean
${\cal N}\left[F(a,a^{\dag})\right]$
 which is
obtained by moving all the annihilation operators $a$ to the right
using the commutation relation of Eq.(\ref{HW}). We stress the
fact that after this  normal ordering procedure, the  operator
remains the same ${\cal N}\left[F(a,a^{\dag})\right]=
F(a,a^{\dag})$. It is only its {\it functional} representation
which changes.
\\
On the contrary the double dot operation operation
$:\!F(a,a^{\dag})\!:$ means the same procedure but {\it without}
taking into account the commutation relation of Eq.(\ref{HW}),
{\it i.e.} moving all annihilation operators $a$ to the right as
if they commute with the creation operators $a^\dag$. We emphasize
that in general $F(a,a^{\dag})\neq\ :\!F(a,a^{\dag})\!:$ . The
equality holds only for operators which are already in  normal
form ({\it e.g.} ${\cal N}\left[F(a,a^{\dag})\right]=\ :{\cal
N}\left[F(a,a^{\dag})\right]:$ ).
\\
Using these two operations we say that the \emph{normal ordering
problem} for $F(a,a^{\dag})$ is solved if we are able to find an
operator $G(a,a^{\dag})$ for which the following equality is
satisfied
\begin{eqnarray}
    F(a,a^{\dag}) ={\cal N}\left[F(a,a^{\dag})\right]\equiv\ :G(a,a^{\dag}):\ .
\end{eqnarray}
This normally ordered form is especially useful in the coherent
state representation  widely used in quantum optics (see Appendix
\ref{Coherent states}). Also calculation of the vacuum expectation
values in quantum field theory is immediate whenever this form is
known.
\\
Here is an example of the above ordering procedures
\begin{eqnarray}\nonumber
\begin{array}{rcc}
aa^\dag aaa^\dag a&\xrightarrow[\ \ \ \ \ {[a,a^\dag]=1}\ \ \ \ \
]{\cal{N}}&
\underbrace{(a^\dag)^2 a^4+4\ a^\dag a^3+2\ a^2}\vspace{2mm}\\
&&\begin{footnotesize}{ a^\dag\ \text{- to the left}\ \ \ a\ \text{- to the right}}\end{footnotesize}\vspace{2mm}\\
aa^\dag aaa^\dag
a&\xrightarrow[\begin{scriptsize}\begin{array}{c}a,a^\dag\ \text{-
commute}\\\text{(like numbers)}\end{array}\end{scriptsize}]{:\ \
:}&\overbrace{a^\dag a^\dag aaa}
\end{array}
\end{eqnarray}
Another simple illustration is the ordering of the product
$a^k(a^\dag)^l$ which is in the so called \emph{anti-normal} form
({\it i.e.} all annihilation operators stand to the left of
creation operators). The double dot operation readily gives
\begin{eqnarray}\nonumber
    :a^k(a^\dag)^l:\ =(a^\dag)^la^k,
\end{eqnarray}
while the normal ordering procedure $\cal{N}$ requires some
exercise in the use of Eq.(\ref{HW}) yielding (proof by induction)
\begin{eqnarray}\label{aa}
a^k(a^\dag)^l={\cal
N}\left[a^k(a^\dag)^l\right]\equiv\sum_{p=0}^k\binom{k}{p}l^{\underline{p}}\
(a^\dag)^{l-p}a^{k-p}.
\end{eqnarray}
These examples explicitly show that these two procedures furnish
completely different results (except for the operators which are
already in  normal form).
\\
There is also a 'practical' difference in their use. That is while
the application of the double dot operation $:\ :$ is almost immediate,
for the normal ordering procedure $\cal{N}$ certain skill in
commuting operators $a$ and $a^\dag$ is needed.
\\
A standard approach to the problem is by the \emph{Wick theorem}.
It reduces the normal ordering procedure $\cal{N}$ to the double
dot operation on the sum over all possible contractions ({\em
contraction} means removal of a pair of annihilation and creation
operators in the expression such that $a$ precedes $a^\dag$). Here
is an example
\begin{eqnarray}\nonumber
    aa^\dag aaa^\dag aaa^\dag&=&\sum\ :\{\ \text{all contractions}\ \}:\\\nonumber\\\nonumber
    &=& :aa^\dag aaa^\dag aaa^\dag:\\\nonumber
    && +:aa^\dag aaa^\dag a\not \!a\not \!a^\dag+aa^\dag aaa^\dag\not \!aa\not \!a^\dag+aa^\dag a\not \!aa^\dag aa\not \!a^\dag+\\\nonumber
    &&\ \ \ \ \ aa^\dag\not \!aaa^\dag aa\not \!a^\dag+\not \!aa^\dag aaa^\dag aa\not \!a^\dag+aa^\dag a\not \!a\not \!a^\dag aaa^\dag+\\\nonumber
    &&\ \ \ \ \ aa^\dag\not \!aa\not \!a^\dag aaa^\dag+\not \!aa^\dag aa\not \!a^\dag aaa^\dag+\not \!a\not \!a^\dag aaa^\dag aaa^\dag:\\\nonumber
    && +:aa^\dag a\not \!a\not \!a^\dag a\not \!a\not \!a^\dag+aa^\dag\not \!aa\not \!a^\dag a\not \!a\not \!a^\dag+\not \!aa^\dag aa\not \!a^\dag a\not \!a\not \!a^\dag+\\\nonumber
    &&\ \ \ \ \ \not \!a\not \!a^\dag aaa^\dag a\not \!a\not \!a^\dag+aa^\dag a\not \!a\not \!a^\dag\not \!aa\not \!a^\dag+aa^\dag\not \!aa\not \!a^\dag\not \!aa\not \!a^\dag+\\\nonumber
    &&\ \ \ \ \ \not \!aa^\dag aa\not \!a^\dag\not \!aa\not \!a^\dag+\not \!a\not \!a^\dag aaa^\dag\not \!aa\not \!a^\dag+aa^\dag\not \!a\not \!a\not \!a^\dag aa\not \!a^\dag+\\\nonumber
    &&\ \ \ \ \ \not \!aa^\dag a\not \!a\not \!a^\dag aa\not \!a^\dag+\not \!a\not \!a^\dag a\not \!aa^\dag aa\not \!a^\dag+aa^\dag\not \!a\not \!a\not \!a^\dag aa\not \!a^\dag+\\\nonumber
    &&\ \ \ \ \ \not \!aa^\dag\not \!aa\not \!a^\dag aa\not \!a^\dag+
    \not \!a\not \!a^\dag\not \!aaa^\dag aa\not \!a^\dag+
    \not \!aa^\dag a\not \!a\not \!a^\dag aa\not \!a^\dag+\\\nonumber
    &&\ \ \ \ \
    \not \!aa^\dag\not \!aa\not \!a^\dag aa\not \!a^\dag+
    \not \!a\not \!a^\dag a\not \!a\not \!a^\dag aaa^\dag+
    \not \!a\not \!a^\dag\not \!aa\not \!a^\dag aaa^\dag:\\\nonumber
    && +:\ \not \!a\not \!a^\dag a\not \!a\not \!a^\dag a\not \!a\not \!a^\dag+\not \!a\not \!a^\dag\not \!aa\not \!a^\dag a\not \!a\not \!a^\dag+\not \!a\not \!a^\dag a\not \!a\not \!a^\dag\not \!aa\not \!a^\dag+\\\nonumber
    &&\ \ \ \ \ \not \!a\not \!a^\dag\not \!aa\not \!a^\dag\not \!aa\not \!a^\dag+\not \!a\not \!a^\dag\not \!a\not \!a\not \!a^\dag aa\not \!a^\dag+\not \!a\not \!a^\dag\not \!a\not \!a\not \!a^\dag aa\not \!a^\dag:\\\nonumber\\\nonumber
    &=&(a^\dag)^3 a^5+9\ (a^\dag)^2 a^4+18\ a^\dag a^3+6\ a^2.
\end{eqnarray}
One  can easily see that the number of contractions may be quite
big. This difficulty for polynomial expressions may be overcome by
using modern computer algebra systems. Nevertheless, for
nontrivial functions (having infinite expansions) the problem
remains open. Also it does not provide the analytic formulas for
the coefficients of the normally ordered terms in the final
expression. A systematic treatment of a large class of such
problems is the subject of this work.
\\
At the  end of this Section we recall some formulas connected with
 operator reordering. The first one is the exponential mapping
formula, sometimes called the Hausdorff transform, which for any
well defined function $F(a,a^\dag)$ yields
\begin{eqnarray}\label{eMe}
\begin{array}{rcl}
    e^{xa}F(a,a^\dag)e^{-xa}&=&F(a,a^\dag+x),\vspace{2mm}\\
    e^{-xa^\dag}F(a,a^\dag)e^{xa^\dag}&=&F(a+x,a^\dag).
\end{array}
\end{eqnarray}
It can be used to derive the following commutators
\begin{eqnarray}\label{af}
\begin{array}{rcl}
{[a,F(a,a^\dag)]}&=&\frac{\displaystyle \partial}{\displaystyle\partial a^\dag}\ F(a,a^\dag),\vspace{2mm}\\
{[a^\dag,F(a,a^\dag)]}&=&-\frac{\displaystyle
\partial}{\displaystyle \partial a}\ F(a,a^\dag).
\end{array}
\end{eqnarray}
The proofs may be found in any book on Quantum Mechanics, {\it
e.g.} \cite{Louisell}.
\\
Also a well known property of the Heisenberg-Weyl algebra of
Eq.(\ref{HW}) is a disentangling formula
\begin{eqnarray}\label{BCH}
    e^{\lambda(a+a^\dag)}=e^{\lambda^2/2}e^{\lambda a^\dag}e^{\lambda a}=
    e^{-\lambda^2/2}e^{\lambda a}e^{\lambda a^\dag},
\end{eqnarray}
which may serve as an example of the normal ordering procedure
\begin{eqnarray}
    e^{\lambda(a+a^\dag)}={\cal{N}}\left[e^{\lambda(a+a^\dag)}\right]
    \equiv\ e^{\lambda^2/2}\ :e^{\lambda (a+a^\dag)}:\ .
\end{eqnarray}
This type  of expressions exploits the Lie structure of the
algebra and uses a simplified form of Baker-Campbell-Hausdorff
formula. For this and other disentangling properties of the
exponential operators, see
\cite{Wilcox}\cite{Witschel1975}\cite{Mufti}\cite{DasGupta}.
\\
Finally,  we must mention  other ordering procedures also used  in
physics, like the anti-normal or the Weyl (symmetric) ordered
form. We note that there exist translation formulas between these
expressions, see {\it e.g.}
\cite{Glauber1969}\cite{Shalitin}.

\section{$X$ and $D$ representation}
\label{DXRep}

The choice of the representation of the algebra may be used  to
simplify the calculations. In the following we apply this,
although, we note that with some  effort one could also manage
using solely the operator properties of  $a$ and $a^\dag$.
\\
There are some common choices of the Heisenberg-Weyl algebra
representation in Quantum Mechanics. One may take {\it e.g.} a
pair of hermitian operators $X$ and $P=i\frac{d}{dx}$ acting on
the dense subspace of  square integrable functions or the infinite
matrix representation of Eq.(\ref{an}) defined in the Fock space.
This choice is connected with a particular interpretation
intimately connected with the quantum mechanical problem to be
solved.
\\
Here we choose the simplest possible representation of the
commutation relation of Eq.(\ref{HW}) which acts in the space of
(formal) polynomials. We make the identification
\begin{eqnarray}
\begin{array}{ccc}
a^\dag&\longleftrightarrow& X\\
a&\longleftrightarrow& D,
\end{array}
\end{eqnarray}
where $X$ and $D$ are formal multiplication and derivative
($D=\frac{d}{dx}$) operators, respectively. They are defined by
their action on monomials
\begin{eqnarray}\label{DXMono}
\begin{array}{lcl}
X x^n&=&x^{n+1},\\
D x^n&=&nx^{n-1}.
\end{array}
\end{eqnarray}
(It also defines the action of these operators on the formal power
series.)
\\
Note that the commutation relation of Eq. (\ref{HW}) remains the
same, {\it i.e.}
\begin{eqnarray}\label{DX}
[D,X]=1.
\end{eqnarray}
For more details see Appendix \ref{Umbral} and Chapter
\ref{Sheffer}.
\\
This choice of representation is most appropriate,  as we note
that the problem we shall be concerned with has a purely algebraic
background. We are interested in the reordering of operators, and
that  only depends on the algebraic properties of the commutator
of Eqs.(\ref{HW}) or (\ref{DX}). We emphasize that the conjugacy
property of the operators or the scalar product do not play any
role in that problem. To get rid of these unnecessary
constructions we choose the representation of the commutator of
Eq.(\ref{HW}) in the space of polynomials defined by
Eq.(\ref{DXMono}) where these properties do not play primary role,
however possible to implement 
(see the Bargmann-Segal representation \cite{Bargmann}\cite{Segal}). 
We benefit from this choice by the
resulting increased simplicity of the  calculations.

%%%%%%%%%%%%%%%%%%%%%%%%%%%%%%%%%%%%%%%%%%%%%%%%%%%%%%%%%%%%%%%%%%%%%%%%%%%%%%%%%%%%%%

\chapter{Generic example. Stirling and Bell numbers.}
\label{Generic}

\begin{chapterabstract}
\rule[1mm]{\textwidth}{.3pt}
We define Stirling and Bell numbers as solutions to the normal
ordering problem for powers and the exponential of the number
operator $N=a^\dag a$ , where $a$ and $a^\dag$ are the annihilation
and creation operators. All their combinatorial properties are
derived using only this definition. Coherent state matrix elements
of $N^n$ and $e^{\lambda N}$ are shown to be the Bell polynomials
and their exponential generating function, respectively.
\\
\rule[1mm]{\textwidth}{.3pt}
\end{chapterabstract}

This Chapter may be thought of as a simple introduction to some
methods and problems encountered later on in the text. We give a
simple example on which all the essential techniques of the
proceeding sections are used. Although Stirling and Bell numbers
have a well established purely combinatorial origin
\cite{Comtet}\cite{Riordan}\cite{Wilf}\cite{Knuth} we choose here a more physical approach.
We define and investigate Stirling and Bell numbers as solutions
to the normal ordering problem
\cite{Katriel}\cite{Katriel2000}\cite{Katriel2002}.

Consider the number operator $N=a^\dag a$ and search for the
normally ordered form of its $n$-th power (iteration). It can be
written as
\begin{eqnarray}\label{Snk}
\left(a^\dag a\right)^n=\sum_{k=1}^n S(n,k) (a^\dag)^k a^k
\end{eqnarray}
where the integers $S(n,k)$ are the so called \emph{Stirling
numbers}.
\\
One may also define  the so called \emph{Bell polynomials}
\begin{eqnarray}\label{B}
B(n,x)=\sum_{k=1}^n S(n,k) x^k
\end{eqnarray}
and \emph{Bell numbers}
\begin{eqnarray}\label{BB}
B(n)=B(n,1)=\sum_{k=1}^n S(n,k)
\end{eqnarray}
For convenience we apply the conventions
\begin{eqnarray}\label{Initial}
S(n,0)&=&\delta_{n,0}\ \ \ \ \ \text{and}\ \ \ \ \ \ S(n,k)=0\ \ \
\ \text{for}\ \ k>n
\end{eqnarray}
and
\begin{eqnarray}
B(0)=B(0,x)=1.
\end{eqnarray}
(It makes the calculations easier, {\it e.g.} one may ignore the
summation limits when changing their order.)
\\
In the following we are concerned with the properties of these
Stirling and Bell numbers.
\\
First we state the \emph{recurrence relation} for Stirling numbers
\begin{eqnarray}\label{Rec}
S(n+1,k)=kS(n,k)+S(n,k-1),
\end{eqnarray}
with initial conditions as in Eq.(\ref{Initial}). The proof 9by induction) can be
deduced from the equalities
\begin{eqnarray}\nonumber
&&\sum_{k=1}^{n+1}\ S(n+1,k)\ (a^\dag)^k a^k=\left(a^\dag
a\right)^{n+1}=a^\dag a\left(a^\dag a\right)^n\\\nonumber
&&=\sum_{k=1}^n S(n,k)\ a^\dag a(a^\dag)^k
a^k\stackrel{(\ref{aa})}{=}\sum_{k=1}^n S(n,k)\ (a^\dag)^k (a^\dag
a+k)a^k\\\nonumber &&=\sum_{k=2}^{n+1} S(n,k-1)\ (a^\dag)^k a^k+
\sum_{k=1}^n k\ S(n,k)\ (a^\dag)^k a^k\\\nonumber
&&\stackrel{(\ref{Initial})}{=}\sum_{k=1}^{n+1} \left(S(n,k-1)+k\ S(n,k)\right)\ (a^\dag)^k a^k.
\end{eqnarray}
We shall not make make use of this recurrence relation, preferring
the simpler analysis based on the Dobi\'nski relation. To this end
we proceed to the essential step which we shall extensively use
later on, {\it i.e.} \emph{change of representation}.
Eq.(\ref{Snk}) rewritten in the $X$ and $D$ representation (see
Section \ref{DXRep}) takes the form
\begin{eqnarray}\label{SXD}
\left(X D\right)^n=\sum_{k=1}^n S(n,k)\ X^k D^k.
\end{eqnarray}
With this trick we shall obtain  all the properties of Stirling
and Bell numbers.
\\
We first  act with this equation on the monomial $x^m$. This gives
for $m$ integer
\begin{eqnarray}\nonumber
m^n=\sum_{k=1}^n S(n,k) m^{\underline{k}},
\end{eqnarray}
where $x^{\underline {k}}=x\cdot(x-1)\cdot...\cdot(x-k+1)$ is the
\emph{falling factorial} ($x^{\underline {0}}=1$). Observing
that a (non zero) polynomial can have only a finite set of zeros
justifies the generalization
\begin{eqnarray}
x^n=\sum_{k=1}^n S(n,k) x^{\underline{k}}.
\end{eqnarray}
This equation can be interpreted as a change of basis in the space
of polynomials. This gives an interpretation of Stirling numbers
as the \emph{connection coefficients} between two bases
$\{x^n\}_{n=0}^\infty$ and $\{x^{\underline{n}}\}_{n=0}^\infty$.
\\
Now  we act with Eq.(\ref{SXD}) on the exponential function $e^x$.
We obtain
\begin{eqnarray}\nonumber
\sum_{k=0}^\infty k^n\frac{x^k}{k!}=e^x\sum_{k=1}^n S(n,k) x^k.
\end{eqnarray}
Recalling the definition of the Bell polynomials Eq.(\ref{B}) we
get
\begin{eqnarray}\label{D}
B(n,x)=e^{-x}\sum_{k=0}^\infty k^n\frac{x^k}{k!},
\end{eqnarray}
which is a celebrated \textit{Dobi\'nski formula} \cite{Wilf}. It is usually
stated for Bell numbers in the form
\begin{eqnarray}\label{DD}
B(n)=\frac{1}{e}\sum_{k=1}^\infty \frac{k^n}{k!}.
\end{eqnarray}
Note that both series are convergent by the d'Alembert criterion.
\\
The most striking property of the Dobi\'nski formula is the fact
that integer numbers $B(n)$ or polynomials $B(n,x)$ can be
represented as nontrivial infinite sums. Here we only mention that
this remarkable property provides also  solutions to the moment
problem (see Section \ref{Generalized coherent states})
\\
In the following we shall exploit the Dobi\'nski formula
Eq.(\ref{D}) to investigate further properties of  Stirling and
Bell numbers.
\\
Applying the Cauchy multiplication rule Eq.(\ref{Cauchy}) to
Eq.(\ref{D}) and comparing coefficients we obtain \emph{explicit
expression} for $S(n,k)$
\begin{eqnarray}\label{Scan}
S(n,k)=\frac{1}{k!}\sum_{j=1}^k\binom{k}{j}(-1)^{k-j}j^n.
\end{eqnarray}
\\
Next, we define the \textit{exponential generating function} of
the polynomials $B(n,x)$ (see Appendix \ref{Umbral}) as
\begin{eqnarray}\label{Gen}
G(\lambda,x)=\sum_{n=0}^\infty B(n,x)\frac{\lambda^n}{n!},
\end{eqnarray}
which contains all the information about the Bell polynomials.
Substituting Eq.(\ref{D}) into Eq.(\ref{Gen}), changing the
summation order and then identifying the expansions of the
exponential functions we obtain
\begin{eqnarray}\nonumber
G(\lambda,x)&=&e^{-x}\sum_{n=0}^\infty \sum_{k=0}^\infty
k^n\frac{x^k}{k!}\frac{\lambda^n}{n!} =e^{-x} \sum_{k=0}^\infty
\frac{x^k}{k!}\sum_{n=0}^\infty k^n\frac{\lambda^n}{n!}\\\nonumber
&=&e^{-x} \sum_{k=0}^\infty \frac{x^k}{k!}e^{\lambda k}=e^{-x}
e^{xe^\lambda}.
\end{eqnarray}
Finally we may write the exponential generating function
$G(\lambda,x)$ in the compact form
\begin{eqnarray}\label{GFunct}
G(\lambda,x)=e^{x(e^\lambda-1)}.
\end{eqnarray}
Sometimes in applications the following exponential generating
function of the Stirling numbers $S(n,k)$ is used
\begin{eqnarray}\label{GenS}
\sum_{n=k}^\infty S(n,k)\frac{\lambda^n}{n!}=\frac{(e^\lambda-1)^k}{k!}.
\end{eqnarray}
It can be derived by comparing expansions in $x$ of
Eqs.(\ref{Gen}) and (\ref{GFunct})
\begin{eqnarray}\nonumber
G(\lambda,x)&=&1+\sum_{n=1}^\infty \left(\sum_{k=1}^n S(n,k)
x^k\right)\frac{\lambda^n}{n!}\\\nonumber &=&1+\sum_{k=1}^\infty
\left(\sum_{n=k}^\infty S(n,k)
\frac{\lambda^n}{n!}\right)x^k,\\\nonumber
G(\lambda,x)&=&e^{x(e^\lambda-1)}=1+\sum_{k=1}^\infty
\frac{(e^\lambda-1)^k}{k!}x^k.
\end{eqnarray}
Bell polynomials share an interesting property called the
\emph{Sheffer identity} (note resemblance to the binomial
identity)
\begin{eqnarray}
    B(n,x+y)=\sum_{k=0}^n\binom{n}{k}B(k,y)B(n-k,x)
\end{eqnarray}
It is the consequence of the following equalities
\begin{eqnarray}\nonumber
\sum_{n=0}^\infty B(n,x+y)\frac{\lambda^n}{n!}&=&e^{(x+y)(e^\lambda-1)}=e^{x(e^\lambda-1)}e^{y(e^\lambda-1)}\\\nonumber
&=&\sum_{n=0}^\infty
B(n,x)\frac{\lambda^n}{n!}\cdot\sum_{n=0}^\infty
B(n,y)\frac{\lambda^n}{n!}\\\nonumber
&\stackrel{(\ref{Cauchy})}{=}&\sum_{n=0}^\infty
\left(\sum_{k=0}^n\binom{n}{k}B(k,y)B(n-k,x)\right)\frac{\lambda^n}{n!}.
\end{eqnarray}
By this identity and also by the characteristic exponential
generating function of Eq.(\ref{GFunct}) Bell polynomials $B(n,x)$
are found to be of Sheffer-type (see Appendix \ref{Umbral}).
\\
Differentiating the exponential generating function of
Eqs.(\ref{Gen}) and (\ref{GFunct}) we have (see Eq.(\ref{DF}))
\begin{eqnarray}\nonumber
    \sum_{n=0}^\infty B(n+1,x)\frac{\lambda^n}{n!}=\frac{\partial}{\partial\lambda} G(\lambda,x)=e^\lambda e^{x(e^\lambda-1)}
    =e^\lambda\sum_{n=0}^\infty B(n,x)\frac{\lambda^n}{n!}
\end{eqnarray}
which by the Cauchy product rule Eq.(\ref{Cauchy}) yields the
\emph{recurrence relation} for the Bell polynomials
\begin{eqnarray}
    B(n+1,x)=\sum_{k=0}^n\binom{n}{k}B(k,x).
\end{eqnarray}
And consequently for the Bell numbers we have
$B(n+1)=\sum_{k=0}^n\binom{n}{k}B(k)$.
\\
By the same token applied to the exponential generating function
of Stirling numbers $S(n,k)$ of Eq.(\ref{GenS}) the recurrence
relation of Eq.(\ref{Rec}) can be also derived.
\\
Finally, using any of the derived properties of Stirling or Bell numbers
one can easily calculate them explicitly. Here are some of them

\begin{eqnarray}
    S(n,k),\ \ \ \ \ \ 1\leq k\leq n\qquad\qquad\qquad\qquad B(n)\ \ \ &&\nonumber\\
    \begin{array}{cl|lllllllllllcc}\cline{3-14}&&&&&&\\
    n=1&&&1&&&&&&&&&&1&\\
        n=2&&&1&1&&&&&&&&&2&\\
        n=3&&&1&3&1&&&&&&&&5&\\
        n=4&&&1&7&6&1&&&&&&&15&\\
        n=5&&&1&15&25&10&1&&&&&&52&\\
    n=6&&&1&31&90&65&15&1&&&&&203&\\
    n=7&&&1&63&301&350&140&21&1&&&&877\\
    n=8&&&1&127&966&1701&1050&266&28&1&&&4140\\
    ...&&&...&...&...&...&...&...&...&...&...&&...
    \end{array}&&\nonumber
\end{eqnarray}
\\

Now we come back to normal ordering. Using the properties of
coherent states (see Appendix \ref{Coherent states}) we conclude
from Eqs.(\ref{Snk}) and (\ref{B}) that diagonal coherent state
matrix elements generate Bell polynomials \cite{Katriel2000}
\begin{eqnarray}
\langle z|(a^\dag a)^n|z\rangle=B(n,|z|^2).
\end{eqnarray}
Moreover, expanding the exponential $e^{\lambda a^\dag a}$ and
taking the diagonal coherent state matrix element we have
\begin{eqnarray}\nonumber
\langle z|e^{\lambda a^\dag a}|z\rangle
=\sum_{n=0}^\infty\langle z|(a^\dag
a)^n|z\rangle\frac{\lambda^n}{n!} =\sum_{n=0}^\infty
B(n,|z|^2)\frac{\lambda^n}{n!}.
\end{eqnarray}
We see that the diagonal coherent state matrix elements of
$e^{\lambda a^\dag a}$ yield the exponential generating function
of the Bell polynomials (see Eqs.(\ref{Gen}) and (\ref{GFunct}))
\begin{eqnarray}
\langle z|e^{\lambda a^\dag a}|z\rangle=e^{|z|^2(e^\lambda-1)}.
\end{eqnarray}
This result allows us to read off the normally ordered form of
$e^{\lambda a^\dag a}$ (see Appendix \ref{Coherent states})
\begin{eqnarray}\label{eaa}
e^{\lambda a^\dag a}={\cal N}\left[e^{\lambda a^\dag
a}\right]\equiv\ :e^{a^\dag a(e^\lambda-1)}\ :.
\end{eqnarray}

These are more or less all properties of Stirling and Bell
numbers. Note that they were defined here as  solutions to the
normal ordering problem for  powers of the number operator
$N=a^\dag a$. The above analysis relied firmly on this definition.
\\
On the other hand these numbers are well known in combinatorial
analysis
\cite{Comtet}\cite{Riordan}\cite{Wilf}\cite{Knuth} where are
called Stirling numbers of the second kind and are usually
explored using only recurrence relation Eq.(\ref{Rec}). Here is
their original interpretation is in terms of partitions of the
set.
\begin{itemize}
\item{Stirling numbers $S(n,k)$ count the number of ways of putting $n$ different objects
into $k$ identical containers (none left empty).}
\item{Bell numbers $B(n)$ count the number of ways of putting $n$ different objects
into $n$ identical containers (some may be left empty).}
\end{itemize}
Some other pictorial representations can be also given {\it e.g.}
in terms of graphs \cite{Mendez} or rook numbers
\cite{Navon}\cite{QTS3}\cite{Varvak}.

%%%%%%%%%%%%%%%%%%%%%%%%%%%%%%%%%%%%%%%%%%%%%%%%%%%%%%%%%%%%%%%%%%%%%%%%%%%%%%%%%%%%%%%%%%%

\chapter{Normal Ordering of Boson Expressions}
\label{Boson Expressions}

\begin{chapterabstract}
\rule[1mm]{\textwidth}{.3pt}
We solve the normal ordering problem for expressions in boson
creation $a^\dag$ and annihilation $a$ operators in the form of a
string
$(a^\dag)^{r_M}a^{s_M}\dots(a^\dag)^{r_2}a^{s_2}(a^\dag)^{r_1}a^{s_1}$.
We are especially concerned with its powers (iterations). Next we
extend the solution to iterated homogeneous boson polynomials,
{\it i.e.} powers of the operators which are sums of boson strings
of the same excess. The numbers obtained in the solutions are
generalizations of Stirling and Bell numbers. Recurrence
relations, closed-form expressions (including Dobi\'nski-type
relations) and generating formulas are derived. Normal ordering of
the exponentials of the the aforementioned operators are also
treated. Some special cases including Kerr-type hamiltonians are
analyzed in detail.
\\
\rule[1mm]{\textwidth}{.3pt}
\end{chapterabstract}

\section{Introduction}

In this chapter we consider expressions in boson creation $a^\dag$
and annihilation $a$ operators (see Section \ref{Boson
operators}). We search for the normally ordered form and show
effective ways of finding combinatorial numbers arising in that
problem. These numbers generalize Stirling and Bell numbers (see
Chapter \ref{Generic}).
\\
The first class of expressions treated in this Chapter are so
called
\textit{boson strings} or \textit{boson monomials}, see Section
\ref{General boson strings}. Here is an example
\begin{eqnarray}\nonumber
a^\dag a a^\dag a^\dag aa a^\dag a^\dag a^\dag a^\dag a^\dag
aa^\dag a^\dag aaaaaa a^\dag a a^\dag aa a^\dag a^\dag a^\dag a
a^\dag,
\end{eqnarray}
which we compactly denote as
\begin{eqnarray}\nonumber
(a^\dag)^{r_M}a^{s_M}\dots(a^\dag)^{r_2}a^{s_2}(a^\dag)^{r_1}a^{s_1}
\end{eqnarray}
with some nonnegative integers $r_m$ and $s_m$. By an
\textit{excess} of a string we call the difference between the
number of creation and annihilation operators, {\it i.e.}
$d=\sum_{m=1}^M (r_m-s_m)$. In the following we assume that the
boson strings are of nonnegative excess (for negative excess see
Section \ref{NegExcess}).
\\
Boson strings (monomials) can be extended further to polynomials.
We treat \textit{homogeneous boson polynomials} which are
combinations of boson monomials of the same excess (possibly with
some coefficients). Here is an example
\begin{eqnarray}\nonumber
a a a^\dag aa^\dag a^\dag aa a^\dag a^\dag a^\dag+a^\dag a a^\dag
a^\dag a + a^\dag,
\end{eqnarray}
which using methods of Section \ref{General boson strings} can be
written in the form
\begin{eqnarray}\nonumber
(a^\dag)^d\ \sum_{k=N_0}^N\alpha_k\ (a^\dag)^k a^k,
\end{eqnarray}
with appropriate $d$, $N$, $N_0$ and $\alpha_k$'s.
\\
Next, in Sections \ref{HomPol} and \ref{IterBosonStr} we give
recipes on how to approach the problem of normal ordering of
\textit{iterated} boson strings and homogeneous boson polynomials, {\it i.e.} their $n$-th powers, like
\begin{eqnarray}\nonumber
\left(a^\dag a a^\dag a^\dag aa a^\dag a^\dag a^\dag a^\dag a^\dag aa^\dag a^\dag
aaaaaa a^\dag a a^\dag aa a^\dag a^\dag a^\dag a a^\dag\right)^n
\end{eqnarray}
and
\begin{eqnarray}\nonumber
\left(a a a^\dag aa^\dag a^\dag aa a^\dag a^\dag a^\dag+a^\dag a a^\dag a^\dag a + a^\dag\right)^n.
\end{eqnarray}
These considerations serve to calculate in Section \ref{GenFunct}
the \textit{exponential generating functions} of generalized
Stirling and Bell numbers. This in turn gives the solution to the
normal ordering problem of the exponential of the aforementioned
boson operators.
\\
In Section \ref{NegExcess} we comment on  operators with negative
excess.
\\
Finally in the last Section \ref{Special} we work out some
examples which include  generalized Kerr-type hamiltonian.
\\
We refer to
\cite{Mendez}\cite{BlasiakJMP2005}\cite{QTS3}\cite{Varvak} for
interpretation of considered combinatorial structures in terms of
graphs and rook polynomials.

\section{General boson strings}
\label{General boson strings}

In this section we define the generalization of ordinary Bell and
Stirling numbers which arise in the solution of the general normal
ordering problem for a boson string (monomial) \cite{Mendez}\cite{Witschel2005}.
Given two sequences of nonnegative integers ${\mathbf
r}=(r_1,r_2,\dots,r_M)$ and ${\mathbf s}=(s_1,s_2,\dots,s_M)$ we
define the operator
\begin{eqnarray}\label{hh}
H_{{\mathbf r},{\mathbf
s}}=(a^{\dagger})^{r_M}a^{s_M}\dots(a^{\dagger})^{r_2}a^{s_2}(a^{\dagger})^{r_1}a^{s_1}.
\end{eqnarray}
We let $S_{{\mathbf r},{\mathbf s}}(k)$ be the nonnegative
integers appearing in the normally ordered expansion
\begin{eqnarray}\label{SSS}
H_{{\mathbf r},{\mathbf
s}}=(a^{\dagger})^{d_M}\sum_{k=s_1}^{s_1+s_2+\dots+s_M}S_{{\mathbf
r},{\mathbf s}}(k)(a^{\dagger})^ka^k,
\end{eqnarray}
where $d_n=\sum_{m=1}^n(r_m-s_m)$,  $n=1,...,M$. Here we assume that
the (\emph{overall}) \emph{excess} of a string is nonnegative
$d_M\geq 0$ (for negative excess see Section \ref{NegExcess}).
\\
Observe that the r.h.s. of Eq.(\ref{SSS}) is already normally
ordered and constitutes a homogeneous boson polynomial being
the solution to the normal ordering problem. We shall provide
explicit formulas for the coefficients $S_{{\mathbf r},{\mathbf
s}}(k)$. This is a step further than the Wick's theorem which is
nonconstructive in that respect.
\\
We call $S_{{\mathbf r},{\mathbf s}}(k)$  \emph{generalized
Stirling number}. The \emph{generalized Bell polynomial} is
defined as
\begin{eqnarray}\label{bx}
B_{{\mathbf r},{\mathbf
s}}(x)=\sum_{k=s_1}^{s_1+s_2+\dots+s_M}S_{{\mathbf r},{\mathbf
s}}(k)x^k,
\end{eqnarray}
and the \emph{generalized Bell number} is the sum
\begin{eqnarray}\label{b}
B_{{\mathbf r},{\mathbf s}}=B_{{\mathbf r},{\mathbf s}}(1)=
\sum_{k=s_1}^{s_1+s_2+\dots+s_M}S_{{\mathbf r},{\mathbf s}}(k).
\end{eqnarray}
Note that sequence of numbers $S_{{\mathbf r},{\mathbf s}}(k)$ is
defined here for $s_1 \leq k \leq s_1+s_2+\dots+s_M$. Initial
terms may vanish (depending on the structure of ${\mathbf r}$ and
${\mathbf s}$), all the next are positive integers and the last
one is equal to one, $S_{{\mathbf r},{\mathbf
s}}(s_1+s_2+\dots+s_M)=1$.
\\
For convenience we apply the convention
\begin{eqnarray}\label{initial}
S_{{\mathbf r},{\mathbf s}}(k)=0\ \ \ \ \ \ \ \ \ \text{for}\ \
k<s_1\ \ \text{or}\ \ k>s_1+s_2+\dots+s_M.
\end{eqnarray}
\\
We introduce the notation ${\mathbf r}\ \uplus\
r_{M+1}=(r_1,r_2,\dots,r_M,r_{M+1})$ and ${\mathbf s}\ \uplus\
s_{M+1}=(s_1,s_2,\dots,s_M,s_{M+1})$ and state the
\emph{recurrence relation} satisfied by generalized Stirling
numbers $S_{{\mathbf r},{\mathbf s}}(k)$
\begin{eqnarray}\label{rec}
S_{{\mathbf r}\uplus r_{M+1},{\mathbf s}\uplus
s_{M+1}}(k)=\sum_{j=0}^{s_{M+1}}
\binom{s_{M+1}}{j}(d_M+k-j)^{\underline {s_{M+1}-j}}\ S_{{\mathbf r},{\mathbf
s}}(k-j).
\end{eqnarray}
\newline
One can deduce the derivation of Eq.(\ref{rec}) from the following
equalities
\begin{eqnarray}\nonumber
&&(a^\dag)^{d_{M+1}}\ \sum_{k=s_1}^{s_1+\dots+s_{M+1}}S_{{\mathbf
r}\uplus r_{M+1},{\mathbf s}\uplus s_{M+1}}(k)\ (a^\dag)^k
a^k\\\nonumber &&\stackrel{(\ref{SSS})}{=}H_{{\mathbf r}\uplus
r_{M+1},{\mathbf s}\uplus
s_{M+1}}\stackrel{(\ref{hh})}{=}(a^{\dagger})^{r_{M+1}}a^{s_{M+1}}H_{{\mathbf
r},{\mathbf s}}\\\nonumber
&&\stackrel{(\ref{SSS})}{=}(a^{\dagger})^{r_{M+1}}a^{s_{M+1}}\
(a^\dag)^{d_{M}}\ \sum_{k=s_1}^{s_1+\dots+s_{M}}S_{{\mathbf
r},{\mathbf s}}(k)\ (a^\dag)^k a^k\\\nonumber
&&=(a^{\dagger})^{r_{M+1}}\
\sum_{k=s_1}^{s_1+\dots+s_{M}}S_{{\mathbf r},{\mathbf s}}(k)\
a^{s_{M+1}}(a^\dag)^{d_M+k} a^k\\\nonumber
&&\stackrel{(\ref{aa})}{=}(a^{\dagger})^{r_{M+1}}\
\sum_{k=s_1}^{s_1+\dots+s_{M}}S_{{\mathbf r},{\mathbf s}}(k)\
\sum_{j=0}^{s_{M+1}}\binom{s_{M+1}}{j}(d_M+k)^{\underline{j}}(a^\dag)^{d_M+k-j}a^{s_{M+1}+k-j}\\\nonumber
&&=(a^{\dagger})^{d_{M+1}}\
\sum_{k=s_1}^{s_1+\dots+s_{M}}S_{{\mathbf r},{\mathbf s}}(k)\
\sum_{j=0}^{s_{M+1}}\binom{s_{M+1}}{j}(d_M+k)^{\underline{j}}(a^\dag)^{s_{M+1}+k-j}a^{s_{M+1}+k-j}\\\nonumber
&&=(a^{\dagger})^{d_{M+1}}\
\sum_{j=0}^{s_{M+1}}\binom{s_{M+1}}{j}\sum_{k=s_1+s_{M+1}-j}^{s_1+\dots+s_{M+1}-j}S_{{\mathbf
r},{\mathbf s}}(k)\
(d_M+k-s_{M+1}+j)^{\underline{j}}(a^\dag)^{k}a^{k}\\\nonumber
&&\stackrel{(\ref{rec})}{=}(a^{\dagger})^{d_{M+1}}\
\sum_{k=s_1}^{s_1\dots+s_{M}}\sum_{j=0}^{s_{M+1}}\binom{s_{M+1}}{j}S_{{\mathbf
r},{\mathbf s}}(k-s_{M+1}+j)\
(d_M+k-s_{M+1}+j)^{\underline{j}}(a^\dag)^{k}a^{k}\\\nonumber
&&=(a^{\dagger})^{d_{M+1}}\
\sum_{k=s_1}^{s_1+\dots+s_{M}}\sum_{j=0}^{s_{M+1}}\binom{s_{M+1}}{j}S_{{\mathbf
r},{\mathbf s}}(k-j)\
(d_M+k-j)^{\underline{s_{M+1}-j}}(a^\dag)^{k}a^{k}\\\nonumber
\end{eqnarray}
The problem stated in Eq.(\ref{SSS}) can be also formulated in
terms of multiplication $X$ and derivative $D$ operators (see
Section
\ref{DXRep})
\begin{eqnarray}\label{XDXD}
X^{r_M}D^{s_M}\dots X^{r_2}D^{s_2}X^{r_1}D^{s_1}=
X^{d_M}\sum_{k=s_1}^{s_1+s_2+\dots+s_M}\ S_{{\mathbf r},{\mathbf
s}}(k)X^kD^k
\end{eqnarray}
Acting with the r.h.s. of Eq.(\ref{XDXD}) on the monomial $x^l$ we
get $x^{l+d_M}\sum_{k=s_1}^{s_1+s_2+\dots+s_M}S_{{\mathbf
r},{\mathbf s}}(k)l^{\underline{k}}$. On the other hand action of
the l.h.s. on $x^l$ gives
$\prod_{m=1}^M(d_{m-1}+l)^{\underline{s_m}}x^{l+d_M}$. Equating
these two results (see Eq.(\ref{XDXD})) we have
\begin{eqnarray}\nonumber
\prod_{m=1}^M(d_{m-1}+l)^{\underline{s_m}}=\sum_{k=s_1}^{s_1+s_2+\dots+s_M}S_{{\mathbf r},{\mathbf
s}}(k)l^{\underline{k}}
\end{eqnarray}
By invoking the fact that the only polynomial with an infinite
number of zeros is the zero polynomial we justify the
generalization
\begin{eqnarray}
\prod_{m=1}^M(d_{m-1}+x)^{\underline{s_m}}=\sum_{k=s_1}^{s_1+s_2+\dots+s_M}S_{{\mathbf r},{\mathbf
s}}(k)x^{\underline{k}}.
\end{eqnarray}
This allows us to interpret the numbers $S_{{\mathbf r},{\mathbf
s}}(k)$ as the \emph{expansion coefficients} of the polynomial
$\{\prod_{m=1}^M(d_{m-1}+x)^{\underline{s_m}}\}$ in the basis
$\{x^{\underline{k}}\}_{k=0}^\infty$.
\\
Now, replacing monomials in the above considerations with the
exponential $e^{x}$ we conclude that
\begin{eqnarray}\nonumber
X^{r_M}D^{s_M}\dots X^{r_2}D^{s_2}X^{r_1}D^{s_1}\
e^x&=&x^{d_M}e^xB_{{\mathbf r},{\mathbf s}}(x)\\\nonumber
X^{d_M}\sum_{k=s_1}^{s_1+s_2+\dots+s_M}S_{{\mathbf r},{\mathbf
s}}(k)X^kD^k\ e^x&=&\sum_{k=s_1}^\infty
\left[\prod_{m=1}^M(d_{m-1}+k)^{\underline{s_m}}\right]\frac{x^{k+d_M}}{k!}
\end{eqnarray}
With these two observations together and Eq.(\ref{XDXD}) we arrive
at the
\emph{Dobi\'nski-type relation} for the generalized Bell polynomial
\begin{eqnarray}\label{dob}
B_{{\mathbf r},{\mathbf s}}(x)=e^{-x}\sum_{k=s_1}^\infty
\left[\prod_{m=1}^M(d_{m-1}+k)^{\underline{s_m}}\right]\frac{x^k}{k!}
\end{eqnarray}
Observe that the d'Alembert criterion assures the convergence of
the series.
\\
Direct multiplication of series in Eq.(\ref{dob}) using the Cauchy
rule of Eq.(\ref{Cauchy}) gives the \emph{explicit formula} for generalized Stirling
numbers $S_{{\mathbf r},{\mathbf s}}(k)$
\begin{eqnarray}\label{sss}
S_{{\mathbf r},{\mathbf
s}}(k)=\frac{1}{k!}\sum_{j=0}^k\binom{k}{j}(-1)^{k-j}\prod_{m=1}^M(d_{m-1}+j)^{\underline{s_m}}\
.
\end{eqnarray}
We finally return  to  normal ordering and observe that the
coherent state matrix element (see Appendix \ref{Coherent states})
of the boson string yields the generalized Bell polynomial (see
Eqs.(\ref{hh}) and (\ref{SSS}))
\begin{eqnarray}
    \langle z|H_{{\mathbf r},{\mathbf
s}}|z\rangle
    =(z^*)^{d_M}B_{{\mathbf r},{\mathbf s}}(|z|^2).
\end{eqnarray}

\section{Iterated homogeneous polynomials in boson operators}
\label{HomPol}

Consider an operator $H_{\boldsymbol{\alpha}}^d$ of the form
\begin{eqnarray}\label{HHH}
H_{\boldsymbol{\alpha}}^d=(a^\dag)^d\ \sum_{k=N_0}^N\alpha_k\
(a^\dag)^k a^k,
\end{eqnarray}
where the $\alpha_k$'s are constant coefficients (for convenience
we take $\alpha_{N_0}\neq 0$ and $\alpha_{N}\neq 0$) and the
nonnegative integer $d$ is the excess of the polynomial. Note that
this is a homogeneous polynomial in boson operators. It is already
in a normal form. If the starting homogeneous boson polynomial is
not normally ordered then the methods of Section
\ref{General boson strings} provide proper tools to put it in the
shape of Eq.(\ref{HHH}).
For boson polynomials of negative degree see Section \ref{NegExcess}.
\\
Suppose we want to calculate the normally ordered form of
$(H_{\boldsymbol{\alpha}}^d)^n$. It can be written as
\begin{eqnarray}\label{NHH}
\left(H_{\boldsymbol{\alpha}}^d\right)^n=(a^\dag)^{nd}\ \sum_{k=N_0}^{nN}S_{\boldsymbol{\alpha}}^d(n,k)\ (a^\dag)^k a^k,
\end{eqnarray}
with $S_{\boldsymbol{\alpha}}(n,k)$ to be determined. These
coefficients \emph{generalize Stirling numbers} (see
Eq.(\ref{Snk})). In the same manner as in Eqs.(\ref{B}) and
(\ref{BB}) we define \emph{generalized Bell polynomials}
\begin{eqnarray}
B_{\boldsymbol{\alpha}}^d(n,x)=\sum_{k=N_0}^{nN}S_{\boldsymbol{\alpha}}^d(n,k)\
x^k,
\end{eqnarray}
and \emph{generalized Bell numbers}
\begin{eqnarray}
B_{\boldsymbol{\alpha}}^d(n)=B_{\boldsymbol{\alpha}}^d(n,1)=\sum_{k=N_0}^{nN}S_{\boldsymbol{\alpha}}^d(n,k).
\end{eqnarray}
Note that $S_{{\mathbf\alpha}}^d(n,k)$,
$B_{{\mathbf\alpha}}^d(n,x)$ and $B_{{\mathbf\alpha}}^d(n)$ depend
on the set of $\alpha_k$'s defined in Eq.(\ref{HHH}). For
convenience we make the conventions
\begin{eqnarray}\label{InitialA}
\begin{array}{ccl}
S_{\boldsymbol{\alpha}}^d(0,0)&=&1,\\
S_{\boldsymbol{\alpha}}^d(n,k)&=&0\ \ \ \text{for}\ \ k>nN,\\
S_{\boldsymbol{\alpha}}^d(n,k)&=&0\ \ \ \text{for}\ \ k<N_0\ \
\text{and}\ \ n>0,
\end{array}
\end{eqnarray}
and
\begin{eqnarray}
B_{\boldsymbol{\alpha}}^d(0)=B_{\boldsymbol{\alpha}}^d(0,x)=1.
\end{eqnarray}
In the following we show how to calculate generalized Stirling and
Bell numbers explicitly. First we state the \emph{recurrence
relation}
\begin{eqnarray}\label{Recurrence}
S_{\boldsymbol{\alpha}}^d(n+1,k)=\sum_{l=N_0}^N\alpha_l\
\sum_{p=0}^l \binom{l}{p}(nd+k-l+p)^{\underline{p}}\ S_{\boldsymbol{\alpha}}^d(n,k-l+p),
\end{eqnarray}
with initial conditions as in Eq.(\ref{InitialA}).
\\
It can be deduced from the following equalities
\begin{eqnarray}\nonumber
&&(a^\dag)^{(n+1)d}\
\sum_{k=N_0}^{(n+1)N}S_{\boldsymbol{\alpha}}^d(n+1,k)\ (a^\dag)^k
a^k
\stackrel{(\ref{NHH})}{=}(H_{\boldsymbol{\alpha}}^d)^{n+1}=H_{\boldsymbol{\alpha}}(H_{\boldsymbol{\alpha}}^d)^n\\\nonumber
&&\stackrel{(\ref{NHH})}{=}(a^\dag)^d\ \sum_{l=N_0}^N\alpha_l\
(a^\dag)^l a^l\ \ (a^\dag)^{nd}\
\sum_{k=N_0}^{nN}S_{\boldsymbol{\alpha}}^d(n,k)\ (a^\dag)^k
a^k\\\nonumber
&&=\sum_{k=N_0}^{nN}S_{\boldsymbol{\alpha}}^d(n,k)\sum_{l=N_0}^N\alpha_l\
(a^\dag)^{d+l} a^l\ (a^\dag)^{nd+k}a^k\\\nonumber
&&\stackrel{(\ref{aa})}{=}\sum_{k=N_0}^{nN}S_{\boldsymbol{\alpha}}^d(n,k)\sum_{l=N_0}^N\alpha_l\
(a^\dag)^{d+l}
\left(\sum_{p=0}^l \binom{l}{p}(nd+k)^{\underline{p}}\ (a^\dag)^{nd+k-p} a^{l-p}\right)a^k\\\nonumber
&&=(a^\dag)^{(n+1)d}\
\sum_{k=N_0}^{nN}S_{\boldsymbol{\alpha}}^d(n,k)\sum_{l=N_0}^N\alpha_l\
\sum_{p=0}^l \binom{l}{p}(nd+k)^{\underline{p}}\ (a^\dag)^{l+k-p}
a^{l+k-p}\\\nonumber &&=(a^\dag)^{(n+1)d}\ \sum_{l=N_0}^N\alpha_l\
\sum_{p=0}^l
\binom{l}{p}\sum_{k=N_0+l-p}^{nN+l-p}(nd+k-l+p)^{\underline{p}}\
S_{\boldsymbol{\alpha}}^d(n,k-l+p)(a^\dag)^k a^k\\\nonumber
&&\stackrel{(\ref{InitialA})}{=}(a^\dag)^{(n+1)d}\
\sum_{k=N_0}^{nN}\sum_{l=N_0}^N\alpha_l\ \sum_{p=0}^l
\binom{l}{p}(nd+k-l+p)^{\underline{p}}\
S_{\boldsymbol{\alpha}}^d(n,k-l+p)(a^\dag)^k a^k.
\end{eqnarray}
\\
Eq.(\ref{NHH}) rewritten in the $X$ and $D$ representation takes
the form
\begin{eqnarray}\label{NHDX}
\left(H_{\boldsymbol{\alpha}}^d(D,X)\right)^n=X^{nd}\ \sum_{k=N_0}^{nN}S_{\boldsymbol{\alpha}}^d(n,k)\ X^k D^k,
\end{eqnarray}
where
\begin{eqnarray}\label{HDX}
H_{\boldsymbol{\alpha}}^d(D,X)=X^d\ \sum_{k=N_0}^N\alpha_k\ X^k
D^k.
\end{eqnarray}
We shall act with both sides of the Eq.(\ref{NHDX}) on the
monomials $x^l$. The r.h.s. yields
\begin{eqnarray}\nonumber
X^{nd}\ \sum_{k=N_0}^{nN}S_{{\boldsymbol\alpha}}^d(n,k)\ X^k D^k\
x^l=\ \sum_{k=N_0}^{nN}S_{{\boldsymbol\alpha}}^d(n,k)\
l^{\underline{k}}x^{l+nd}.
\end{eqnarray}
On the other hand iterating the  action of Eq.(\ref{HDX}) on
monomials $x^l$ we have
\begin{eqnarray}\nonumber
\left(H_{\boldsymbol{\alpha}}^d(D,X)\right)^n\ x^l=\left[\prod_{i=1}^n\sum_{k=N_0}^N\alpha_k\ (l+(i-1)d)^{\underline{k}}\right]x^{l+nd}.
\end{eqnarray}
Then when substituted in Eq.(\ref{NHDX}) we get
\begin{eqnarray}\nonumber
\prod_{i=1}^n\sum_{k=N_0}^N\alpha_k\ (l+(i-1)d)^{\underline{k}}=\ \sum_{k=N_0}^{nN}S_{{\boldsymbol\alpha}}^d(n,k)\ l^{\underline{k}}.
\end{eqnarray}
Recalling the fact that a nonzero polynomial can have only a
finite set of zeros we obtain
\begin{eqnarray}
\prod_{i=1}^n\sum_{k=N_0}^N\alpha_k\ (x+(i-1)d)^{\underline{k}}=\ \sum_{k=N_0}^{nN}S_{{\boldsymbol\alpha}}^d(n,k)\ x^{\underline{k}}.
\end{eqnarray}
This provides interpretation of the generalized Stirling numbers
$S_{{\boldsymbol\alpha}}^d(n,k)$ as the \emph{connection
coefficients} between two sets of polynomials
$\{x^{\underline{k}}\}_{k=0}^\infty$ and
$\{\prod_{i=1}^n\sum_{k=N_0}^N\alpha_k\
(x+(i-1)d)^{\underline{k}}\}_{n=0}^\infty$.
\\
Now acting with both sides of Eq.(\ref{NHDX}) on the exponential
function we get
\begin{eqnarray}\nonumber
\left(H_{\boldsymbol{\alpha}}^d(D,X)\right)^n\ e^x=e^x x^{nd}\ \sum_{k=N_0}^{nN}S_{{\boldsymbol\alpha}}^d(n,k)\ X^k=e^x x^{nd}\ B_{{\boldsymbol\alpha}}^d(n,x)\\\nonumber
\left(H_{\boldsymbol{\alpha}}^d(D,X)\right)^n\ e^x=x^{nd}\sum_{l=0}^\infty\left[\prod_{i=1}^n\sum_{k=N_0}^N\alpha_k\ (l+(i-1)d)^{\underline{k}}\right]\frac{x^l}{l!}
\end{eqnarray}
Taking these equations together we arrive at the
\emph{Dobi\'nski-type relation}
\begin{eqnarray}\label{HDob}
B_{{\boldsymbol\alpha}}^d(n,x)=e^{-x}\sum_{l=0}^\infty\left[\prod_{i=1}^n\sum_{k=N_0}^N\alpha_k\
(l+(i-1)d)^{\underline{k}}\right]\frac{x^l}{l!}.
\end{eqnarray}
Note that by the d'Alembert criterion this series is convergent.
\\
Now multiplying the series in Eq.(\ref{HDob}) and using the Cauchy
multiplication rule of Eq.(\ref{Cauchy}) we get the explicit
expression for generalized Stirling numbers
\begin{eqnarray}\label{Sa}
S_{{\boldsymbol\alpha}}^d(n,k)=\frac{1}{k!}\sum_{j=0}^k\binom{k}{j}(-1)^{k-j}\prod_{i=1}^n\sum_{l=N_0}^N\alpha_l\
(j+(i-1)d)^{\underline{l}}.
\end{eqnarray}
\\
We end this section observing that the diagonal coherent state
matrix elements (see Appendix \ref{Coherent states}) of
$\left(H_{\boldsymbol{\alpha}}\right)^n$ generate generalized Bell
polynomials (see Eq.(\ref{NHH}))
\begin{eqnarray}\label{zH}
\langle z|\left(H_{\boldsymbol{\alpha}}^d\right)^n|z\rangle=(z^*)^{nd}\ B_{{\boldsymbol\alpha}}^d(n,|z|^2).
\end{eqnarray}
Other properties of generalized Stirling and Bell numbers will be
investigated in subsequent Sections.

\section{Iterated boson string}
\label{IterBosonStr}

In Section \ref{General boson strings} it was indicated that every
boson string can be put in  normally ordered form which is a
homogeneous polynomial in $a$ and $a^\dag$. (Note, the reverse
statement is evidently not  true, {\it i.e.} 'most' homogeneous
polynomials in $a$ and $a^\dag$ can not be written as a boson
string.) This means that the results of Section \ref{HomPol} may
be applied to calculate the normal form of iterated boson string.
We take an operator
\begin{eqnarray}
H_{\mathbf{r},\mathbf{s}}=(a^\dag)^{r_M}a^{s_M}\dots(a^\dag)^{r_2}a^{s_2}(a^\dag)^{r_1}a^{s_1}
\end{eqnarray}
where ${\mathbf r}=(r_1,r_2,\dots,r_M)$ and ${\mathbf
s}=(s_1,s_2,\dots,s_M)$ are fixed integer vectors and search for
its $n$-th power normal form
\begin{eqnarray}
\left(H_{\mathbf{r},\mathbf{s}}\right)^n=(a^{\dagger})^{nd}\sum_{k=s_1}^{n(s_1+s_2+\dots+s_M)}S_{{\mathbf
r},{\mathbf s}}(n,k)(a^{\dagger})^ka^k,
\end{eqnarray}
where $d=\sum_{i=1}^M(r_m-s_m)$.
\\
Then the procedure is straightforward. In the first step we use
results of Section \ref{General boson strings} to obtain
\begin{eqnarray}\nonumber
H_{\mathbf{r},\mathbf{s}}=(a^\dag)^d\sum_{k=s_1}^{s_1+s_2+\dots+s_M}S_{{\mathbf
r},{\mathbf s}}(k)(a^\dag)^ka^k.
\end{eqnarray}
One can use any of the formulas in the Eqs.(\ref{rec}),
(\ref{dob}) or (\ref{sss}).
\\
Once the numbers $S_{{\mathbf r},{\mathbf s}}(k)$ are calculated
they play the role of the coefficients in the homogeneous
polynomial in $a^\dag$ and $a$ which was treated in preceding
Section. Taking $\alpha_k=S_{{\mathbf r},{\mathbf s}}(k)$,
$N_0=s_1$ and $N=s_1+s_2+...+s_M$ in Eq.(\ref{HHH}) we arrive at
the normal form using methods of Section \ref{HomPol}.
\\
We note that analogous scheme can be applied to a homogeneous
boson polynomial which is not originally in the normally ordered
form.

\section{Generating functions}
\label{GenFunct}

In the preceding Sections we have considered generalized Stirling
numbers as solutions to the normal ordering problem. We have given
recurrence relations and closed form expressions including
Dobi\'nski-type formulas. Another convenient way to describe them
is through their generating functions. We define
\textit{exponential generating functions} of generalized Bell
polynomials $B_{\boldsymbol{\alpha}}^d(n,x)$ as
\begin{eqnarray}\label{Ga}
G_{\boldsymbol{\alpha}}^d(\lambda,x)=\sum_{n=0}^\infty
B_{\boldsymbol{\alpha}}^d(n,x)\frac{\lambda^n}{n!}
\end{eqnarray}
They are usually formal power series. Because
$B_{\boldsymbol{\alpha}}^d(n)$ grows too rapidly with $n$ these
series are divergent (except in the case $\alpha_k=0$ for $k>1$
treated in Chapters \ref{Generic} and \ref{Sheffer}). In Section \ref{Special} we
give a detailed discussion and show how to improve the convergence
by the use of
\textit{hypergeometric generating functions}.
\\
In spite of this 'inconvenience', knowing that formal power series
can be rigorously handled (see Appendix \ref{Umbral}), in the
following we give some useful (formal) expressions.
\\
Substituting the Dobi\'nski-type formula of Eq.(\ref{HDob}) into
definition Eq.(\ref{Ga})
\begin{eqnarray}\nonumber
G_{\boldsymbol{\alpha}}^d(\lambda,x) =e^{-x}\sum_{n=0}^\infty
\sum_{l=0}^\infty\left[\prod_{i=1}^n\sum_{k=N_0}^N\alpha_k\
(l+(i-1)d)^{\underline{k}}\right]\frac{x^l}{l!}\frac{\lambda^n}{n!}
\end{eqnarray}
which  changing the summation order  yields
\begin{eqnarray}\label{G1}
G_{\boldsymbol{\alpha}}^d(\lambda,x)=e^{-x}\sum_{l=0}^\infty\frac{x^l}{l!}\sum_{n=0}^\infty
\left[\prod_{i=1}^n\sum_{k=N_0}^N\alpha_k\
(l+(i-1)d)^{\underline{k}}\right]\frac{\lambda^n}{n!}.
\end{eqnarray}
Using the Cauchy multiplication rule of Eq.(\ref{Cauchy}) we get
the expansion in $x$:
\begin{eqnarray}\nonumber
G_{\boldsymbol{\alpha}}^d(\lambda,x)
&=&\sum_{j=0}^\infty\frac{x^j}{j!}\sum_{l=0}^j\binom{j}{l}
(-1)^{j-l}
\sum_{n=0}^\infty\left[\prod_{i=1}^n\sum_{k=N_0}^N\alpha_k\ (l+(i-1)d)^{\underline{k}}\right]\frac{\lambda^n}{n!},\\\label{G2}
\end{eqnarray}
which by comparison with
\begin{eqnarray}\nonumber
G_{\boldsymbol{\alpha}}^d(\lambda,x)=1+\sum_{n=1}^\infty
\left(\sum_{k=1}^n S_{\boldsymbol{\alpha}}^d(n,k)
x^k\right)\frac{\lambda^n}{n!} =1+\sum_{k=N_0}^\infty
\left(\sum_{n=k}^\infty S_{\boldsymbol{\alpha}}^d(n,k)
\frac{\lambda^n}{n!}\right)x^k
\end{eqnarray}
gives the exponential generating function of the generalized
Stirling numbers $S_{\boldsymbol{\alpha}}^d(n,k)$ in the form
(compare with Eq.(\ref{Sa}))
\begin{eqnarray}\nonumber
\sum_{n=k}^\infty S_{\boldsymbol{\alpha}}^d(n,k) \frac{\lambda^n}{n!}
=\frac{1}{k!}\sum_{l=0}^k\binom{k}{l} (-1)^{k-l}
\sum_{n=0}^\infty\left[\prod_{i=1}^n\sum_{k=N_0}^N\alpha_k\ (l+(i-1)d)^{\underline{k}}\right]\frac{\lambda^n}{n!}.\\\label{G3}
\end{eqnarray}
We note that for specific cases Eqs.(\ref{G1}), (\ref{G2}) and
(\ref{G3}) simplify a lot, see Section \ref{Special} for some
examples.
\\
This kind of generating functions are connected with the diagonal
coherent state elements of the operator $e^{\lambda
H_{\boldsymbol{\alpha}}^d}$. Using observation Eq.(\ref{zH}) we
get
\begin{eqnarray}
\langle z|e^{\lambda H_{\boldsymbol{\alpha}}^d}|z\rangle
=\sum_{n=0}^\infty\langle
z|\left(H_{\boldsymbol{\alpha}}^d\right)^n|z\rangle\frac{\lambda^n}{n!}
=\sum_{n=0}^\infty \
B_{\boldsymbol{\alpha}}^d(n,|z|^2)\frac{\left((z^*)^d\lambda\right)^n}{n!}
\end{eqnarray}
which yields
\begin{eqnarray}
\langle z|e^{\lambda H_{\boldsymbol{\alpha}}^d}|z\rangle
=G_{\boldsymbol{\alpha}}^d\left((z^*)^d\lambda,|z|^2\right).
\end{eqnarray}
This in turn lets us write (see Appendix \ref{Coherent states})
the normally ordered expression
\begin{eqnarray}\label{Gd::}
e^{\lambda H_{\boldsymbol{\alpha}}^d} =\
:G_{\boldsymbol{\alpha}}^d((a^\dag)^d\lambda,a^\dag a):\ .
\end{eqnarray}
Observe that the exponential generating functions
$G_{\boldsymbol{\alpha}}^d(\lambda,x)$ in general are not
analytical around $\lambda=0$ (they converge only in the case
$\alpha_k=0$ for $k>1$, see Chapters \ref{Generic} and \ref{Sheffer}). Nevertheless,
{\it e.g.} for negative values of $\lambda\alpha_N$ and $d=0$ the
expressions in Eqs.(\ref{G1}) or (\ref{G2}) converge and may be
used for explicit calculations. The convergence properties can be
handled with the d'Alembert or Cauchy criterion.
\\
We also note that the number state matrix elements of $e^{\lambda
H_{\boldsymbol{\alpha}}^d}$ are finite (the operator is well
defined on the dense subset generated by the number states). It
can be seen and explicitly calculated from Eqs.(\ref{Gd::}),
(\ref{G1}) or (\ref{G2}).

\section{Negative excess}
\label{NegExcess}

We have considered so far the generalized Stirling and Bell
numbers arising in the normal ordering problem of boson
expressions and their powers (iterations) with nonnegative
excess. Actually when the excess is negative, {\it i.e.} there are
more annihilation $a$ than creation $a^\dag$ operators in a
string, the problem is dual and does not lead to new integer
sequences.
\\
To see this  we first take a string defined by two vectors
${\mathbf r}=(r_1,r_2,\dots,r_M)$ and ${\mathbf
s}=(s_1,s_2,\dots,s_M)$ with a negative (overall) excess
$d_M=\sum_{m=1}^M(r_m-s_m)<0$. The normal ordering procedure
extends the definition of the numbers $S_{{\mathbf r},{\mathbf
s}}(k)$ for the case of negative excess through
\begin{eqnarray}\nonumber
H_{{\mathbf r},{\mathbf
s}}=(a^{\dagger})^{r_M}a^{s_M}\dots(a^{\dagger})^{r_2}a^{s_2}(a^{\dagger})^{r_1}a^{s_1}=
\sum_{k=r_M}^{r_1+r_2+\dots+r_M}S_{{\mathbf
r},{\mathbf s}}(k)(a^{\dagger})^ka^k\ a^{-d_M}.\\\label{H-}
\end{eqnarray}
Definition of $B_{{\mathbf r},{\mathbf s}}(x)$ and $B_{{\mathbf
r},{\mathbf s}}$ is analogous to Eqs.(\ref{bx}) and (\ref{b}):
\begin{eqnarray}
B_{{\mathbf r},{\mathbf
s}}(x)=\sum_{k=r_M}^{r_1+r_2+\dots+r_M}S_{{\mathbf r},{\mathbf
s}}(k)x^k,
\end{eqnarray}
and
\begin{eqnarray}
B_{{\mathbf r},{\mathbf s}}=B_{{\mathbf r},{\mathbf s}}(1)=
\sum_{k=r_M}^{r_1+r_2+\dots+r_M}S_{{\mathbf r},{\mathbf s}}(k).
\end{eqnarray}
Note that the limits in the above equations are different from
those  in Section \ref{General boson strings}.
\\
By taking the hermitian conjugate we have
\begin{eqnarray}\nonumber
&&\sum_{k=r_M}^{r_1+r_2+\dots+r_M}S_{{\mathbf r},{\mathbf
s}}(k)(a^{\dagger})^ka^k\
a^{-d_M}\stackrel{(\ref{H-})}{=}\left(\left((a^{\dagger})^{r_M}a^{s_M}\dots(a^{\dagger})^{r_2}a^{s_2}(a^{\dagger})^{r_1}a^{s_1}\right)^\dag\right)^\dag\\\nonumber
&&=\left((a^{\dagger})^{s_1}a^{r_1}(a^{\dagger})^{s_2}a^{r_2}\dots(a^{\dagger})^{s_M}a^{r_M}\right)^\dag{}
\stackrel{(\ref{SSS})}{=}\left((a^\dag)^{-d_M}\ \sum_{k=r_M}^{r_1+r_2+\dots+r_M}S_{{\overline{\mathbf s}},{\overline{\mathbf
r}}}(k)(a^{\dagger})^ka^k\right)^\dag\\\nonumber
&&=\sum_{k=r_M}^{r_1+r_2+\dots+r_M}S_{{\overline{\mathbf
s}},{\overline{\mathbf r}}}(k)(a^{\dagger})^ka^k\ a^{-d_M},
\end{eqnarray}
where ${\overline{\mathbf s}}=(s_M,\dots,s_2,s_1)$ and
${\overline{\mathbf r}}=(r_M,\dots,r_2,r_1)$ correspond to the
string $H_{{\overline{\mathbf s}},{\overline{\mathbf r}}}$ with
positive overall excess equal to $-d_M>0$.
\\
This leads to the symmetry property
\begin{eqnarray}\label{symmetry1}
S_{{\mathbf r},{\mathbf s}}(k)=S_{{\overline{\mathbf
s}},{\overline{\mathbf r}}}(k)
\end{eqnarray}
as well as
\begin{eqnarray}\label{symmetry2}
B_{{\mathbf r},{\mathbf s}}(n,x)=B_{{\overline{\mathbf
s}},{\overline{\mathbf r}}}(n,x)\ \ \ \ \ \text{and}\ \ \
B_{{\mathbf r},{\mathbf s}}(n)=B_{{\overline{\mathbf
s}},{\overline{\mathbf r}}}(n)
\end{eqnarray}
for any integer vectors ${\mathbf r}$ and ${\mathbf s}$.
\\
Formulas in Section \ref{General boson strings} may be used to
calculate the generalized Stirling and Bell numbers in this case
if the symmetry properties Eqs.(\ref{symmetry1}) and
(\ref{symmetry2}) are taken into account.
We conclude by writing explicitly the coherent state matrix
element of a string for $d_M<0$:
\begin{eqnarray}
    \langle z|(a^{\dagger})^{r_M}a^{s_M}\dots(a^{\dagger})^{r_2}a^{s_2}(a^{\dagger})^{r_1}a^{s_1}|z\rangle
    =B_{{\mathbf r},{\mathbf s}}(|z|^2)z^{-d_M}.
\end{eqnarray}
\\
Now we proceed to homogeneous polynomials of Section \ref{HomPol}.
In the case of the negative excess we define
\begin{eqnarray}\label{h}
H_{\boldsymbol{\alpha}}^{-d}=\sum_{k=N_0}^N\alpha_k\ (a^\dag)^k
a^k\ a^d\ ,
\end{eqnarray}
with some constant coefficients $\alpha_k$'s (for convenience we
take $\alpha_{N_0}\neq 0$ and $\alpha_{N}\neq 0$) and nonnegative
integer $d$. The excess of this homogeneous boson polynomial is
$-d$. We search for the numbers
$S_{\boldsymbol{\alpha}}^{-d}(n,k)$ defined by the normal form of
the  $n$-th power of Eq.(\ref{h}) as
\begin{eqnarray}
\left(H_{\boldsymbol{\alpha}}^{-d}\right)^n=\sum_{k=N_0}^{nN}S_{\boldsymbol{\alpha}}^{-d}(n,k)\
(a^\dag)^k a^k\ a^{nd}.
\end{eqnarray}
Generalized Bell polynomials $B_{\boldsymbol{\alpha}}^{-d}(n,x)$ and Bell numbers
$B_{\boldsymbol{\alpha}}^{-d}(n)$ are defined in the usual manner.
\\
In the same way, taking  the hermitian conjugate, we have
\begin{eqnarray}\nonumber
&&\sum_{k=N_0}^{nN}S_{\boldsymbol{\alpha}}^{-d}(n,k)\ (a^\dag)^k
a^k\ a^{nd}\
\stackrel{(\ref{h})}{=}\left(\left(H_{\boldsymbol{\alpha}}^{-d})^n\right)^\dag\right)^\dag
=\left(\left((H_{\boldsymbol{\alpha}}^{-d})^\dag\right)^n\right)^\dag\\\nonumber
&&=\left(\left( (a^\dag)^d\ \sum_{k=N_0}^N\alpha_k^*\ (a^\dag)^k
a^k\right)^n\right)^\dag
\stackrel{(\ref{NHH})}{=}
\left(\left(H_{\boldsymbol{\alpha^*}}^d\right)^n\right)^\dag\\\nonumber
&&=\sum_{k=N_0}^{nN}S_{\boldsymbol{\alpha}}^d(n,k)\ (a^\dag)^k
a^k\ a^{nd}
\end{eqnarray}
(Note that
$(S_{\boldsymbol{\alpha}^*}^d)^*=S_{\boldsymbol{\alpha}}^d$ , see
any formula of Section \ref{HomPol}.)
\\
Consequently the following symmetry property holds true
\begin{eqnarray}
S_{\boldsymbol{\alpha}}^{-d}(n,k)=S_{\boldsymbol{\alpha}}^d(n,k)
\end{eqnarray}
and consequently
\begin{eqnarray}
B_{\boldsymbol{\alpha}}^{-d}(n,x)=B_{\boldsymbol{\alpha}}^{d}(n,x)\
\ \ \ \ \text{and}\ \ \
B_{\boldsymbol{\alpha}}^{-d}(n)=B_{\boldsymbol{\alpha}}^{d}(n)
\end{eqnarray}
for any vector $\boldsymbol{\alpha}$ and integer $d$.
\\
Again any of the formulas of Section \ref{HomPol} may be used in
calculations for the case of negative excess. Coherent state matrix
element of Eq.(\ref{h}) takes the form
\begin{eqnarray}
\langle z|\left(H_{\boldsymbol{\alpha}}^{-d}\right)^n|z\rangle
    =B_{\boldsymbol{\alpha}}^{-d}(n,|z|^2)z^{d}.
\end{eqnarray}

\section{Special cases}
\label{Special}

The purpose of this section is to illustrate the above formalism
with some examples. First we consider iteration of a simple string
of the form $(a^\dag)^r a^s$ and work out some special cases
in detail. Next we proceed to  expressions involving homogeneous
boson polynomials with excess zero $d=0$. This provides the
solution to the normal ordering of a generalized Kerr-type
hamiltonian.

\subsection{Case $\left((a^\dag)^r a^s\right)^n$}\label{rs}

Here we consider the boson string in the form
\begin{eqnarray}
    H_{r,s}=(a^\dag)^r a^s
\end{eqnarray}
for $r\geq s$ (the symmetry properties provide the opposite case,
see Section \ref{NegExcess}).
\\
It corresponds in the previous notation to $H_{{\mathbf
r},{\mathbf s}}$ with $({\mathbf r},{\mathbf s})=(r,s)$ or
$H_{\boldsymbol{\alpha}}^{d}$ with $d=r-s$ and $\alpha_s=1$ (zero
otherwise).
\\
The $n$-th power in the normally ordered form defines generalized
Stirling numbers $S_{r,s}(n,k)$ as
\begin{eqnarray}\label{Srs}
    \left(H_{r,s}^d\right)^n=(a^\dag)^{n(r-s)} \sum_{k=s}^{ns} S_{r,s}(n,k)(a^\dag)^ka^k.
\end{eqnarray}
Consequently we define the generalized Bell polynomials
$B_{r,s}(n,x)$ and generalized Bell numbers  $B_{r,s}(n)$ as
\begin{eqnarray}
    B_{r,s}(n,x)=\sum_{k=s}^{ns} S_{r,s}(n,k)x^k
\end{eqnarray}
and
\begin{eqnarray}\label{Brs}
    B_{r,s}(n)=B_{r,s}(n,1)=\sum_{k=s}^{ns} S_{r,s}(n,k).
\end{eqnarray}
The following convention is assumed:
\begin{eqnarray}\label{Init}
\begin{array}{ccl}
S_{r,s}(0,0)&=&1,\\
S_{r,s}(n,k)&=&0\ \ \text{for}\ k>ns,\\
S_{r,s}(n,k)&=&0\ \ \text{for}\ k<s\ \text{and}\ n>0,
\end{array}
\end{eqnarray}
and
\begin{eqnarray}
B_{r,s}(0)=B_{r,s}(0,x)=1.
\end{eqnarray}
\\
As pointed out these numbers and polynomials are  special cases of
those considered in Section \ref{HomPol}, {\it i.e.}
\begin{eqnarray}\nonumber
  &&S_{r,s}(n,k)=S_{\boldsymbol{\alpha}}^{(d)}(n,k)\\\nonumber
    &&B_{r,s}(n,x)=B_{\boldsymbol{\alpha}}^{(d)}(n,x)\\\nonumber
    &&B_{r,s}(n)=S_{\boldsymbol{\alpha}}^{(d)}(n)
\end{eqnarray}
for $d=r-s$ and $\alpha_s=1$ (zero otherwise).
\\
In the following the formalism of Section \ref{HomPol} is applied
to investigate the numbers in this case. By this we demonstrate
that the formulas simplify considerably when the special case is considered. The
discussion also raises some new points concerning the connection
with special functions.
\\
Below we choose the way of increasing simplicity, {\it i.e.} we
start with $r>s$ and then restrict our attention to $r=s$ and
$s=1$. For $r=1$ and $s=1$ we end up with conventional Stirling
and Bell numbers (see Chapter \ref{Generic}). Where possible we
just state the results without comment.
\\
Before proceeding to the  program as sketched, we state the
essential  results which are the same for each case. We define the
exponential generating function as
\begin{eqnarray}\label{ggg}
    G_{r,s}(\lambda,x)=\sum_{n=0}^\infty B_{r,s}(n,x)\frac{\lambda^n}{n!}.
\end{eqnarray}
The coherent state matrix elements can be read off as
\begin{eqnarray}\label{z}
    \langle z|(H_{r,s})^n|z\rangle=B_{r,s}(\lambda (z^*)^{r-s},|z|^2)
\end{eqnarray}
and
\begin{eqnarray}\label{ez}
    \langle z|e^{\lambda H_{r,s}}|z\rangle=G_{r,s}(\lambda (z^*)^{r-s},|z|^2).
\end{eqnarray}
This provides the normally ordered forms
\begin{eqnarray}\label{nnn}
    (H_{r,s})^n=\ :B_{r,s}(\lambda (a^\dag)^{r-s},a^\dag a):
\end{eqnarray}
and
\begin{eqnarray}\label{en}
    e^{\lambda H_{r,s}}=\ :G_{r,s}(\lambda (a^\dag)^{r-s},a^\dag a):\ .
\end{eqnarray}
We shall see that usually the generating function of
Eq.(\ref{ggg}) is formal and Eq.(\ref{ez}) is not analytical
around $\lambda=0$.
No matter of the convergence subtleties operator equations Eqs.(\ref{nnn}) and (\ref{en}) hold
true in the occupation number representation.

\subsubsection{r$>$s}
Recurrence relation
\begin{eqnarray}
    S_{r,s}(n+1,k)=\sum_{p=0}^s\binom{s}{p}
    (n(r-s)+k-r+p)^{\underline{p}}\ S_{r,s}(n,k-s+p)
\end{eqnarray}
with initial conditions as in Eq.(\ref{Init}).
\\
Connection property
\begin{eqnarray}
    \prod_{j=1}^n(x+(j-1)(r-s))^{\underline{s}}
    =\sum_{k=s}^{ns}S_{r,s}(n,k)x^{\underline{k}}.
\end{eqnarray}
The Dobi\'nski-type relation
\begin{eqnarray}\label{Aaa}
    B_{r,s}(n,x)&=&
 e^{-x}\sum_{k=s}^\infty\prod_{j=1}^n\left[k+(j-1)(r-s)\right]^{\underline{s}}\frac{x^k}{k!}
 \\\label{AaaA}
    &=&(r-s)^{s(n-1)}e^{-x}\sum_{k=0}^\infty\left[\prod_{j=1}^s
    \frac{\Gamma(n+\frac{k+j}{r-s})}{\Gamma(1+\frac{k+j}{r-s})}\right]\frac{x^k}{k!}.
\end{eqnarray}
Explicit expression
\begin{eqnarray}
    S_{r,s}(n,k)
    =\frac{(-1)^k}{k!}\sum_{p=s}^k(-1)^p\binom{k}{p}
    \prod_{j=1}^n\left(p+(j-1)(r-s)\right)^{\underline{s}}.
\end{eqnarray}
The ``non-diagonal'' generalized Bell numbers $B_{r,s}(n)$ can
always be expressed as special values of generalized
hypergeometric functions $_pF_q$ , {\it e.g.} the series
$B_{2r,r}(n)$ can be written down in a compact form using the
confluent hypergeometric function of Kummer:
\begin{eqnarray}\label{YY}
    B_{2r,r}(n)=\frac{(rn)!}{e\cdot r!}{_1F_1}(rn+1,r+1;1)
\end{eqnarray}
and still more general family of sequences has the form
($p,r=1,2\ldots$):
\begin{eqnarray}\nonumber
    B_{pr+p,pr}(n)&=&\frac{1}{e}\left[\prod_{j=1}^r\frac{(p(n-1)+j)!}{(pj)!}\right]\cdot\\
    \nonumber
    &&\cdot{_r F_r}(pn+1,pn+1+p,\ldots,pn+1+p(r-1);1+p,1+2p,\ldots,1+rp;1),
\end{eqnarray}
etc...
\\
The exponential generating function takes the form
\begin{eqnarray}\nonumber
G_{r,s}(\lambda,x)&=&
e^{-x}\sum_{l=0}^\infty\frac{x^l}{l!}\sum_{n=0}^\infty
\left[\prod_{i=1}^n\
(l+(i-1)(r-s))^{\underline{s}}\right]\frac{\lambda^n}{n!}\\\nonumber
&=&1+\sum_{j=1}^\infty\frac{x^j}{j!}\sum_{l=0}^j\binom{j}{l}
(-1)^{j-l}
\sum_{n=1}^\infty\left[\prod_{i=1}^n(l+(i-1)(r-s))^{\underline{s}}\right]\frac{\lambda^n}{n!}.
\end{eqnarray}
For $s>1$ it is purely formal series (not analytical around
$\lambda=0$).
\\
A well-defined and convergent procedure for such sequences is to
consider what we call
\emph{hypergeometric generating functions},
which are the exponential generating function for the ratios
$B_{r,s}/(n!)^t$, where $t$ is an appropriately chosen integer. A
case in point is the series $B_{3,2}(n)$ which may be written
explicitly from  Eq.(\ref{Aaa}) as:
\begin{eqnarray}\nonumber
    B_{3,2}(n)=\frac{1}{e}\sum_{k=0}^\infty
    \frac{(n+k)!(n+k+1)!}{k!(k+1)!(k+2)!}.
\end{eqnarray}
Its hypergeometric generating function $\tilde{G}_{3,2}(\lambda)$
is then:
\begin{eqnarray}\nonumber
    \tilde{G}_{3,2}(\lambda)=\sum_{n=0}^\infty\left[\frac{B_{3,2}(n)}{n!}\right]\frac{\lambda^n}{n!}
    =\frac{1}{e}\sum_{k=0}^\infty\frac{1}{(k+2)!}{_2F_1(k+2,k+1;1;\lambda)}.
\end{eqnarray}
Similarly for $G_{4,2}(\lambda)$ one obtains:
\begin{eqnarray}\nonumber  \tilde{G}_{4,2}(\lambda)=\frac{1}{e}\sum_{k=0}^\infty\frac{1}{(k+2)!}{_2F_1\left(\frac{k+2}{2},\frac{k}{2}+1;1;4\lambda\right)}.
\end{eqnarray}
 Eq.(\ref{YY}) implies more generally:
\begin{eqnarray}\nonumber &&\tilde{G}_{2r,r}(\lambda)=\sum_{n=0}^\infty\left[\frac{B_{2r,r}(n)}{(n!)^{r-1}}\right]\frac{\lambda^n}{n!}\\\nonumber &&=\left\{\begin{array}{ll}\frac{1}{1!e}\sum_{k=0}^\infty\frac{1}{(k+1)!}{_2F_1\left(\frac{k+1}{2},\frac{k}{2}+1;1;4\lambda\right)},&r=2,\\
\frac{1}{2!e}\sum_{k=0}^\infty\frac{1}{(k+2)!}{_3F_2\left(\frac{k+1}{3},\frac{k+2}{3},\frac{k+3}{3};1,1;27\lambda\right)},&r=3,\\
\frac{1}{3!e}\sum_{k=0}^\infty\frac{1}{(k+3)!}{_4F_3\left(\frac{k+1}{4},\ldots,\frac{k+4}{4};1,1,1;256\lambda\right)},&r=4\ldots
\end{array}
\right.\\\nonumber
&&\text{etc...}
\end{eqnarray}
See \cite{PensonJIntSeq2001} for other instances where this type
of hypergeometric generating functions  appear.

\subsubsection{r$=$s}
Recurrence relation
\begin{eqnarray}\label{Srr-rec}
    S_{r,r}(n+1,k)=\sum_{p=0}^r\binom{r}{p}
    (k-r+p)^{\underline{p}}\ S_{r,r}(n,k-r+p)
\end{eqnarray}
with initial conditions as in Eq.(\ref{Init}).
\\
Connection property
\begin{eqnarray}
    [x^{\underline{r}}]^n=\sum_{k=r}^{nr}S_{r,r}(n,k)x^{\underline{k}}.
\end{eqnarray}
The Dobi\'nski-type relation has the form:
\begin{eqnarray}
    B_{r,r}(n,x)&=&
    e^{-x}\sum_{k=s}^\infty k^{\underline{r}}\frac{x^k}{k!}\\
    &=&e^{-x}\sum_{k=0}^\infty \left[\frac{(k+r)!}{k!}\right]^{n-1}\frac{x^k}{k!}.
\end{eqnarray}
Explicit expression
\begin{eqnarray}\label{Srr-expl}
    S_{r,r}(n,k)
    =\frac{(-1)^k}{k!}\sum_{p=r}^{k}(-1)^p\binom{k}{p}
    \left[p^{\underline{r}}\right]^n.
\end{eqnarray}
\\
Using Eq.(\ref{Srr-expl}) we can find the following exponential
generating function of $S_{r,r}(n,k)$:
\begin{eqnarray}
    \sum_{n=\lceil k/r\rceil }\frac{x^n}{n!}S_{r,r}(n,k)=
    \frac{(-1)^k}{k!}\sum_{p=r}^k(-1)^p\binom{k}{p}
    \left(e^{xp(p-1)\ldots(p-r+1)}-1\right),
\end{eqnarray}
We refer to Section \ref{Kerr} for considerations of the exponential generating functions.

\subsubsection{s$=$1}
Recurrence relation
\begin{eqnarray}
    S_{r,1}(n+1,k)=(n(r-1)+k-r+1)S_{r,1}(n,k)+S_{r,1}(n,k-1)
\end{eqnarray}
with initial conditions as in Eq.(\ref{Init}).
\\
Connection property
\begin{eqnarray}
    \prod_{j=1}^n(x+(j-1)(r-s))
    =\sum_{k=s}^{ns}S_{r,1}(n,k)x^{\underline{k}}.
\end{eqnarray}
The Dobi\'nski-type relation has the form:
\begin{eqnarray}\label{r1}
    B_{r,1}(n,x)=
    e^{-x}\sum_{k=1}^\infty\prod_{j=1}^n\left[k+(j-1)(r-1)\right]\frac{x^k}{k!}.
\end{eqnarray}
Explicit expression
\begin{eqnarray}
    S_{r,1}(n,k)
    =\frac{(-1)^k}{k!}\sum_{p=s}^k(-1)^p\binom{k}{p}
    \prod_{j=1}^n\left(p+(j-1)(r-1)\right).
\end{eqnarray}
The generalized Bell numbers $B_{r,1}(n)$ can always be expressed
as a combination of $r-1$ different hypergeometric functions of
type $_1F_{r-1}(\ldots;x)$, each of them evaluated at the same
value of argument $x=(r-1)^{1-r}$; here are some lowest order
cases:
\begin{eqnarray}
    B_{2,1}(n)&=&\frac{n!}{e}{_1F_1}(n+1;2;1)=(n-1)!L_{n-1}^{(1)}(-1),\label{JJJ}\\\nonumber
    B_{3,1}(n)&=&\frac{2^{n-1}}{e}\left(\frac{2\Gamma(n+\frac{1}{2})}{\sqrt{\pi}}
    {_1F_2}\left(n+\frac{1}{2};\frac{1}{2},\frac{3}{2};\frac{1}{4}\right)+
    n!{_1F_2}\left(n+1;\frac{3}{2},2;\frac{1}{4}\right)\right),\\\nonumber
    B_{4,1}(n)&=&\frac{3^{n-1}}{2e}\left(\frac{3^{3/2}\Gamma(\frac{2}{3})\Gamma(n+\frac{1}{3})}{\pi}
    {_1F_3}\left(n+\frac{1}{3};\frac{1}{3},\frac{2}{3},\frac{4}{3};\frac{1}{27}\right)\right.+\\\nonumber
    &&\left.\frac{3\Gamma(n+\frac{2}{3})}{\Gamma(\frac{2}{3})}
    {_1F_3}\left(n+\frac{2}{3};\frac{2}{3},\frac{4}{3},\frac{5}{3};\frac{1}{27}\right)
    +n!{_1F_3}\left(n+1;\frac{4}{3},\frac{5}{3},2;\frac{1}{27}\right)\right),\\\nonumber
    \text{etc...}
\end{eqnarray}
In Eq.(\ref{JJJ}) $L_m^{(\alpha)}(y)$ is the associated Laguerre
polynomial.
\\
In this case the exponential generating function of Eq.(\ref{ggg})
converges (see also Chapter \ref{Sheffer})
\begin{eqnarray}\label{gr1}
    G_{r,1}(\lambda,x)=
    e^{x\left(\frac{1}{\sqrt[r-1]{1-(r-1)\lambda}}-1\right)}.
\end{eqnarray}
The exponential generating function of $S_{r,1}(n,k)$ takes the form
\begin{eqnarray}
    \sum_{n=\lceil k/r\rceil }^\infty \frac{x^n}{n!}S_{r,1}(n,k)=\frac{1}{k!}
    \left[\left(1-(r-1)x\right)^{-\frac{1}{r-1}}-1 \right]^k.
\end{eqnarray}
See also Chapter \ref{Sheffer} for detailed discussion of this case.

\newpage
We end this section by writing down some triangles of generalized
Stirling and Bell numbers, as defined by Eqs.(\ref{Srs}) and
(\ref{Brs}).
\\\\
\begin{small}

\underline{\bf r=1, s=1}
\begin{eqnarray}
    S_{1,1}(n,k),\ 1\leq k\leq n\qquad B_{1,1}(n)&&\nonumber\\
    \begin{array}{cl|lllllllllllcc}\cline{3-14}&&&&&&\\
    n=1&&&1&&&&&&&&&&1&\\
        n=2&&&1&1&&&&&&&&&2&\\
        n=3&&&1&3&1&&&&&&&&5&\\
        n=4&&&1&7&6&1&&&&&&&15&\\
        n=5&&&1&15&25&10&1&&&&&&52&\\
    n=6&&&1&31&90&65&15&1&&&&&203&
    \end{array}&&\nonumber
\end{eqnarray}
\newline

\underline{\bf r=2, s=1}
\begin{eqnarray}
    S_{2,1}(n,k),\ 1\leq k\leq n\qquad\qquad\qquad B_{2,1}(n)&&\nonumber\\
    \begin{array}{cl|lllllllllllccc}\cline{3-14}&&&&&&\\
    n=1&&&1&&&&&&&&&&1&\\
        n=2&&&2&1&&&&&&&&&3&\\
        n=3&&&6&6&1&&&&&&&&13&\\
        n=4&&&24&36&12&1&&&&&&&73&\\
        n=5&&&120&240&120&20&1&&&&&&501&\\
    n=6&&&720&1800&1200&300&30&1&&&&&4051&
    \end{array}&&\nonumber
\end{eqnarray}
\newline

\underline{\bf r=3, s=1}
\begin{eqnarray}
    S_{3,1}(n,k),\ 1\leq k\leq n\ \quad\qquad\qquad\qquad B_{3,1}(n)&&\nonumber\\
    \begin{array}{cl|lllllllllllccc}\cline{3-14}&&&&&&\\
    n=1&&&1&&&&&&&&&&1\\
        n=2&&&3&1&&&&&&&&&4\\
        n=3&&&15&9&1&&&&&&&&25\\
        n=4&&&105&87&18&1&&&&&&&211\\
        n=5&&&945&975&285&30&1&&&&&&2236\\
    n=6&&&10395&12645&4680&705&45&1&&&&&28471\\
    \end{array}&&\nonumber
\end{eqnarray}
\newline
\end{small}

\begin{landscape}

\begin{scriptsize}

\underline{\bf r=2, s=2}
\begin{eqnarray}
    S_{2,2}(n,k),\ 2\leq k\leq 2n\ \quad\qquad\qquad\qquad\qquad\qquad\qquad\quad\qquad\qquad\qquad\qquad\qquad B_{2,2}(n)\ &&\nonumber\\
    \begin{array}{cl|lllllllllllcccccccc}\cline{3-19}&&&&&&\\
    n=1&&&1&&&&&&&&&&&&&&&1\\
        n=2&&&2&4&1&&&&&&&&&&&&&7\\
        n=3&&&4&32&38&12&1&&&&&&&&&&&87\\
        n=4&&&8&208&652&576&188&24&1&&&&&&&&&1657\\
        n=5&&&16&1280&9080&16944&12052&3840&580&40&1&&&&&&&43833\\
    n=6&&&32&7744&116656&412800&540080&322848&98292&16000&1390&60&1&&&&&1515903\\
    \end{array}&&\nonumber
\end{eqnarray}
\newline

\underline{\bf r=3, s=2}
\begin{eqnarray}
    S_{3,2}(n,k),\ 2\leq k\leq 2n\ \quad \qquad\quad\qquad\qquad\qquad\qquad\qquad\qquad\qquad\qquad\qquad\quad\qquad\qquad\qquad\qquad\qquad B_{3,2}(n)\ \ &&\nonumber\\
    \begin{array}{cl|lllllllllllcccccccc}\cline{3-19}&&&&&&\\
    n=1&&&1&&&&&&&&&&&&&&&1\\
        n=2&&&6&6&1&&&&&&&&&&&&&13\\
        n=3&&&72&168&96&18&1&&&&&&&&&&&355\\
        n=4&&&1440&5760&6120&2520&456&36&1&&&&&&&&&16333\\
        n=5&&&43200&259200&424800&285120&92520&15600&1380&60&1&&&&&&&1121881\\
    n=6&&&1814400&15120000&34776000&33566400&16304400&4379760&682200&62400&3270&90&1&&&&&106708921\\
    \end{array}&&\nonumber
\end{eqnarray}
\newline

\underline{\bf r=3, s=3}
\begin{eqnarray}
    S_{3,3}(n,k),\ 3\leq k\leq 3n\ \ \quad\qquad\qquad\qquad\quad\qquad\qquad\qquad\qquad\qquad\qquad\qquad\qquad\qquad\quad\qquad\qquad\qquad\qquad\qquad B_{3,3}(n)\ \ \ &&\nonumber\\
    \begin{array}{cl|lllllllllllccccc}\cline{3-18}&&&&&&\\
    n=1&&&1&&&&&&&&&&&&&&1\\
        n=2&&&6&18&9&1&&&&&&&&&&&34\\
        n=3&&&36&540&1242&882&243&27&1&&&&&&&&2971\\
        n=4&&&216&13608&94284&186876&149580&56808&11025&1107&54&1&&&&&513559\\
    n=5&&&1296&330480&6148872&28245672&49658508&41392620&18428400&4691412&706833&63375&3285&90&\ 1&&149670844
    \end{array}&&\nonumber
\end{eqnarray}

\end{scriptsize}

\end{landscape}

\subsection{Generalized Kerr Hamiltonian}
\label{Kerr}

In this section we consider the case of the homogeneous boson
polynomial with excess zero ($d=0$). This illustrates the
formalism of Sections \ref{HomPol} and \ref{GenFunct} on the 
example which constitutes the solution to the normal
ordering problem for the generalized Kerr medium.
\\
The Kerr medium is described by the hamiltonian $H=a^\dag a^\dag a
a$, \cite{Mandel}. By generalization we mean the hamiltonian in
the form
\begin{eqnarray}
    H_{\boldsymbol{\alpha}}=\sum_{k=N_0}^N \alpha_k\ (a^\dag)^k a^k.
\end{eqnarray}
Observe that this is exactly the form of the operators (with
excess $d=0$) considered in Section \ref{HomPol}, {\it i.e.}
\begin{eqnarray}
    H_{\boldsymbol{\alpha}}\equiv H_{\boldsymbol{\alpha}}^0.
\end{eqnarray}
To simplify the notation we skip the index $d=0$ in
$S_{\boldsymbol{\alpha}}^d(n,k)$,
$B_{\boldsymbol{\alpha}}^d(n,x)$, $B_{\boldsymbol{\alpha}}^d(n)$
and $G_{\boldsymbol{\alpha}}^d(\lambda,x)$.
\\
We are now ready  to write down the solution to the normal
ordering problem in terms of generalized Stirling numbers
\begin{eqnarray}\label{NH}
(H_{\boldsymbol{\alpha}})^n=\sum_{k=N_0}^{nN}S_{\boldsymbol{\alpha}}(n,k)\
(a^\dag)^k a^k,
\end{eqnarray}
generalized Bell polynomials
\begin{eqnarray}
B_{\boldsymbol{\alpha}}(n,x)=\sum_{k=N_0}^{nN}S_{\boldsymbol{\alpha}}(n,k)\
x^k,
\end{eqnarray}
and Bell numbers
\begin{eqnarray}
B_{\boldsymbol{\alpha}}(n)=B_{\boldsymbol{\alpha}}^d(n,1)=\sum_{k=N_0}^{nN}S_{\boldsymbol{\alpha}}(n,k).
\end{eqnarray}
Recurrence relation
\begin{eqnarray}
S_{\boldsymbol{\alpha}}(n+1,k)=\sum_{l=N_0}^N\alpha_l\
\sum_{p=0}^l \binom{l}{p}(k-l+p)^{\underline{p}}\ S_{\boldsymbol{\alpha}}(n,k-l+p),
\end{eqnarray}
with initial conditions as in Eq.(\ref{InitialA}).
\\
Connection property
\begin{eqnarray}
\left(\sum_{k=N_0}^N\alpha_k\ x^{\underline{k}}\right)^n=\ \sum_{k=N_0}^{nN}S_{{\boldsymbol\alpha}}(n,k)\ x^{\underline{k}}.
\end{eqnarray}
The Dobi\'nski-type relation has the form:
\begin{eqnarray}
B_{{\boldsymbol\alpha}}(n,x)=e^{-x}\sum_{l=0}^\infty\left(\sum_{k=N_0}^N\alpha_k\
l^{\underline{k}}\right)^n\frac{x^l}{l!}.
\end{eqnarray}
Explicit expression
\begin{eqnarray}
S_{{\boldsymbol\alpha}}(n,k)=\frac{1}{k!}\sum_{j=0}^k\binom{k}{j}(-1)^{k-j}\left(\sum_{l=N_0}^N\alpha_l\
j^{\underline{l}}\right)^n.
\end{eqnarray}
Exponential generating function
\begin{eqnarray}\label{gg11}
G_{\boldsymbol{\alpha}}(\lambda,x)&=&\sum_{n=0}^\infty B_{{\boldsymbol\alpha}}(n,x)\frac{\lambda^n}{n!}\\\label{gg22}
&=&e^{-x}\sum_{n=0}^\infty e^{\lambda \sum_{r=N_0}^N\alpha_r n^{\underline r}}\frac{x^n}{n!}\\
&=&\sum_{m=0}^\infty\left[
\sum_{l=0}^m\binom{m}{l}(-1)^{m-l}e^{\lambda\sum_{r=N_0}^N\alpha_rl^{\underline
r}}\right]\frac{x^m}{m!}.
\end{eqnarray}
Again, the exponential generating function of Eq.(\ref{gg11}) is formal around $\lambda=0$ (for $N>1$). Although we note that by the
d'Alembert criterion for $\lambda\alpha_N<0$ the the series of Eq.(\ref{gg22})
converges.
\\
The generating function of the Stirling numbers is
\begin{eqnarray}
\sum_{n=k}^\infty S_{\boldsymbol{\alpha}}(n,k) \frac{\lambda^n}{n!}
=\frac{1}{k!}\sum_{l=0}^k\binom{k}{l} (-1)^{k-l}
e^{\lambda\sum_{k=N_0}^N\alpha_k\ l^{\underline{k}}}.
\end{eqnarray}
Now, returning to the coherent state representation we have
\begin{eqnarray}
\langle z|e^{\lambda H_{\boldsymbol{\alpha}}}|z\rangle
=G_{\boldsymbol{\alpha}}(\lambda,|z|^2).
\end{eqnarray}
This provides the normally ordered form of the exponential
\begin{eqnarray}
e^{\lambda H_{\boldsymbol{\alpha}}} =\
:G_{\boldsymbol{\alpha}}(\lambda,a^\dag a):\ .
\end{eqnarray}
These considerations may serve as a tool in investigations in
quantum optics whenever the coherent state representation is
needed, {\it e.g.} they allow one to calculate the Husimi
functions or other phase space pictures for thermal states \cite{BlasiakKerr}.

%%%%%%%%%%%%%%%%%%%%%%%%%%%%%%%%%%%%%%%%%%%%%%%%%%%%%%%%%%%%%%%%%%%%%%%%%%%%%%%%%%%%%%

\chapter{Monomiality principle and normal ordering}
\label{Sheffer}

\begin{chapterabstract}
\rule[1mm]{\textwidth}{.3pt}
We solve the boson normal ordering problem for
$\left(q(a^\dag)a+v(a^\dag)\right)^n$ with arbitrary functions
$q(x)$ and $v(x)$. This consequently provides the solution for the
exponential $e^{\lambda(q(a^\dag)a+v(a^\dag))}$ generalizing the
shift operator. In the course of these considerations we define
and explore the monomiality principle and find its
representations. We exploit the properties of Sheffer-type
polynomials which constitute the inherent structure of this
problem. In the end we give some examples illustrating the utility
of the method.
\\
\rule[1mm]{\textwidth}{.3pt}
\end{chapterabstract}

\section{Introduction}

In Chapter \ref{Boson Expressions} we treated the normal ordering 
problem
 of powers and exponentials of boson strings and
homogeneous polynomials. Here we shall extend these results in a
very particular direction. We consider operators linear in the
annihilation $a$ or creation $a^\dag$ operators. More
specifically, we consider  operators which, say for linearity in
$a$, have the form
\begin{eqnarray}\nonumber
q(a^\dag)a+v(a^\dag)
\end{eqnarray}
where $q(x)$ and $v(x)$ are arbitrary functions. Passage to
operators linear in $a^\dag$ is immediate through conjugacy
operation.
\\
We shall find the normally ordered form of the $n$-th power
(iteration)
\begin{eqnarray}\nonumber
(q(a^\dag)a+v(a^\dag))^n
\end{eqnarray}
and then the exponential
\begin{eqnarray}\nonumber
e^{\lambda (q(a^\dag)a+v(a^\dag))}.
\end{eqnarray}
This is a far reaching generalization of the results of
\cite{Mikhailov1983}\cite{Mikhailov1985}\cite{Katriel1983} where a
special case of the operator $a^\dag a+a^r$ was considered.
\\
In this approach we use methods which are based on the
monomiality principle \cite{BlasiakJPA2005}. First, using the methods of
umbral calculus, we find a wide
class of representations of the canonical commutation relation
Eq.(\ref{DX}) in the space of polynomials. 
This establishes the link with Sheffer-type
polynomials. Next the specific matrix elements of the above
operators are derived and then, with the help of coherent state
theory, extended to the general form. Finally we obtain the
normally ordered expression for these operators. It turns out that
the exponential generating functions in the case of linear
dependence on the annihilation (or creation) operator are of
Sheffer-type, and  that assures their convergence.
\\
In the end we give some examples with special emphasis on their
Sheffer-type origin.
We also refer to Section \ref{SubsTheorem} for other application
of derived formulas.

\section{Monomiality principle}

Here we introduce the concept of monomiality which arises from the
action of the multiplication and derivative operators on
monomials. Next we provide a wide class of representations of that
property in the framework of Sheffer-type polynomials. Finally we
establish the correspondence to the occupation number
representation.

\subsection{Definition and general properties}

Let us consider the Heisenberg-Weyl algebra satisfying the
commutation relation
\begin{eqnarray}\label{PM}
[P,M]=1.
\end{eqnarray}
In Section \ref{DXRep} we have already mentioned that a convenient
representation of Eq.(\ref{PM}) is the derivative $D$ and
multiplication $X$ representation defined in the space of
polynomials. Action of these operators on the monomials is given
by Eq.(\ref{DXMono}). Here we extend this framework.
\\
Suppose one wants to find the representations of Eq.(\ref{PM})
such that the action of $M$ and $P$ on certain polynomials
$s_n(x)$ is analogous to the action of $X$ and $D$ on monomials.
More specifically one searches for $M$ and $P$ and their
associated polynomials $s_n(x)$ (of degree $n$, $n=0,1,2,...$)
which satisfy
\begin{eqnarray}\label{Monomiality}
\begin{array}{l}
Ms_n(x)=s_{n+1}(x),\\
Ps_n(x)=n\ s_{n-1}(x).
\end{array}
\end{eqnarray}
The rule embodied in Eq.(\ref{Monomiality}) is called the {\em
monomiality principle}. The polynomials $s_n(x)$ are then called
{\it quasi-monomials} with respect to operators $M$ and $P$. These
operators can be immediately recognized as raising and lowering
operators acting on the $s_n(x)$'s.
\\
The definition Eq.(\ref{Monomiality}) implies some general
properties. First the operators $M$ and $P$ obviously satisfy
Eq.(\ref{PM}). Further consequence of Eq.(\ref{Monomiality}) is
the eigenproperty of $MP$, {\it i.e.}
\begin{eqnarray}
MPs_n(x)=ns_n(x).
\end{eqnarray}
The polynomials $s_n(x)$ may be obtained through the action of
$M^n$ on $s_0$
\begin{eqnarray}\label{Mn}
s_n(x)=M^ns_0
\end{eqnarray}
and consequently the exponential generating function of $s_n(x)$'s
is
\begin{eqnarray}\label{G}
G(\lambda,x)\equiv \sum_{n=0}^\infty
s_n(x)\frac{\lambda^n}{n!}=e^{\lambda M}s_0.
\end{eqnarray}
Also, if we write the quasimonomial $s_n(x)$ explicitly as
\begin{eqnarray}\label{S00}
s_n(x)=\sum_{k=0}^n s_{n,k}\ x^k,
\end{eqnarray}
then
\begin{eqnarray}\label{SX}
s_n(x)=\left(\sum_{k=0}^n s_{n,k}\ X^k\right)1.
\end{eqnarray}
Several types of such polynomial sequences were studied recently
using this monomiality principle
\cite{DattoliNuovoCim}\cite{Dattoli1997}\cite{Dattoli1999}\cite{Dattoli2001}\cite{Cesarano2000}.

\subsection{Monomiality principle representations: Sheffer-type polynomials}
\label{representations}
Here we  show that if $s_n(x)$ are of \emph{Sheffer-type} then it
is possible to give explicit representations of $M$ and $P$.
Conversely, if  $M=M(X,D)$ and $P=P(D)$ then $s_n(x)$ of
Eq.(\ref{Monomiality}) are necessarily of Sheffer-type.
\\
Properties of Sheffer-type polynomials are naturally handled
within the so called {\it umbral calculus}
\cite{Roman}\cite{Rota}\cite{DiBucchianico} (see Appendix
\ref{Umbral}). Here we put special emphasis on their ladder
structure. Suppose we have a polynomial sequence $s_n(x)$,
$n=0,1,2,...$ ($s_n(x)$ being a polynomial of degree $n$). It is
called of a Sheffer A-type zero \cite{Sheffer},\cite{Rainville}
(which we shall call Sheffer-type) if there exists a function
$f(x)$ such that
\begin{eqnarray}\label{S0}
f(D)s_n(x)=ns_{n-1}(x).
\end{eqnarray}
Operator $f(D)$ plays the role of the lowering operator. This
characterization is not unique, {\it i.e.} there are many
Sheffer-type sequences $s_n(x)$ satisfying Eq.(\ref{S0}) for given
$f(x)$. We can further classify them by postulating the existence
of the associated raising operator. A general theorem
\cite{Roman}\cite{Cheikh} states that a polynomial sequence
$s_n(x)$ satisfying the monomiality principle
Eq.(\ref{Monomiality}) with an operator $P$ given as a function of
the derivative operator only $P=P(D)$ is {\it uniquely}
determined by two functions $f(x)$ and $g(x)$ such that $f(0)=0$,
$f^{'}(0)\neq0$ and $g(0)\neq 0$. The exponential generating
function of $s_n(x)$ is then equal to
\begin{eqnarray}\label{egf}
G(\lambda,x)=\sum_{n=0}^\infty
s_n(x)\frac{\lambda^n}{n!}=\frac{1}{g(f^{-1}(\lambda))}\
e^{xf^{-1}(\lambda)},
\end{eqnarray}
and their associated raising and lowering operators of
Eq.(\ref{Monomiality}) are given by
\begin{eqnarray}\label{pm}
\begin{array}{l}
P=f(D),\\
M=\left[X-\frac{g'(D)}{g(D)}\right]\frac{1}{f'(D)}\ .
\end{array}
\end{eqnarray}
Observe the important fact that $X$ enters $M$ only linearly. Note
also that the order of $X$ and $D$ in $M(X,D)$ matters.
\\
The above also holds true for $f(x)$ and $g(x)$ which are formal
power series.
\\
By direct calculation one may check that any pair $M$, $P$ from
Eq.(\ref{pm}) automatically satisfies Eq.(\ref{PM}). The detailed
proof can be found in {\it e.g.} \cite{Roman}, \cite{Cheikh}.
\\
Here are some examples we have  obtained of representations of the
monomiality principle Eq.(\ref{Monomiality}) and their associated
Sheffer-type polynomials:

\begin{itemize}
\item[a)]{$M(X,D)=2X-D$, \ \ \ \ $P(D)=\frac{1}{2}D$,

$s_n(x)=H_n(x)$ - Hermite polynomials;

$G(\lambda,x)=e^{2\lambda x-\lambda^2}$.}

\item[b)]{$M(X,D)=-XD^2+(2X-1)D-X-1$, \ \ \ \ $P(D)=-\sum_{n=1}^\infty D^n$,

$s_n(x)=n!L_n(x)$ - where $L_n(x)$ are  Laguerre polynomials;

$G(\lambda,x)=\frac{1}{1-\lambda}e^{x\frac{\lambda}{\lambda -
1}}$.}

\item[c)]{$M(X,D)=X\frac{1}{1-D}$, \ \ \ \ $P(D)=-\frac{1}{2}D^2+D$,

$s_n(x)=P_n(x)$ - Bessel polynomials \cite{Grosswald};

$G(\lambda,x)=e^{x(1-\sqrt{1-2\lambda})}$.}

\item[d)]{$M(X,D)=X(1+D)$, \ \ \ \ $P(D)=\ln(1+D)$,

$s_n(x)=B_n(x)$ - Bell polynomials;

$G(\lambda,x)=e^{x(e^\lambda-1)}$.}

\item[e)]{$M(X,D)=Xe^{-D}$, \ \ \ \ $P(D)=e^D-1$,

$s_n(x)=x^{\underline{n}}$ - the lower factorial
polynomials \cite{Turbiner};

$G(\lambda,x)=e^{x\ln (1+\lambda )}$.\vspace{2mm}}

\item[f)]{$M(X,D)=(X-\tan (D))\cos^2(D)$, \ \ \ \ $P(D)=\tan (D)$,

$s_n(x)=R_n(x)$ - Hahn polynomials \cite{BenderHahn};

$G(\lambda,x)=\frac{1}{\sqrt{1+\lambda^2}}e^{x\arctan(\lambda)}$.}

\item[g)]{$M(X,D)=X\frac{1+W_L(D)}{W_L(D)}D$, \ \ \ \ $P(D)=W_L (D)$,

where $W_L(x)$ is the Lambert $W$ function \cite{KnuthW};

$s_n(x)=I_n(x)$ - the idempotent polynomials \cite{Comtet};

$G(\lambda,x)=e^{x\lambda e^\lambda}$.}
\end{itemize}

\subsection{Monomiality vs Fock space representations}
\label{correspondence}
We have already called operators $M$ and $P$ satisfying
Eq.(\ref{Monomiality}) the rising and lowering operators. Indeed,
their action rises and lowers index $n$ of the quasimonomial
$s_n(x)$. This resembles the property of  creation $a^\dag$ and
annihilation $a$ operators in the Fock space (see Section
\ref{Boson operators}) given by
\begin{eqnarray}
\begin{array}{rcr}
a |n\rangle&=&\sqrt{n}\ |n-1\rangle\\
a^\dag|n\rangle&=&\sqrt{n+1}\ |n+1\rangle.
\end{array}
\end{eqnarray}
These relations are almost the same as Eq.(\ref{Monomiality}).
There is a difference in coefficients. To make them analogous it
is convenient to redefine the number states $|n\rangle$ as
\begin{eqnarray}
\widetilde{|n\rangle}=\sqrt{n!}\ |n\rangle.
\end{eqnarray}
(Note that $\widetilde{|0\rangle}\equiv|0\rangle$).
\\
Then the creation and annihilation operators act as
\begin{eqnarray}
\begin{array}{rcl}
a \widetilde{|n\rangle}&=&n\ \widetilde{|n-1\rangle}\\
a^\dag\widetilde{|n\rangle}&=&\widetilde{|n+1\rangle}.
\end{array}
\end{eqnarray}
Now this exactly mirrors the the relation in
Eq.(\ref{Monomiality}). So, we make the correspondence
\begin{eqnarray}
\begin{array}{ccl}
P&\ \ \longleftrightarrow \ \ &a \\
M&\ \ \longleftrightarrow \ \ &a^\dag\\
s_n(x)&\ \ \longleftrightarrow \ \ &\widetilde{|n\rangle}\ ,\ \ \ \ \ n=0,1,2,...\ \ . \\
\end{array}
\end{eqnarray}
We note that this identification is purely algebraic, {\it i.e.}
we are concerned here only with the commutation relation
Eqs.(\ref{PM}), (\ref{DX}) or (\ref{HW}). We do not impose the
scalar product in the space of polynomials nor consider the
conjugacy properties of the operators. These properties are
irrelevant for our proceeding discussion. We note only that they
may be rigorously introduced, see {\it e.g.} \cite{Roman}.

\section{Normal ordering via monomiality}

In this section we shall exploit the correspondence of Section
\ref{correspondence} to obtain the normally ordered expression of
powers and exponential of the operators $a^\dag q(a)+v(a)$ and (by
the conjugacy property) $q(a^\dag)a+v(a^\dag)$. 
To this end shall apply the
results of Section \ref{representations} to calculate some
specific coherent state matrix elements of operators in question
and then through the exponential mapping property we will extend
it to a general matrix element. In conclusion we shall also comment on
other forms of linear dependence on $a$ or $a^\dag$.
\\
We use the correspondence of Section \ref{correspondence}
cast in the simplest form for $M=X$, $P=D$ and $s_n(x)=x^n$, {\it
i.e.}
\begin{eqnarray}\label{XDa}
\begin{array}{ccl}
D&\ \ \longleftrightarrow \ \ &a \\
X&\ \ \longleftrightarrow \ \ &a^\dag\\
x^n&\ \ \longleftrightarrow \ \ &\widetilde{|n\rangle}\ ,\ \ \ \ \ n=0,1,2,...\ \ . \\
\end{array}
\end{eqnarray}
Then we recall the representation Eq.(\ref{pm}) of operators $M$
and $P$ in terms of $X$ and $D$. Applying the correspondence of
Eq.(\ref{XDa}) it takes the form
\begin{eqnarray}\label{PMa}
\begin{array}{l}
P(a)=f(a),\\
M(a,a^\dag)=\left[a^\dag-\frac{g'(a)}{g(a)}\right]\frac{1}{f'(a)}\
.
\end{array}
\end{eqnarray}
Recalling Eqs.(\ref{Mn}),(\ref{S00}) and (\ref{SX}) we get
\begin{eqnarray}
\left[M(a,a^\dag)\right]^n|0\rangle=\sum_{k=0}^n s_{n,k}\ (a^\dag)^k|0\rangle.
\end{eqnarray}
In the coherent state representation it yields
\begin{eqnarray}\label{M0}
\langle z|\left[M(a,a^\dag)\right]^n|0\rangle=s_n(z^*)\langle z|0\rangle.
\end{eqnarray}
Exponentiating $M(a,a^\dag)$ and using Eq.(\ref{egf}) we obtain
\begin{eqnarray}\label{EM0}
\langle z|e^{\lambda M(a,a^\dag)}|0\rangle=\frac{1}{g(f^{-1}(\lambda))}\ e^{z^*f^{-1}(\lambda)}\langle z|0\rangle.
\end{eqnarray}
By the same token one obtains closed form expressions for the
following matrix elements ( $|l\rangle$ is the $l$-th number
state, $l=0,1,2,...$)
\begin{eqnarray}\label{Ml}
\langle z|\left[M(a,a^\dag)\right]^n|l\rangle=\frac{1}{\sqrt{l!}}s_{n+l}(z^*)\langle z|0\rangle,
\end{eqnarray}
and when property Eq.(\ref{DF}) is applied
\begin{eqnarray}\label{El}
\langle z|e^{\lambda M(a,a^\dag)}|l\rangle=\frac{1}{\sqrt{l!}}
\frac{d^l}{d\lambda^l}\left[\frac{1}{g(f^{-1}(\lambda))}\ e^{z^*f^{-1}(\lambda)}\right]\langle z|0\rangle.
\end{eqnarray}
Observe that in both Eqs.(\ref{M0}) and (\ref{Ml}) we obtain
Sheffer-type polynomials (modulus coherent states overlapping
factor $\langle z|0\rangle$). Also Eqs.(\ref{EM0}) and (\ref{El})
reveal that property through the Sheffer-type generating
function. This connection will be explored in detail in Section
\ref{sheffernormal}.
\\
The result of Eq.(\ref{EM0}) can be further extended to the
general matrix element $\langle z|e^{\lambda
M(a,a^\dag)}|z'\rangle$. To this end recall Eq.(\ref{z0}) and
write
\begin{eqnarray}\nonumber
\langle z|e^{\lambda M(a,a^\dag)}|z'\rangle&=&e^{-\frac{1}{2}|z'|^2}\langle z|e^{\lambda M(a,a^\dag)}e^{z'a^\dag}|0\rangle\\\nonumber
&=&e^{-\frac{1}{2}|z'|^2}\langle
z|e^{z'a^\dag}e^{-z'a^\dag}e^{\lambda
M(a,a^\dag)}e^{z'a^\dag}|0\rangle
\\\nonumber &=&e^{z^*z'-\frac{1}{2}|z'|^2}\langle z|e^{-z'a^\dag}e^{\lambda M(a,a^\dag)}e^{z'a^\dag}|0\rangle.
\end{eqnarray}
Next, using the exponential mapping property Eq.(\ref{eMe}) we
arrive at
\begin{eqnarray}\nonumber
\langle z|e^{\lambda M(a,a^\dag)}|z'\rangle&=&e^{z^*z'-\frac{1}{2}|z'|^2}\langle z|e^{\lambda M(a+z',a^\dag)}|0\rangle\\\nonumber
&=&e^{z^*z'-\frac{1}{2}|z'|^2}\langle
z|e^{\lambda\left(a^\dag-\frac{g'(a+z')}{g(a+z')}\right)
\frac{1}{f'(a+z')}}|0\rangle.
\end{eqnarray}
Now we are almost ready to apply Eq.(\ref{EM0}) to evaluate the
matrix element on the r.h.s. of the above equation. Before doing
so we have to appropriately redefine functions $f(x)$ and $g(x)$
in the following way ($z'$ - fixed)
\begin{eqnarray}\nonumber
    f(x)&\to& \tilde{f}(x)=f(x+z')-f(z'),\\\nonumber
    g(x)&\to& \tilde{g}(x)=g(x+z')/g(z').
\end{eqnarray}
Then $\tilde{f}(0)=0$, $\tilde{f}'(0)\neq 0$ and $\tilde{g}(0)=1$
as required by Sheffer property for $\tilde{f}(x)$ and
$\tilde{g}(x)$. Observe that these conditions are not fulfilled by
$f(x+z')$ and $g(x+z')$. This step imposes (analytical)
constraints on $z'$, {\it i.e.} it is valid whenever
$\tilde{f}'(z')\neq 0$ (although, we note that for formal power
series approach this does not present any difficulty). Now we can
write
\begin{eqnarray}\nonumber
\langle z|e^{\lambda\left(a^\dag-\frac{g'(a+z')}{g(a+z')}\right)
\frac{1}{f'(a+z')}}|0\rangle&=&\langle z|e^{\lambda\left(a^\dag-\frac{\tilde{g}'(a)}{\tilde{g}(a)}\right)
\frac{1}{\tilde{f}'(a)}}|0\rangle\\\nonumber
&\stackrel{(\ref{EM0})}{=}&\frac{1}{\tilde{g}(\tilde{f}^{-1}(\lambda))}\
e^{z^*\tilde{f}^{-1}(\lambda)}\langle z|0\rangle.
\end{eqnarray}
By going back to the initial functions $f(x)$ and $g(x)$ this
readily gives the final result
\begin{eqnarray}\label{Ze}
\langle z|e^{\lambda M(a,a^\dag)}|z'\rangle
=\frac{g(z')}{g(f^{-1}(\lambda+f(z')))}e^{z^*[f^{-1}(\lambda+f(z'))-z']}\langle
z|z'\rangle,
\end{eqnarray}
where $\langle
z|z'\rangle=e^{z^*z'-\frac{1}{2}|z^{'}|^2-\frac{1}{2}|z|^2}$ is
the coherent states overlapping factor (see Eq.(\ref{zz})).
\\
To obtain the normally ordered form of $e^{\lambda M(a,a^\dag)}$ we
apply the crucial property of the coherent state representation of
Eqs.(\ref{n}) and (\ref{N}). Then Eq.(\ref{Ze}) provides the
central result
\begin{eqnarray}\label{Normal}
e^{\lambda M(a,a^\dag)}=\
:e^{a^\dag[f^{-1}(\lambda+f(a))-a]}\frac{g(a)}{g(f^{-1}(\lambda+f(a)))}:\
.
\end{eqnarray}
Let us point out again that $a^\dag$ appears linearly in
$M(a,a^\dag)$, see Eq.(\ref{PMa}). We also note that the
constraints for functions $f(x)$ and $g(x)$, {\it i.e.} $f(0)=0$,
$f'(0)\neq0$ and $g(0)\neq 0$,  play no important role.
For convenience (simplicity) we put
\begin{eqnarray}\nonumber
    q(x)&=&\frac{1}{f'(x)},\\\nonumber
    v(x)&=&\frac{g'(x)}{g(x)}\frac{1}{f'(x)},
\end{eqnarray}
and define
\begin{eqnarray}\nonumber
    T(\lambda,x)&=&f^{-1}(\lambda+f(x)),\\\nonumber
    G(\lambda,x)&=&\frac{g(x)}{g(T(\lambda,x))}.
\end{eqnarray}
This allows us to rewrite the main normal ordering formula of
Eq.(\ref{Normal}) as
\begin{eqnarray}\label{normal}
e^{\lambda \left(a^\dag q(a)+v(a)\right)}=
\ :e^{a^\dag[T(\lambda,a)-a]}\ G(\lambda,a):
\end{eqnarray}
where the functions $T(\lambda,x)$ and $G(\lambda,x)$ fulfill the
following differential equations
\begin{eqnarray}\label{t}
\frac{\partial T(\lambda,x)}{\partial\lambda} =  q(T(\lambda,x))\ , &~~~~~~~~~~~~~~&T(0,x) = x\ ,
\end{eqnarray}
\begin{eqnarray}
\label{g}
\frac{\partial G(\lambda,x)}{\partial\lambda} =  v(T(\lambda,x))\cdot G(\lambda,x)\ , &~~~~&G(0,x) = 1\ .
\end{eqnarray}
In the coherent state representation it takes the form
\begin{eqnarray}\label{zQz}
\langle z'|e^{\lambda \left(a^\dag q(a)+v(a)\right)}|z\rangle=\langle z'|z\rangle\
e^{z'^*[T(\lambda,z)-z]} G(\lambda,z).
\end{eqnarray}
We conclude by making a comment on other possible forms of linear
dependence on $a$ or $a^\dag$.
\\
By hermitian conjugation of Eq.(\ref{normal}) we obtain the
expression for the normal form of $e^{\lambda
\left(q(a^\dag)a+v(a)\right)}$. This amounts to the formula
\begin{eqnarray}\label{aqa}
e^{\lambda \left(q(a^\dag)a+v(a)\right)}=
\ :G(\lambda,a^\dag)e^{[T(\lambda,a^\dag)-a^\dag]a}:\
\end{eqnarray}
with the same differential equations Eqs.(\ref{t}) and (\ref{g})
for functions $T(\lambda,x)$ and $G(\lambda,x)$. In the coherent
state representation it yields
\begin{eqnarray}\label{zqz}
\langle z'|e^{\lambda \left(q(a^\dag)a+v(a)\right)}|z\rangle=\langle z'|z\rangle\
 G(\lambda,z'^*)e^{[T(\lambda,z'^*)-z'^*]z}
\end{eqnarray}
We also note that all other operators linearly dependent on $a$ or
$a^\dag$ may be written in just considered forms with the use of
Eq.(\ref{af}), {\it i.e.}
$aq(a^\dag)+v(a^\dag)=q(a^\dag)a+q'(a^\dag)+v(a^\dag)$ and
$q(a)a^\dag+v(a)=a^\dag q(a)+q'(a)+v(a)$.
\\
Observe that from the analytical point of view certain limitations 
on the domains of $z$, $z'$ and $\lambda$ should be put in some specific cases 
(locally around zero all the formulas hold true). Also we point out
the fact that functions $q(x)$ and $v(x)$ (or equivalently $f(x)$ and $g(x)$)
may be taken as the formal power series. 
Then one stays on the ground of formal power expansions.

In the end we refer to Section \ref{SubsTheorem} where these
results are applied to derive the substitution formula. We note
that the reverse process, {\it i.e.} derivation of the normally
ordered form from the substitution theorem, is also possible, see
\cite{BlasiakPLA2005}.

\section{Sheffer-type polynomials and normal ordering: Examples}
\label{ShefferExamples}

We now proceed to examples. We will put special emphasis on their
Sheffer-type origin.

\subsection{Examples}

We start with enumerating some examples of the evaluation of the
coherent state matrix elements of Eqs.(\ref{M0}) and (\ref{zQz}). We
choose the $M(a,a^\dag)$'s as in the list a) -  g) in Section
\ref{representations}:

\begin{itemize}

\item[a)]{$\langle z|(-a+2a^\dag)^n|0\rangle =H_n(z^*)\langle z|0\rangle$, Hermite polynomials;

$\langle z|e^{\lambda (-a+2a^\dag)}|z'\rangle=e^{\lambda(2 z^*-
z')-\lambda^2}\langle z|z'\rangle$.}

\item[b)]{$\langle z|\left[-a^\dag a+(2a^\dag-1)a-a^\dag+1\right]^n|0\rangle =n!L_{n-1}(z^*)\langle z|0\rangle$,

Laguerre polynomials;

$\langle z|e^{\lambda \left[-a^\dag
a+(2a^\dag-1)a-a^\dag+1\right]}|z'\rangle
=\frac{1}{1-\lambda(z'-1)
}e^{z^*\lambda\frac{(1-z')^2}{\lambda(1-z')-1}} \langle
z|z'\rangle$.}

\item[c)]{$\langle z|\left(a^\dag\frac{1}{1-a}\right)^n|0\rangle =P_n(z^*)\langle z|0\rangle$, Bessel polynomials;

$\langle z|e^{\lambda \left(a^\dag\frac{1}{1-a}\right)}|z'\rangle
=e^{z^*[1-\sqrt{1-2(\lambda+z'-\frac{1}{2}z'^2)}-z']}\langle
z|z'\rangle$.}

\item[d)]{$\langle z|(a^\dag a+a^\dag)^n|0\rangle =B_n(z^*)\langle z|0\rangle$, Bell polynomials;

$\langle z|e^{\lambda (a^\dag
a+a^\dag)}|z'\rangle=e^{z^*(z'+1)(e^\lambda-1)}\langle
z|z'\rangle$.}

\item[e)]{$\langle z|(a^\dag e^{-a})^n|0\rangle =(z^*)^{\underline{n}}\langle z|0\rangle$, the lower factorial polynomials;

$\langle z|e^{\lambda (a^\dag
e^{-a})}|z'\rangle=e^{z^*[\ln(e^{z'}+\lambda)-z']}\langle
z|z'\rangle$.}

\item[f)]{$\langle z|\left[(a^\dag-\tan (a))\cos^2(a)\right]^n|0\rangle =R_n(z^*)\langle z|0\rangle$, Hahn polynomials;

$\langle z|e^{\lambda (a^\dag-\tan (a))\cos^2(a)}|z'\rangle
=\frac{\cos[\arctan
(\lambda+\tan(z'))]}{cos(z')}e^{z^*[\arctan(\lambda\tan(z'))-z']}\langle
z|z'\rangle$.}

\item[g)]{$\langle z|\left[a^\dag\frac{1+W_L(a)}{W_L(a)}a\right]^n|0\rangle =I_n(z^*)\langle z|0\rangle$, the idempotent polynomials;

$\langle z|e^{\lambda
a^\dag\frac{1+W_L(a)}{W_L(a)}a}|z'\rangle=e^{z^*[\lambda
e^{\lambda+W_L(z')}+z'(e^\lambda-1)]}\langle z|z'\rangle$.}
\end{itemize}
Note that for $z'=0$ we obtain the exponential generating
functions of appropriate polynomials multiplied by the coherent
states overlapping factor $\langle z|0\rangle$, see Eq.(\ref{Ze}).
\\
These examples show how the Sheffer-type polynomials and their
exponential generating functions arise in the coherent state
representation. This generic structure is the consequence of
Eqs.(\ref{M0}) and (\ref{Ze}) or in general Eqs.(\ref{zQz}) or
(\ref{zqz}) and it will be investigated in more detail now.
Afterwords we shall provide more examples of combinatorial origin.

\subsection{Sheffer polynomials and normal ordering}
\label{sheffernormal}

First recall the definition of the family of Sheffer-type
polynomials $s_n(z)$ defined (see Section \ref{Umbral}) through
the exponential generating function as
\begin{eqnarray}\label{egfsheffer}
G(\lambda,x)=\sum_{n=0}^\infty
s_n(z)\frac{\lambda^n}{n!}=A(\lambda)\ e^{zB(\lambda)}
\end{eqnarray}
where functions $A(\lambda)$ and $B(\lambda)$ satisfy: $A(0)\neq
0$ and $B(0)=0$, $B'(0)\neq 0$.
\\
Returning to normal ordering, recall that the coherent state
expectation value of Eq.(\ref{aqa}) is given by Eq.(\ref{zqz}).
When one
\underline{fixes} $z'$ and takes $\lambda$ and $z$ as indeterminates, then the r.h.s. of Eq.(\ref{zqz})
may be read off as an exponential generating function of
Sheffer-type polynomials defined by Eq.(\ref{egfsheffer}). The
correspondence is given by
\begin{eqnarray}
\label{AB1}
A(\lambda)=g(\lambda,z'^*),\\
\label{AB2}
B(\lambda)=\left[T(\lambda,z'^*)-z'^*\right].
\end{eqnarray}
This allows us to make the statement that the coherent state
expectation value $\langle z'|...|z\rangle$ of the operator
$\exp\left[\lambda(q(a^\dag)a+v(a^\dag))\right]$ for any fixed
$z'$ yields (up to the overlapping factor $\langle z'|z\rangle$)
the exponential generating function of a certain sequence of
Sheffer-type polynomials in the variable $z$ given by
Eqs.(\ref{AB1}) and (\ref{AB2}). The above construction
establishes the connection between the coherent state
representation of the operator
$\exp\left[\lambda(q(a^\dag)a+v(a^\dag))\right]$ and a family of
Sheffer-type polynomials $s^{(q,v)}_n(z)$ related to $q$ and $v$
through
\begin{eqnarray}
\label{shefferseq}
\langle z^{\prime}|e^{\lambda\left[q(a^{\dag})a + v(a^{\dag})\right]}|z\rangle =
\langle z^{\prime}|z\rangle\left( 1+\sum_{n=1}^\infty s_n^{(q,v)}(z)\frac{\lambda^n}{n!}\right),
\end{eqnarray}
where explicitly (again for $z'$ fixed):
\begin{equation}
\label{shefferseq2}
\begin{array}{rcl}
s_n^{(q,v)}(z)=\langle z^{\prime}|z\rangle^{-1}\langle
z^{\prime}|\left[q(a^{\dag})a + v(a^{\dag})\right]^n|z\rangle.
\end{array}
\end{equation}
We observe that Eq.(\ref{shefferseq2}) is an extension of the seminal formula of J. Katriel
\cite{Katriel},\cite{Katriel2000} where $v(x)=0$ and $q(x)=x$. The Sheffer-type polynomials are
in this case Bell polynomials expressible through the Stirling
numbers of the second kind (see Section \ref{Generic}).
\\
Having established relations leading from the normal ordering
problem to Sheffer-type polynomials we may consider the reverse
approach. Indeed, it turns out that for any Sheffer-type sequence
generated by $A(\lambda)$ and $B(\lambda)$ one can find functions
$q(x)$ and $v(x)$ such that the coherent state expectation value
$\langle
z'|\exp\left[\lambda(q(a^\dag)a+v(a^\dag))\right]|z\rangle$
results in a corresponding exponential generating function of
Eq.(\ref{egfsheffer}) in indeterminates $z$ and $\lambda$ (up to
the overlapping factor $\langle z'|z\rangle$ and $z'$ fixed).
Appropriate formulas can be derived from Eqs.(\ref{AB1}) and
(\ref{AB2}) by substitution into Eqs.(\ref{t}) and (\ref{g}):
\begin{eqnarray}\label{111}
q(x)&=&B'(B^{-1}(x-z'^*)),\\\label{2} v(x)&=&\frac{A'(B^{-1
}(x-z'^*))}{A(B^{-1 }(x-z'^*))}.
\end{eqnarray}
\noindent One can check that this choice of $q(x)$ and $v(x)$, if inserted into
Eqs. (\ref{t}) and (\ref{g}), results in
\begin{eqnarray}\label{3}
T(\lambda,x)&=&B(\lambda+B^{-1 }(x-z'^*))+z'^*,\\\label{4}
g(\lambda,x)&=&\frac{A(\lambda+B^{-1}(x-z'^*))}{A(B^{-1}(x-z'^*))},
\end{eqnarray}
which reproduce
\begin{eqnarray}\label{zABz}
\langle z^{\prime}|e^{\lambda\left[q(a^{\dag})a + v(a^{\dag})\right]}|z\rangle = \langle z^{\prime}|z\rangle A(\lambda)e^{zB(\lambda)}.
\end{eqnarray}
\noindent The result summarized in Eqs.(\ref{AB1}) and (\ref{AB2}) and in their 'dual' forms Eqs.(\ref{111})-(\ref{4})
provide us with a considerable flexibility in conceiving and
analyzing a large number of examples.

\subsection{Combinatorial examples}
In this section we will work out examples illustrating how the
exponential generating function $G(\lambda)=\sum_{n=0}^\infty
a_n\frac{x^n}{n!}$ of certain combinatorial sequences
$(a_n)_{n=0}^\infty$  appear naturally in the context of boson
normal ordering. To this end we shall assume specific forms of
$q(x)$ and $v(x)$ thus specifying the operator that we
exponentiate. We then give solutions to Eqs.(\ref{t}) and
(\ref{g}) and subsequently through Eqs.(\ref{AB1}) and (\ref{AB2})
we write the exponential generating function of combinatorial
sequences whose interpretation will be given.

\begin{itemize}
\item[a)]{ Choose $q(x)=x^r$, $r>1$ (integer), $v(x)=0$ (which implies
$g(\lambda,x)=1$). Then $T(\lambda,x) = x\left[1 - \lambda(r -
1)x^{r-1}\right]^{\frac{1}{1-r}}$. This gives
\begin{eqnarray}\nonumber
{\cal N}\left[e^{\lambda (a^\dag)^ra}\right] \equiv
\ : \exp\left[\left(\frac{a^\dag}{\left(1 - \lambda(r - 1)(a^\dag)^{r-1}\right)^{\frac{1}{r-1}}}-1\right)a\right]:\
\end{eqnarray}
as the normally ordered form. Now we take $z^{'}=1$ in
Eqs.(\ref{AB1}) and (\ref{AB2}) and from Eq.(\ref{zABz}) we obtain
\begin{eqnarray}\nonumber
\langle 1|z\rangle^{-1}\langle 1|e^{\lambda (a^\dag)^ra}|z\rangle  =\ \exp\left[z\left(\frac{1}{\left(1 - \lambda(r - 1)\right)^{\frac{1}{r-1}}}-1\right)\right]\ ,
\end{eqnarray}
which for $z=1$ generates the following sequences:
\begin{eqnarray}\nonumber
\begin{array}{lcl}
r=2:&&a_n=1,1,3,13,73,501,4051,...\\
r=3:&&a_n=1,1,4,25,211,2236,28471,...\ \ \ \ \ \ \ \ ,\ {\rm etc.}
\end{array}
\end{eqnarray}
These sequences enumerate $r$-ary forests
\cite{Sloane}\cite{Stanley}\cite{Flajolet}.}

\item[b)]{ For $q(x)=x\ln(ex)$ and $v(x)=0$ (implying $g(\lambda,x)=1$)
we have $T(\lambda,x)=e^{e^\lambda-1}x^{e^\lambda }$. This
corresponds to
\begin{eqnarray}\nonumber
{\cal N}\left[e^{\lambda a^\dag\ln(ea^\dag)a}\right] \equiv\ :
\exp\left[\left(e^{e^\lambda-1}(a^\dag)^{e^\lambda}-1\right)a\right]:\
,
\end{eqnarray}
whose coherent state matrix element with $z^{'}=1$ is equal to
\begin{eqnarray}\nonumber
\langle 1|z\rangle^{-1}\langle 1|e^{\lambda a^\dag\ln(ea^\dag)a}|z\rangle  = \exp\left[z\left(e^{e^\lambda-1}-1\right)\right]\ ,
\end{eqnarray}
which for $z=1$ generates $a_n=1,1,3,12,60,385,2471,...$
enumerating partitions of partitions \cite{Stanley},
\cite{Sloane}, \cite{Flajolet}.}
\end{itemize}
The following two examples  refer to the reverse procedure, see
Eqs.(\ref{111})-(\ref{4}). We choose first a Sheffer-type
exponential generating function and deduce $q(x)$ and $v(x)$
associated with it.

\begin{itemize}
\item[c)]{ $A(\lambda)=\frac{1}{1-\lambda}$, $B(\lambda)=\lambda$, see
Eq.(\ref{egfsheffer}). This exponential generating function for
$z=1$ counts the number of arrangements
$a_n=n!\sum_{k=0}^n\frac{1}{k!}=1,2,5,65,326,1957,...$ of the set
of $n$ elements \cite{Comtet}. The solutions of Eqs.(\ref{111}) and
(\ref{2}) are: $q(x)=1$ and $v(x)=\frac{1}{2-x}$. In terms of
bosons it corresponds to
\begin{eqnarray}\nonumber
{\cal N}\left[e^{\lambda \left(a+\frac{1}{2-a^\dag}\right)}\right]
\equiv\ :\frac{2-a^\dag}{2-a^\dag-\lambda}e^{\lambda a}:\
=\frac{2-a^\dag}{2-a^\dag-\lambda}e^{\lambda a}.
\end{eqnarray}}

\item[d)]{ For $A(\lambda)=1$ and $B(\lambda)=1-\sqrt{1-2\lambda}$ one
gets the exponential generating function of the Bessel polynomials \cite{Grosswald}. For $z=1$
they enumerate  special paths on a lattice \cite{Pittman}. The
corresponding sequence is $a_n=1,1,7,37,266,2431,...\ $. The
solutions of Eqs.(\ref{111}) and (\ref{2}) are: $q(x)=\frac{1}{2-x}$
and $v(x)=0$. It corresponds to
\begin{eqnarray}\nonumber
{\cal N}\left[e^{\lambda \frac{1}{2-a^\dag}a}\right] \equiv\
:e^{\left(1-\sqrt{(2-a^\dag)-2\lambda}\right)a}:\
\end{eqnarray}
in the boson formalism.}
\end{itemize}
These examples show that any combinatorial structure which can be
described by a Sheffer-type exponential generating function can be
cast in  boson language. This gives rise to a large number of
formulas of the above type which put them in a quantum mechanical
setting.

%%%%%%%%%%%%%%%%%%%%%%%%%%%%%%%%%%%%%%%%%%%%%%%%%%%%%%%%%%%%%%%%%%%%%%%%%%%%%%%%%%%%%

\chapter{Miscellany: Applications}
\label{Miscellany}

\begin{chapterabstract}
\rule[1mm]{\textwidth}{.3pt}
We give three possible extensions of the above formalism. First we
show how this approach is modified when one considers
deformations of the canonical boson algebra. Next we construct and
analyze the generalized coherent states based on some number
series which arise in the ordering problems for which we also
furnish the solution to the moment problem. We end by using
results of Chapter \ref{Sheffer} to derive a substitution formula.
\\
\rule[1mm]{\textwidth}{.3pt}
\end{chapterabstract}

%%%%%%%%%%%%%%%%%%%%%%%%%%%%%%%%%%%%%%%%%%%%%%%%%%%%%%%%%%%%%%%%%%%%

\section{Deformed bosons}
\label{Deformed bosons}

We solve the normal ordering problem for $(A^{\dag}A)^n$ where $A$
and $A^\dag$ are one mode deformed boson ladder operators
($[A,A^\dag]=[N+1]-[N]$). The solution generalizes results known
for canonical bosons (see Chapter \ref{Generic}). It involves combinatorial polynomials in the
number operator $N$ for which the generating function and explicit
expressions are found. Simple deformations provide illustration of the method.

\subsubsection{Introduction}

We consider the general deformation of the boson algebra
\cite{SolomonPLA} in the form
\begin{eqnarray}
\label{algebra}
\begin{array}{l}
{[A,N]}=A,\\
{[A^\dag,N]}=-A^\dag,\vspace{2mm}\\
{[A,A^\dag]}={[N+1]-[N]}.
\end{array}
\end{eqnarray}
In the above $A$ and $A^\dag$ are the (deformed) annihilation and
creation operators, respectively, while the number operator $N$
counts particles. It is defined in the occupation number basis as
$N|n\rangle=n|n\rangle$ and commutes with $A^\dag A$ (which is the
consequence of the first two commutators in Eq.({\ref{algebra})).
Because of that in any representation of (\ref{algebra}) $N$ can
be written in the form $A^\dag A = [N]$, where $[N]$ denotes an
arbitrary function of $N$, usually called the ``box'' function.
Note that this is a generalization of the canonical boson algebra
of Eqs.(\ref{HW}) and (\ref{NN}) described in Section \ref{Boson
operators} (it is recovered for $[N]=N+ const$).
\\
For general considerations we do not assume any  realization of
the number operator $N$ and we treat it as an independent element
of the algebra.  Moreover, we do not assume any particular form of
the ``box'' function $[N]$. Special cases, like the $so(3)$ or
$so(2,1)$ algebras,  provide us with examples showing how such a
general approach simplifies if an algebra and its realization are
chosen.
\\
In the following we give the solution to the problem of normal
ordering of a monomial $(A^\dag A)^n$ in deformed annihilation and
creation operators. It is an extension of the problem for
canonical pair $[a,a^\dag]=1$ described in Chapter \ref{Generic}
where we have considered Stirling numbers $S(n,k)$ arising from
\begin{eqnarray}
\label{S}
(a^\dag a)^n=\sum_{k=1}^n S(n,k) (a^\dag)^k a^k\ \ \ \ \ \ \ \
{\rm{(for\ canonical\ bosons)}}.
\end{eqnarray}
In the case of deformed bosons we cannot express the monomial
$(A^\dag A)^n$ as a combination of normally ordered expressions in
$A^\dag$ and $A$ only. This was found for $q$-deformed bosons some
time ago,
\cite{KatrielKibler}\cite{Katriel2002}\cite{Duchamp1995}\cite{Schork}, and recently
for the R-deformed Heisenberg algebra related to the Calogero
model, \cite{Burdik}. The number operator $N$ occurs in the final
formulae because on commuting creation operators to the left we
cannot get rid of $N$ if it is assumed to be an independent
element of the algebra. In general we can look for a solution of
the form
\begin{eqnarray}
\label{P}
(A^\dag A)^n=\sum_{k=1}^n \mathcal{S}_{n,k}(N)\ (A^\dag)^k A^k,
\end{eqnarray}
where coefficients $\mathcal{S}_{n,k}(N)$ are functions of the
number operator $N$ and their shape depend on the box function
$[N]$. Following \cite{KatrielKibler} we will call them
\emph{operator-valued deformed Stirling numbers} or just
\emph{deformed Stirling numbers}.
\\
In the sequel we give a comprehensive analysis of this
generalization. We shall give recurrences satisfied by
$\mathcal{S}_{n,k}(N)$ of (\ref{P}) and shall construct their
generating functions. This will enable us to write down
$\mathcal{S}_{n,k}(N)$ explicitly and to demonstrate how the
method works on examples of simple Lie-type deformations of the
canonical case.

\subsubsection{Recurrence relations}

One checks by induction that for each $k\geq 1$ the following
relation holds
\begin{eqnarray}\label{a}
[A^k,A^\dag]=([N+k]-[N])A^{k-1}.
\end{eqnarray}
Using this relation it is easy to check by induction that deformed
Stirling numbers satisfy the recurrences\footnote[1]{The
recurrence relation Eq.(\ref{recurrence}) holds for all $n$ and
$k$ if one puts the following ``boundary conditions'':
$\mathcal{S}_{i,j}(N)=0$ for $i=0$ or $j=0$ or $i<j$, except
$\mathcal{S}_{0,0}(N)=1$.}
\begin{eqnarray}
\label{recurrence}
\mathcal{S}_{n+1,k}(N)=\mathcal{S}_{n,k-1}(N)+([N]-[N-k])\mathcal{S}_{n,k}(N)\ \ \ \ \ \textstyle{{\rm for}\ 1<k<n}
\end{eqnarray}
with initial values
\begin{eqnarray}
\label{initialdef}
\mathcal{S}_{n,1}(N)=([N]-[N-1])^{n-1}, \ \ \ \ \ \ \ \mathcal{S}_{n,n}(N)=1.
\end{eqnarray}
The proof can be deduced from the equalities
\begin{eqnarray}\nonumber
&&\sum_{k=1}^{n+1}\mathcal{S}_{n+1,k}(N)\ (A^\dag)^k A^k=(A^\dag
A)^{n+1}=(A^\dag A)^n \ A^\dag A\\\nonumber &&=\sum_{k=1}^n
\mathcal{S}_{n,k}(N)\ (A^\dag)^k\ A^k A^\dag\
A\stackrel{(\ref{a})}{=}
\sum_{k=1}^n \mathcal{S}_{n,k}(N)\ (A^\dag)^k\left(A^\dag A+[N+k]-[N]\right)A^k
\\\nonumber
&&\stackrel{(\ref{algebra})}{=}\sum_{k=2}^{n+1}\mathcal{S}_{n,k-1}(N)\
(A^\dag)^k A^k+\sum_{k=1}^n([N]-[N-k])\mathcal{S}_{n,k}(N)\
(A^\dag)^k A^k.
\end{eqnarray}
It can be shown that each $\mathcal{S}_{n,k}(N)$ is a homogeneous
polynomial of order $n-k$ in variables $[N],...,[N-k]$. For a
polynomial box function  one obtains  $\mathcal{S}_{n,k}(N)$ as a
polynomial in $N$.  As we shall show, a simple example of the
latter is the deformation given by the $so(3)$ or $so(2,1)$ Lie
algebras.
\\
Note that for the ``box'' function $[N]=N+const$ ({\it {\it
i.e.}}, for the canonical algebra) we get the conventional
Stirling numbers of Chapter \ref{Generic}.

\subsubsection{Generating functions and general expressions}

We define the set of \emph{ordinary generating functions} of
polynomials $\mathcal{S}_{n,k}(N)$, for $k\geq 1$, in the form
\begin{eqnarray}\label{Pk}
P_{k}(x,N):=\sum_{n=k}^\infty \mathcal{S}_{n,k}(N)\ x^n.
\end{eqnarray}
The initial conditions (\ref{initialdef}) for $k=1$ give
\begin{eqnarray}\label{AAAA}
P_1(x,N)=\frac{x}{1-([N]-[N-1])\ x}.
\end{eqnarray}
Using the recurrencies (\ref{recurrence}) completed with
(\ref{initialdef}) one finds the relation
\begin{eqnarray}\label{BBBB}
P_k(x,N)=\frac{x}{1-([N]-[N-k])\ x}\ P_{k-1}(x,N),\ \ \ \ \ {\rm
for}\ k>1,
\end{eqnarray}
whose proof is provided by the equalities
\begin{eqnarray}
\nonumber &&P_k(x,N)\stackrel{(\ref{recurrence})}{=}\sum_{n=k}^\infty\left(\mathcal{S}_{n-1,k-1}(N)+([N]-[N-k])\ \mathcal{S}_{n-1,k}(N)\right)\ x^n
\\\nonumber
&&=x\sum_{n=k}^\infty \mathcal{S}_{n-1,k-1}(N)\ x^{n-1}+([N]-[N-k])\
x\sum_{n=k}^\infty \mathcal{S}_{n-1,k}(N)\ x^{n-1}\nonumber
\\\nonumber
&&=x\ P_{k-1}(x,N)+([N]-[N-k])\ x\ P_k (x,N).
\end{eqnarray}
The expressions (\ref{AAAA}) and (\ref{BBBB}) give explicit formula for
the ordinary generating function
\begin{eqnarray}\label{OGF}
P_k(x,N)=\prod_{j=1}^k \frac{x}{1-([N]-[N-j])\ x}.
\end{eqnarray}
For the canonical algebra it results in the ordinary generating
function for Stirling numbers $S(n,k)$:
\begin{eqnarray}
P_k(x)=\frac{x}{(1-x)(1-2x)\cdot...\cdot(1-kx)}.
\end{eqnarray}
Explicit knowledge of the ordinary generating functions
(\ref{OGF}) enables us to find the $\mathcal{S}_{n,k}(N)$ in a
compact form. As a rational function of $x$ it can be expressed as
a sum of partial fractions
\begin{eqnarray}
P_k(x,N)=x^k\ \sum_{r=1}^k\frac{\alpha_r}{1-([N]-[N-r])\ x},
\end{eqnarray}
where
\begin{eqnarray}
\label{alphas}
\alpha_r=\frac{1}{\prod_{j=1,j\neq r}^k\left(1-\frac{[N]-[N-j]}{[N]-[N-r]}\right)}.
\end{eqnarray}
From the definition (\ref{Pk}) we have that $\mathcal{S}_{n,k}(N)$
is the coefficient multiplying $x^n$ in the formal Taylor
expansion of $P_k(x,N)$. Expanding the fractions in above
equations and collecting the terms we get
$\mathcal{S}_{n,k}(N)=\sum_{r=1}^k\alpha_r([N]-[N-r])^{n-k}$.
Finally,  using (\ref{alphas}), we arrive at
\begin{eqnarray}
\mathcal{S}_{n,k}(N)=\sum_{r=1}^k\frac{([N]-[N-r])^{n-1}}{\prod_{j=1,j\neq r}^k([N-j]-[N-r])}
\end{eqnarray}
where monotonicity of $[N]$ was assumed.
\\
For the canonical case it yields the conventional Stirling numbers
$S(n,k)=\frac{1}{k!}\sum_{j=1}^k\binom{k}{j}(-1)^{k-j}j^n$, see
Eq.(\ref{Scan}).
\\
The first four families of general deformed Stirling numbers
(polynomials) defined by (\ref{P}) read
\begin{eqnarray}\nonumber
    \begin{array}{l}
\mathcal{S}_{1,1}(N)=1,\\
\\
\mathcal{S}_{2,1}(N)=[N]-[N-1],\\
\mathcal{S}_{2,2}(N)=1,\\
\\
\mathcal{S}_{3,1}(N)=([N]-[N-1])^2,\\
\mathcal{S}_{3,2}(N)=2[N]-[N-1]-[N-2],\\
\mathcal{S}_{3,3}(N)=1,\\
\\
\mathcal{S}_{4,1}(N)=([N]-[N-1])^3,\\
\mathcal{S}_{4,2}(N)=3[N]^2-3[N][N-1]-3[N][N-2]+[N-1]^2+[N-2][N-1]+[N-2]^2,\\
\mathcal{S}_{4,3}(N)=3[N]-[N-1]-[N-2]-[N-3],\\
\mathcal{S}_{4,4}(N)=1.
  \end{array}
\end{eqnarray}

\subsubsection{Examples of simple deformations}

If we fix the ''box'' function as $[N]=\mp\frac{N(N-1)}{2}$ then
(\ref{algebra}) become structural relations of $so(3)$ and
$so(2,1)$ algebras, respectively. The first four families of
polynomials,
satisfying $\mathcal{S}_{n,k}^{so(2,1)}(N)=(-1)^{n-k}\ \mathcal{S}_{n,k}^{so(3)}(N)$, are% because of (\ref{recurrence}) and (\ref{initialdef}), are
\begin{eqnarray}\nonumber
    \begin{array}{lccl}
    \textbf{so(3)}&&&\textbf{so(2,1)}\\
    \\
    \mathcal{S}_{1,1}(N)=1&&&\mathcal{S}_{1,1}(N)=1\\
    \\
    \mathcal{S}_{2,1}(N)=-N+1&&&\mathcal{S}_{2,1}(N)=N-1\\
    \mathcal{S}_{2,2}(N)=1&&&\mathcal{S}_{2,2}(N)=1\\
    \\
    \mathcal{S}_{3,1}(N)=(N-1)^2&&&\mathcal{S}_{3,1}(N)=(N-1)^2\\
    \mathcal{S}_{3,2}(N)=-3N+4&&&\mathcal{S}_{3,2}(N)=3N-4\\
    \mathcal{S}_{3,3}(N)=1&&&\mathcal{S}_{3,3}(N)=1\\
    \\
    \mathcal{S}_{4,1}(N)=-(N-1)^3&&&\mathcal{S}_{4,1}(N)=(N-1)^3\\
    \mathcal{S}_{4,2}(N)=7N^2-19N+13&&&\mathcal{S}_{4,2}(N)=7N^2-19N+13\\
    \mathcal{S}_{4,3}(N)=-6N+10&&&\mathcal{S}_{4,3}(N)=6N-10\\
    \mathcal{S}_{4,4}(N)=1&&&\mathcal{S}_{4,4}(N)=1\\
    \end{array}
\end{eqnarray}
Up to now we have been considering $A$, $A^\dag$, and $N$ as
independent  elements of the algebra. Choosing the representation
and expressing $N$ in terms of $A$ and $A^\dag$ we arrive at
expressions like (\ref{P}) which can be normally ordered further.
We may consider a realization of the $so(2,1)$ in terms of
canonical operators $a$ and $a^\dag$
\begin{eqnarray}
\label{so21}
A = \frac{1}{2\sqrt{2}}aa
\ \ \ \ \ \
A^\dag= \frac{1}{2\sqrt{2}}a^\dag a^\dag
\ \ \ \ \ \
N = \frac{1}{2}\left(a^\dag a+\frac{1}{2}\right).
\end{eqnarray}
The solution of the normal ordering problem for
$\left((a^\dag)^2a^2\right)^{n}$ was already considered in Section
\ref{rs}, and leads to
\begin{eqnarray}
\label{genstirling1}
\left((a^\dag)^2a^2\right)^n &=& \sum\limits_{k=2}^{2n} S_{2,2}(n,k)a^{\dag k}a^k,
\end{eqnarray}
where $S_{2,2}(n,k)$ are generalized Stirling numbers for which we
have (see Eqs.(\ref{Srr-rec}) and (\ref{Srr-expl}))

\begin{eqnarray}
\label{genstirling2}
S_{2,2}(n,k)&=&\frac{(-1)^k}{ k!}\sum\limits_{p=2}^{k}(-1)^{p}
\binom{k}{p}\left[p(p-1)\right]^n
\\
&=& \sum_{l=0}^n (-1)^{l}\binom{n}{l}S(2n-l,k).
\end{eqnarray}
Terms in (\ref{genstirling1}) with $k$ even are directly
expressible by $A$ and $A^{\dag}$. If $k$ is odd then the factor
$a^{\dag}a$ may be commuted to the left and we get (the convention
of Eq.(\ref{Init}) is used)
\begin{eqnarray}\nonumber
\left((a^\dag)^2a^2\right)^n = \sum\limits_{k=1}^{n} \left(S_{2,2}(n,2k) + S_{2,2}(n,2k+1)(a^\dag a - 2k)\right)a^{\dag2k}a^{2k}.
\end{eqnarray}
Finally, we get
\begin{eqnarray}\nonumber
(A^\dag A)^n=\sum\limits_{k=1}^{n} 8^{k-n}\left(S_{2,2}(n,2k) +
S_{2,2}(n,2k+1)(2N - 2k - 1/2)\right)A^{\dag k}A^{k}.
\end{eqnarray}
This indicates that the general solution involving higher order
polynomials in $N$ can be simplified when a particular realization
of the algebra is postulated.
\newpage

%%%%%%%%%%%%%%%%%%%%%%%%%%%%%%%%%%%%%%%%%%

\section{Generalized coherent states}
\label{Generalized coherent states}

We construct and analyze a family of coherent states built on
sequences of integers originating from the solution of the boson
normal ordering problem investigated in Section \ref{rs}. These
sequences generalize the conventional combinatorial Bell numbers
and are shown to be moments of positive functions. Consequently,
the resulting coherent states automatically satisfy the resolution
of unity condition. In addition they display such non-classical
fluctuation properties as super-Poissonian statistics and
squeezing.

\subsubsection{Introduction}

Since their introduction in quantum optics many generalizations of
standard coherent states have been proposed (see Appendix
\ref{Coherent states}). The main purpose of such generalizations
is to account for a full description of interacting quantum
systems. The conventional coherent states provide a correct
description of a typical non-interacting system, the harmonic
oscillator. One formal approach to this problem is to redefine the
standard boson creation $a^\dag$ and annihilation $a$ operators,
satisfying $[a,a^\dag]=1$, to $A=af(a^\dag a)$, where the function
$f(N)$, $N=a^\dag a$, is chosen to adequately describe the
interacting problem. Any deviation of $f(x)$ from $f(x)=const$
describes a non-linearity in the system. This amounts to
introducing the modified (deformed) commutation relations
\cite{SolomonPLA}\cite{Manko1}\cite{Manko2} (see also Section
\ref{Deformed bosons})
\begin{eqnarray}\label{0}
[A,A^\dag]=[N+1]-[N],
\end{eqnarray}
where the ``box'' function $[N]$ is defined as $[N]=Nf^2(N)>0$.
Such a way of generalizing the boson commutator naturally leads to
generalized ''nonlinear'' coherent states in the form ($[n]!=[0][1]\ldots[n]$,
$[0]=1$)
\begin{eqnarray}\label{1}
|z\rangle = \mathcal{N}^{-1/2}(|z|^2)\sum_{n=0}^\infty
\frac{z^n}{\sqrt{[n]!}}|n\rangle,
\end{eqnarray}
which are eigenstates of the ``deformed'' boson annihilation
operator $A$
\begin{eqnarray}
A|z\rangle =z|z\rangle .
\end{eqnarray}
It is worth pointing out that such nonlinear coherent states have
been successfully applied to a large class of physical problems in
quantum optics
\cite{Vogel1}\cite{Vogel2}. The comprehensive treatment of coherent states of
the form of Eq. (\ref{1}) can be found in
\cite{SolomonPLA}\cite{Manko1}\cite{Manko2}.
\\
An essential ingredient in the definition of coherent states is
the completeness property (or the resolution of unity condition)
\cite{Klauder}\cite{KlauderJMP1}\cite{KlauderJMP2}\cite{KPS1}. A
guideline for the construction of coherent states in general has
been put forward in \cite{KlauderJMP1} as a minimal set of
conditions. Apart from the conditions of normalizability and
continuity in the complex label $z$, this set reduces to
satisfaction of the  resolution of unity condition. This implies
the existence of a positive function $\tilde{W}(|z|^2)$ satisfying
\cite{KPS1}
\begin{eqnarray}\label{A}
\int_{\mathbb{C}}d^2z\ |z\rangle \tilde{W}(|z|^2)\langle z|=I=\sum_{n=0}^\infty|n\rangle \langle n|,
\end{eqnarray}
which reflects the completeness of the set $\{|z\rangle\}$.
\\
In Eq.(\ref{A}) $I$ is the unit operator and $|n\rangle$ is a
complete set of orthonormal eigenvectors. In a general approach
one chooses strictly positive parameters $\rho (n),\ n=0,1,\ldots$
such that the state $|z\rangle$ which is normalized, $\langle
z|z\rangle =1$, is given by
\begin{eqnarray}\label{I}
|z\rangle =\mathcal{N}^{-1/2}(|z|^2)\sum_{n=0}^\infty
\frac{z^n}{\sqrt{\rho(n)}}|n\rangle,
\end{eqnarray}
with normalization
\begin{eqnarray}\label{J}
\mathcal{N}(|z|^2)=\sum_{n=0}^\infty \frac{|z|^{2n}}{\rho(n)}>0,
\end{eqnarray}
which we assume here to be a convergent series in $|z|^2$ for all
$z\in\mathbb{C}$. In view of Eqs. (\ref{0}) and (\ref{1}) this
corresponds to $\rho(n)=[n]!$ or $[n]=\rho(n)/\rho(n-1)$ for
$n=1,2,\ldots$.
\\
Condition (\ref{A}) can be shown to be equivalent to the following
infinite set of equations \cite{KPS1}:
\begin{eqnarray}\label{Bb}
\int_0^\infty x^n\left[\pi \frac{\tilde{W}(x)}{\mathcal{N}(x)}\right]dx=\rho(n),\ \ n=0,1,\ldots,
\end{eqnarray}
which is a Stieltjes moment problem for
$W(x)=\pi\frac{\tilde{W}(x)}{\mathcal{N}(x)}$.
\\
Recently considerable progress was made in finding explicit
solutions of Eq.(\ref{Bb}) for a large set of $\rho(n)$'s,
generalizing the conventional choice $|z\rangle_c$ for which
$\rho_c(n)=n!$ with $\mathcal{N}_c(x)=e^x$ (see Refs.
\cite{KPS1}\cite{PS}\cite{KPS2}\cite{PS2}\cite{Sixdeniers} and references therein),
thereby extending the known families of coherent states. This
progress was facilitated by the observation that when the moments
form certain combinatorial sequences a solution of the associated
Stieltjes moment problem may be obtained explicitly
\cite{PSQTS}\cite{BlasiakJPA2003}\cite{BlasiakJPA2004}.

In the following we make contact with the combinatorial sequences
appearing in the solution of the boson normal ordering problem
considered in  Section \ref{rs}. These sequences have the very
desirable property of being moments of Stieltjes-type measures and
so automatically fulfill the resolution of unity requirement which
is a consequence of the Dobi\'nski-type relations. It is therefore
natural to use these sequences for the coherent states
construction, thereby providing a link between the quantum states
and the combinatorial structures.

\subsubsection{Moment problem}

We recall the definition of the generalized Stirling numbers
$S_{r,1}(n,k)$ of Section \ref{rs},
 defined through ($n,r>0$ integers):
\begin{eqnarray}\label{C}
[(a^\dag)^ra]^n=(a^\dag)^{n(r-1)}\sum_{k=1}^{n}S_{r,1}(n,k)(a^\dag)^ka^k,
\end{eqnarray}
as well as generalized Bell numbers $B_{r,1}(n)$
\begin{eqnarray}\label{Dd}
B_{r,1}(n)=\sum_{k=1}^{n}S_{r,1}(n,k).
\end{eqnarray}
For both $S_{r,1}(n,k)$ and $B_{r,1}(n)$ exact and explicit
formulas have been given in Section \ref{rs}.
\\
For series $B_{r,1}(n)$ a convenient infinite series representation
may be given by the Dobi\'nski-type relations (see Eqs.(\ref{r1}), (\ref{AaaA}) and (\ref{DD})
)
\begin{eqnarray}\label{E}
B_{r,1}(n)=\frac{(r-1)^{n-1}}{e}\sum_{k=0}^\infty
\frac{1}{k!}\frac{\Gamma(n+\frac{k+1}{r-1})}{\Gamma(1+\frac{k+1}{r-1})},\ \ r>1,
\end{eqnarray}
and
\begin{eqnarray}\label{M}
B_{1,1}(n)=\frac{1}{e}\sum_{k=0}^\infty\frac{k^n}{k!}.
\end{eqnarray}
From Eqs.(\ref{C}) and (\ref{Dd}) one directly sees that
$B_{r,1}(n)$ are integers. Explicitly:
$B_{1,1}(n)=1,2,5,15,52,203,\ldots$;
$B_{2,1}(n)=1,3,13,73,501,4051\ldots$;
$B_{3,1}(n)=1,4,25,211,2236,28471,\ldots$;
$B_{4,1}(n)=1,5,41,465,6721,117941\ldots$ etc.
\\
It is essential for our purposes to observe that the integer
$B_{r,1}(n+1)$, $n=0,1,\ldots$ is the $n$-th moment of a positive
function $W_{r,1}(x)$ on the positive half-axis. For $r=1$ we see
from Eq.(\ref{M}) that $B_{1,1}(n+1)$ is the $n$-th moment of a
discrete distribution $W_{1,1}(x)$ located at positive integers, a
so-called {\em Dirac comb}:
\begin{eqnarray}\label{Y}
B_{1,1}(n+1)=\int_0^\infty x^n\left[\frac{1}{e}\sum_{k=1}^\infty
\frac{\delta(x-k)}{(k-1)!}\right]\ dx.
\end{eqnarray}
For every $r>1$ a continuous distribution $W_{r,1}(x)$ will be
obtained by excising $(r-1)^n\ \Gamma(n+\frac{k+1}{r-1})$ from Eq.(\ref{E}), 
performing the inverse Mellin transform on it and
inserting the result back in the sum of Eq.(\ref{E}), see
\cite{KPS1}\cite{Sixdeniers} for details. (Note that $B_{r,1}(0)=\frac{e-1}{e}$,
$r=2,3,\ldots$ is no longer integral). In this way we obtain
\begin{eqnarray}\label{Nnn}
B_{r,1}(n+1)=\int_0^\infty x^nW_{r,1}(x)\ dx,
\end{eqnarray}
which yields for $r=2,3,4$:
\begin{eqnarray}
&&W_{2,1}(x)=e^{-x-1}\sqrt{x}\ I_1(2\sqrt{x}),\label{F}\\\label{Gggg}
&&W_{3,1}(x)=\frac{1}{2}\sqrt{\frac{x}{2}}e^{-\frac{x}{2}-1}\left(\frac{2}{\sqrt{\pi}}\
{_0F_2}(\frac{1}{2},\frac{3}{2};\frac{x}{8})+\frac{x}{\sqrt{2}}\ {_0F_2}(\frac{3}{2},2;\frac{x}{8})\right),\\\label{H}
&&W_{4,1}(x)=\frac{1}{18\pi\Gamma(\frac{2}{3})}e^{-\frac{x}{2}-1}\left(3^{\frac{13}{6}}\Gamma^2(\frac{2}{3})x^{\frac{1}{3}}\
{_0F_3}(\frac{1}{3},\frac{2}{3},\frac{4}{3};\frac{x}{81})+\right.\\
&&\ \ \ \ \ \ \ \ \ \ \ \ \ \ \left.3^{\frac{4}{3}}\pi
x^{\frac{2}{3}}\
{_0F_3}(\frac{2}{3},\frac{4}{3},\frac{5}{3};\frac{x}{81})+\pi\Gamma(\frac{2}{3})x\
{_0F_3}(\frac{4}{3},\frac{5}{3},2;\frac{x}{81})\right).\nonumber
\end{eqnarray}
In Eqs.(\ref{F}), (\ref{Gggg}) and (\ref{H}) $I_\nu (y)$ and
$_0F_p(\ldots;y)$ are modified Bessel and hypergeometric
functions, respectively. Other $W_{r,1}(x)>0$ for $r>4$ can be
generated by essentially the same procedure.\\
In Figure \ref{FigA} we display the weight functions $W_{r,1}(x)$
for $r=1\ldots4$; all of them are normalized to one. In the inset
\begin{figure}[t]
\vspace{1cm}
\begin{center}\resizebox{10cm}{!}{\includegraphics{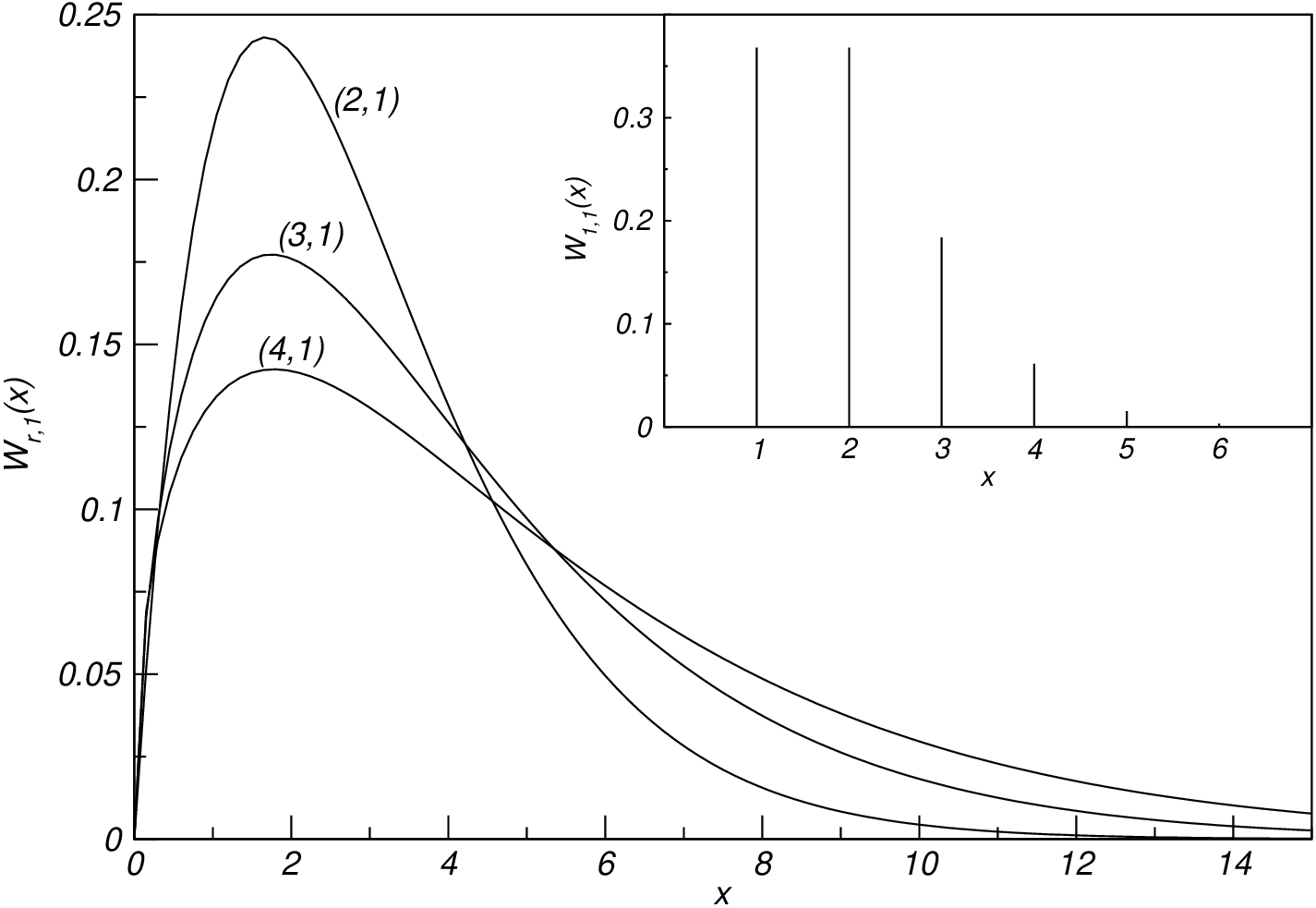}}
\caption{The weight functions $W_{r,1}(x)$, ($x=|z|^2$), in
the resolution of unity for $r=2,3,4$ (continuous curves) and for
$r=1$, a  Dirac's comb (in the inset), as a function of
$x$.}\label{FigA}
\end{center}
\end{figure}the height of the vertical line at $x=k$ symbolizes the strength
of the delta function $\delta(x-k)$, see Eq.(\ref{Y}). For further
properties of $W_{1,1}(x)$ and more generally of $W_{r,r}(x)$
associated with Eq.(\ref{Dd}), see \cite{BlasiakJPA2004}.
\begin{figure}[t]
\vspace{1cm}
\begin{center}\resizebox{10cm}{!}{\includegraphics{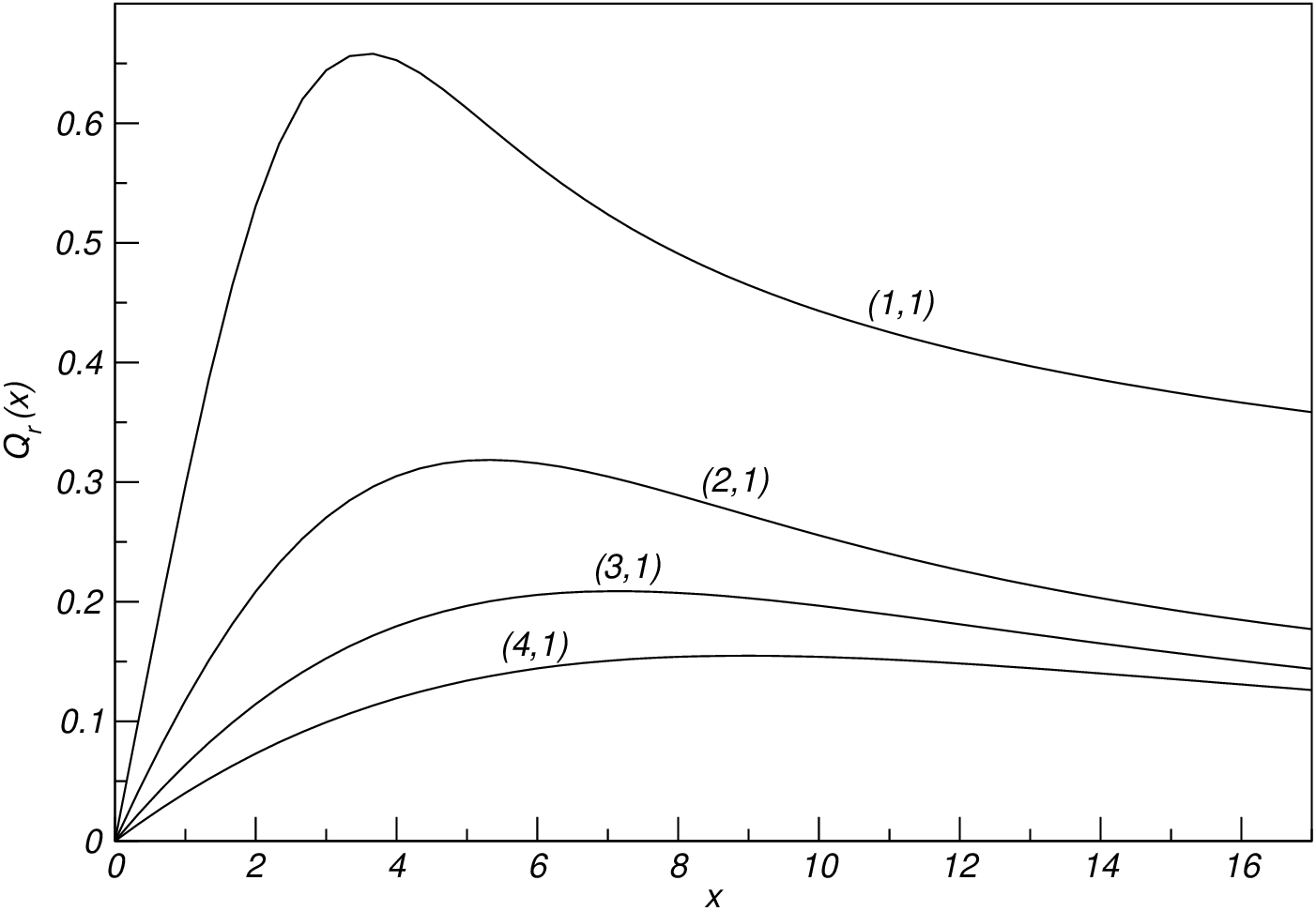}}
\caption{Mandel parameters $Q_r(x)$, for
$r=1\ldots4$, as a function of $x=|z|^2$, see
Eq.(\ref{Q}).}\label{FigB}
\end{center}
\end{figure}

\subsubsection{Construction and properties}

A comparison of Eqs.(\ref{I}), (\ref{J}) and (\ref{Nnn}) indicates
that the normalized states defined through $\rho(n)=B_{r,1}(n+1)$
as
\begin{eqnarray}\label{Z}
|z\rangle_r =\mathcal{N}_r^{-1/2}(|z|^2)\sum_{n=0}^\infty
\frac{z^n}{\sqrt{B_{r,1}(n+1)}}|n\rangle,
\end{eqnarray}
with normalization
\begin{eqnarray}
\mathcal{N}_r(x)=\sum_{n=0}^\infty \frac{x^n}{B_{r,1}(n+1)}>0,
\end{eqnarray}
automatically satisfy the resolution of unity condition of
Eq.(\ref{A}), since for
$W_{r,1}(x)=\pi\frac{\tilde{W}_{r,1}(x)}{\mathcal{N}_r(x)}$:
\begin{eqnarray}
\int_{\mathbb{C}}d^2z\ |z\rangle_r \tilde{W}_{r,1}(|z|^2){_r\langle z|}=I=\sum_{n=0}^\infty|n\rangle \langle n|.
\end{eqnarray}
Note that Eq.(\ref{Z}) is equivalent to Eq.(\ref{1}) with the
definition $[n]_r=B_{r,1}(n+1)/B_{r,1}(n)$, $n=0,1,2,\ldots$.
\\
Having satisfied the completeness condition with the functions
$W_{r,1}(x)$, $r=1,2,\ldots$ we now proceed to examine the
quantum-optical fluctuation properties of the states
$|z\rangle_r$. From now on we consider the $|n\rangle$'s to be
eigenfunctions of the boson number operator $N=a^\dag a$, {\it
i.e.} $N|n\rangle=n|n\rangle$. The Mandel parameter \cite{KPS1}
\begin{eqnarray}\label{Q}
Q_r(x)=x\left(\frac{{\mathcal{N}_r}^{\prime\prime}
(x)}{{\mathcal{N}_r}^\prime (x)}-\frac{{\mathcal{N}_r}^\prime
(x)}{\mathcal{N}_r(x)}\right),
\end{eqnarray}
allows one to distinguish between the sub-Poissonian (antibunching
effect, $Q_r<0$) and super-Poissonian (bunching effect, $Q_r>0$)
statistics of the beam. In Figure \ref{FigB} we display $Q_r(x)$
for $r=1\ldots4$. It can be seen that all the states $|z\rangle_r$
in question are super-Poissonian in nature, with the deviation
from $Q_r=0$, which characterizes the conventional coherent
states, diminishing for $r$ increasing.
\begin{figure}[t]
\vspace{1cm}
\begin{center}\resizebox{10cm}{!}{\includegraphics{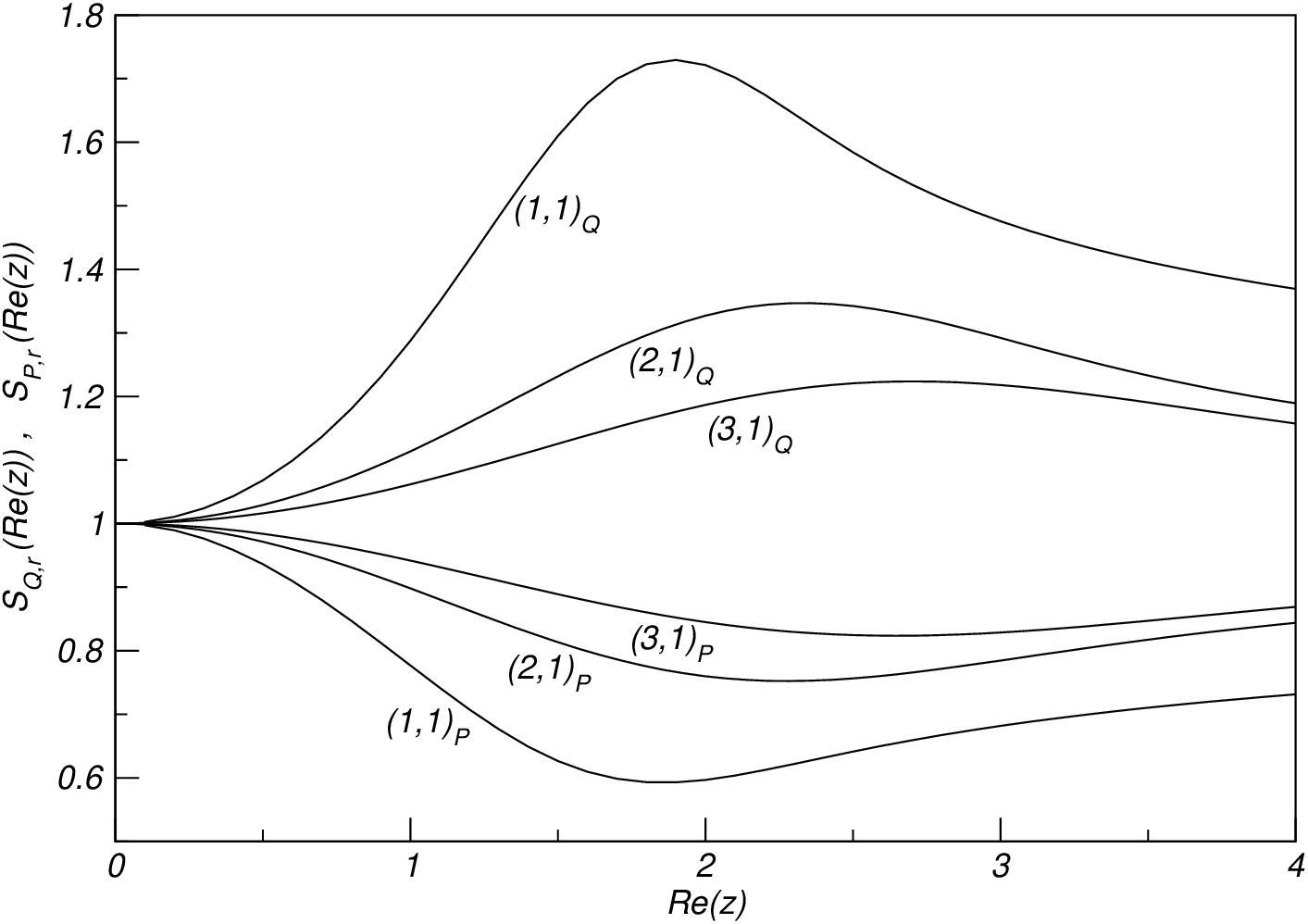}}
\caption{The squeezing parameters of Eqs.(\ref{SQ}) and (\ref{SP}) for the
coordinate $Q$ (three upper curves) and for the momentum $P$
(three lower curves) for different $r$, as a function of $Re(z)$,
for $r=1,2,3$.}\label{FigC}
\end{center}
\end{figure}
\\
In Figure \ref{FigC} we show the behavior of
\begin{eqnarray}\label{SQ}
S_{Q,r}(z)=\frac{{_r\langle z|(\Delta Q)^2|z\rangle_r}}{2},
\end{eqnarray}
and
\begin{eqnarray}\label{SP}
S_{P,r}(z)=\frac{{_r\langle z|(\Delta P)^2|z\rangle_r}}{2},
\end{eqnarray}
which are the measures of squeezing in the coordinate and
momentum quadratures respectively. In the display we have chosen
the section along $Re(z)$. All the states $|z\rangle_r$ are
squeezed in the momentum $P$ and dilated in the coordinate $Q$.
The degree of squeezing and dilation diminishes with increasing
$r$. By introducing the imaginary part in $z$ the curves of
$S_Q(z)$ and $S_P(z)$ smoothly transform into one another, with
the identification $S_Q(i\alpha)=S_P(\alpha)$ and
$S_P(i\alpha)=S_Q(\alpha)$ for any positive $\alpha$.
\begin{figure}[t]
\vspace{1cm}
\begin{center}\resizebox{10cm}{!}{\includegraphics{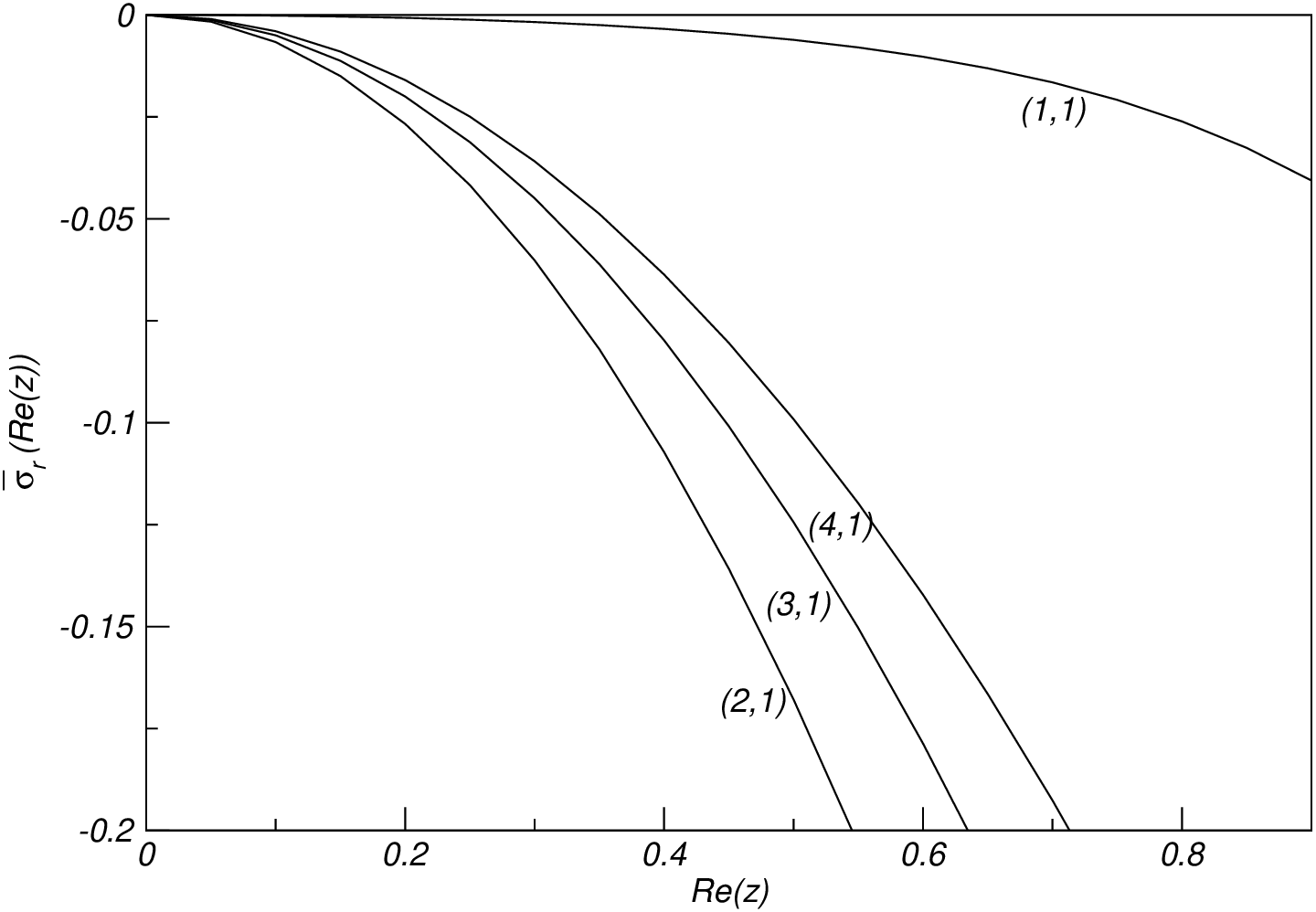}}
\caption{The signal-to-quantum noise ratio relative
to its value in the standard coherent states $\bar{\sigma}_r$, see
Eq.(\ref{O}), as a function of $Re(z)$, for
$r=1\ldots4$.\bigskip}\label{FigD}
\end{center}
\end{figure}
\\
In Figure \ref{FigD} we show the signal-to-quantum noise ratio
\cite{Yuen} relative to $4[{_c\langle z|N|z\rangle_c}]=4|z|^2$,
its value in conventional coherent states; i.e the quantity
$\bar{\sigma}_r=\sigma_r-4[{_c\langle z|N|z\rangle_c}]$, where
\begin{eqnarray}\label{O}
\sigma_r=\frac{[{_r\langle z|Q|z\rangle_r}]^2}{(\Delta Q)^2},
\end{eqnarray}
with $(\Delta Q)^2={_r\langle z|Q^2|r\rangle_r}-({_r\langle
z|Q|r\rangle_r})^2$. Again only the section $Re(z)$ is shown. We
conclude from Figure \ref{FigD} that the states $|z\rangle_r$ are
more ``noisy'' than the standard coherent states with
$\rho_c(n)=n!$ .
\\
In Figure \ref{FigE} we give the metric factors
\begin{eqnarray}\label{W}
\omega_r(x)=\left[x\frac{{\mathcal{N}_r}^\prime (x)}{\mathcal{N}_r(x)}\right]^\prime ,
\end{eqnarray}
which describe the geometrical properties of embedding the surface
of coherent states in Hilbert space, or equivalently a measure of
a distortion of the complex plane induced by the coherent states
\cite{KPS1}. Here, as far as $r$ is concerned, the state
$|z\rangle_1$ appears to be most distant from the $|z\rangle_c$
coherent states for which $\omega_c=1$.
\begin{figure}[t]
\vspace{1cm}
\begin{center}
\resizebox{10cm}{!}{\includegraphics{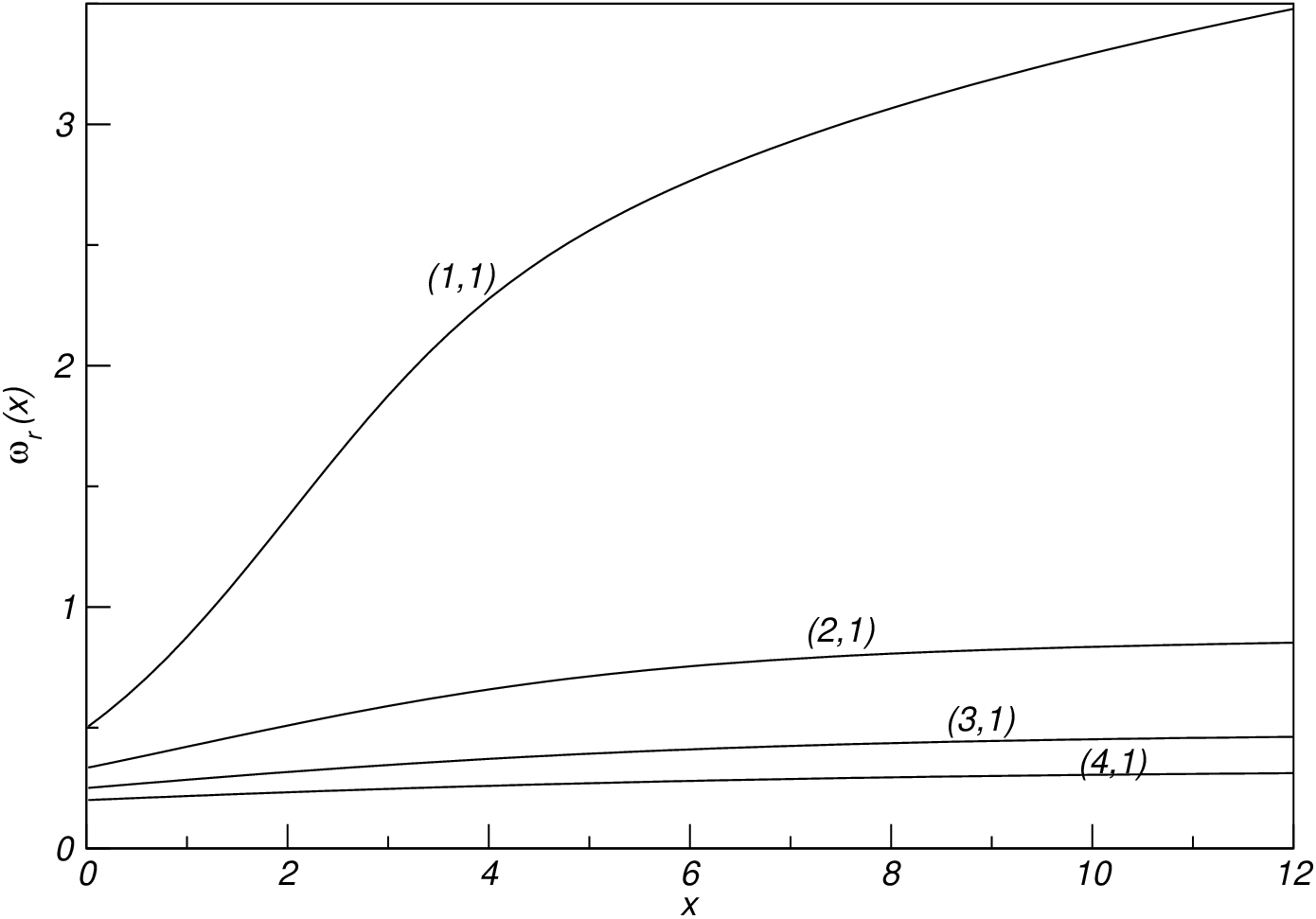}}
\caption{Metric factor $\omega_r(x)$, calculated with
Eq.(\ref{W}) as a function of $x=|z|^2$, for
$r=1\ldots4$.}\label{FigE}
\end{center}
\end{figure}

\subsubsection{Remarks}

The use of sequences $B_{r,1}(n)$ to construct coherent states is
not limited to the case exemplified by Eq.(\ref{Z}). In fact, any
sequence of the form $B_{r,1}(n+p)$, $p=0,1,...$ will also
define a set of coherent states, as then their respective weight
functions will be $V_{r,1}^{(p)}(x)=x^{p-1}W_{r,1}(x)>0$. Will the
physical properties of coherent states defined with
$\rho_p(n)=B_{r,1}(n+p)$  depend sensitively on $p$ ?
A case in point is that of $p=0$ for which
qualitative differences from Figure \ref{FigB} ($p=1$) appear. In
Figure \ref{FigF} we present the Mandel parameter for these
states. Whereas for $r=1$ the state is still super-Poissonian, for
$r=2,3,4$ one observes novel behavior, namely a crossover from
sub- to super-Poissonian statistics for finite values of $x$.
Also, as indicated in Figure \ref{FigG}, we note a cross-over
between squeezing and dilating behavior for different $r$. These
curious features merit further investigation.
\begin{figure}[p]
\vspace{1cm}
\begin{center}\resizebox{10cm}{!}{\includegraphics{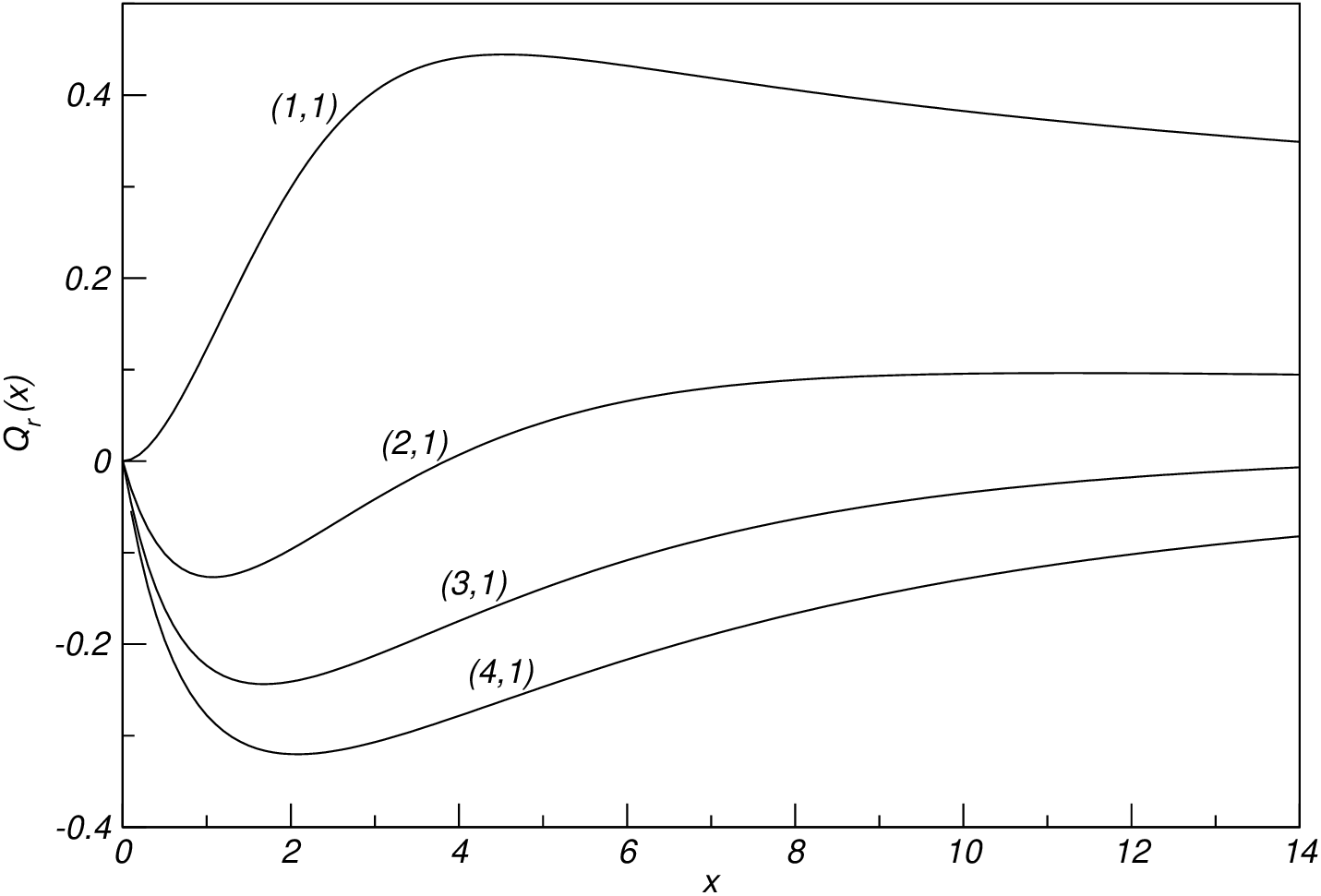}}
\caption{Mandel parameters $Q_r(x)$ for
$r=1\ldots4$, as a function of $x=|z|^2$, for states with
$\rho(n)=B_{r,1}(n)$.\bigskip}\label{FigF}
\vspace{1cm}
\resizebox{10cm}{!}{\includegraphics{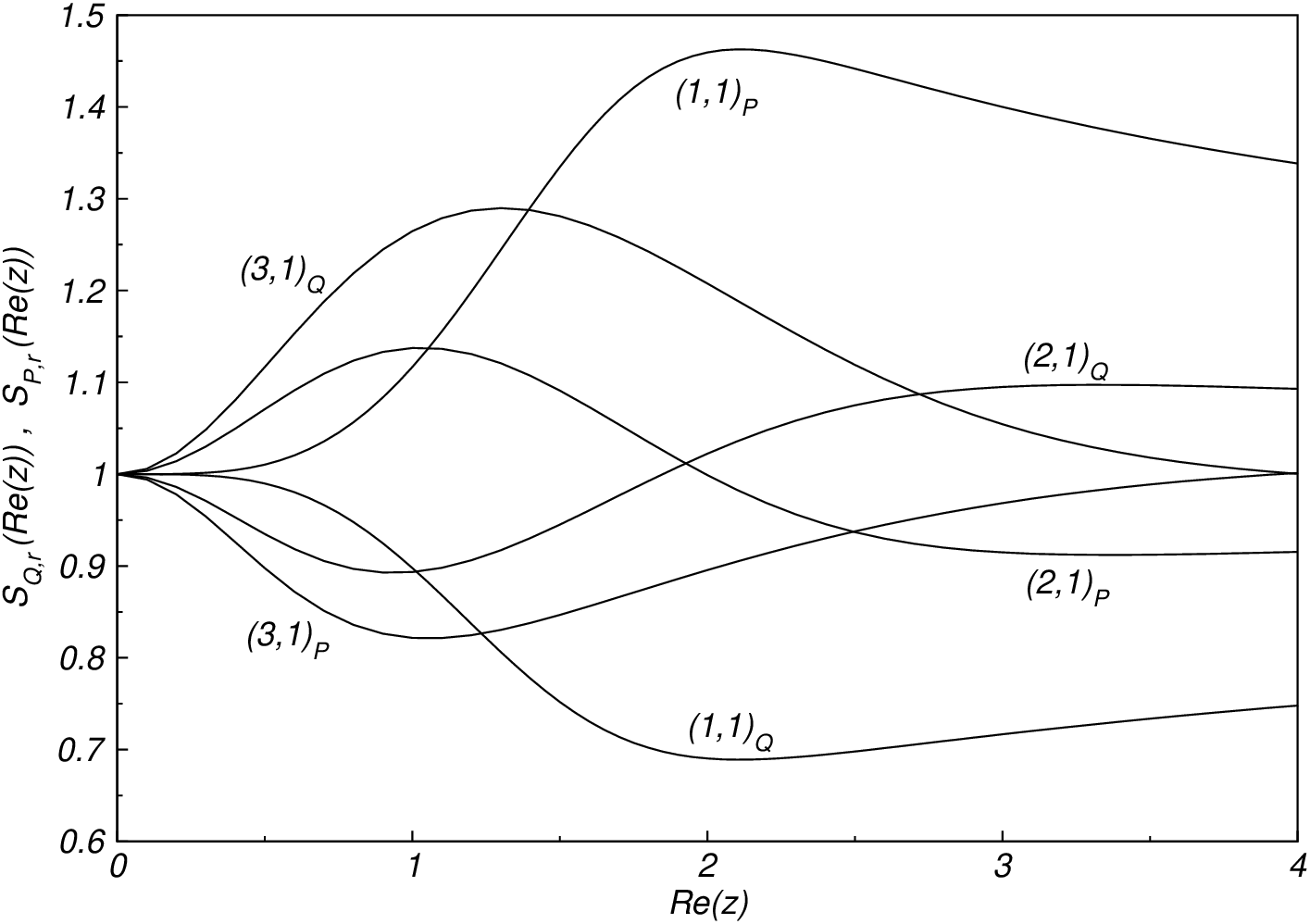}}
\caption{Squeezing parameters for the states with
$\rho(n)=B_{r,1}(n)$ for $r=1,2,3$, as a function of $x=|z|^2$.
The subscripts $P$ and $Q$ refer to momentum and coordinate
variables respectively.}\label{FigG}
\end{center}
\end{figure}
\newpage

%%%%%%%%%%%%%%%%%%%%%%%%%%%%%%%%%%%%%%%%%%%%%%%%%%%%%%%%%%%%%%%%%%%%%%%%%%%%%%%%

\section{Substitution theorem}
\label{SubsTheorem}

In Chapter \ref{Sheffer} we have demonstrated how to find the
normally ordered form of the exponential of the operator linear in
the annihilation or creation operator. This representation is of
fundamental meaning in the coherent state representation used in physics.
\\
As a demonstration of possible other utility of such formulas we
show the direct proof of the \emph{substitution theorem}. More
specifically we will demonstrate that for any function $F(x)$ the
following equality holds
\begin{eqnarray}\label{substitution}
e^{\lambda\left(q(X)D + v(X)\right)}F(x) = G(\lambda, x)
\cdot F(T(\lambda,x))
\end{eqnarray}
where functions $T(\lambda,x)$ and $G(\lambda,x)$ may be found
from the following equations
\begin{eqnarray}\label{T}
\frac{\partial T(\lambda,x)}{\partial\lambda} =  q(T(\lambda,x))\ , &~~~~~~~~~~~~~~&T(0,x) = x\ ,
\end{eqnarray}
\begin{eqnarray}
\label{GGG}
\frac{\partial G(\lambda,x)}{\partial\lambda} =  v(T(\lambda,x))\cdot G(\lambda,x)\ , &~~~~&G(0,x) = 1\ .
\end{eqnarray}
First observe from Eq.(\ref{substitution}) that the action of
$e^{\lambda\left(q(X)D + v(X)\right)}$ on a function $F(x)$
amounts to:
\\
a) change of argument $x\to T(\lambda,x)$ in $F(x)$ which is in
fact a substitution;
\\
b) multiplication by a prefactor $G(\lambda,x)$ which we call a
{\it prefunction}.
\\
We also see from Eq.(\ref{GGG}) that $G(\lambda,x) = 1$ for $v(x) =
0$. Finally, note that $e^{\lambda\left(q(X)D + v(X)\right)}$ with
$\lambda$ real generates an abelian, one-parameter group,
implemented by Eq.(\ref{substitution}); this gives  the following
group composition law for $T(\lambda,x)$ and $G(\lambda,x)$:
\begin{eqnarray}\label{group}
\begin{array}{ccl}
T(\lambda + \theta,x) &=& T(\theta,T(\lambda,x)),
\\
G(\lambda + \theta,x) &=& G(\lambda,x)\cdot
G(\theta,T(\lambda,x)).
\end{array}
\end{eqnarray}
Now we proceed to the proof of
Eqs.(\ref{substitution})-(\ref{GGG}). Using the multiplication $X$
and derivative $D$ representation we may rewrite Eq.(\ref{aqa}) as
\begin{eqnarray}
e^{\lambda \left(q(X)D+v(X)\right)}=
\ :G(\lambda,X)\cdot e^{[T(\lambda,X)-X]D}:\
\end{eqnarray}
with accompanying differential Eqs.(\ref{t}) and (\ref{g}) for
functions $T(\lambda,x)$ and $G(\lambda,x)$ which are exactly the
same as Eqs.(\ref{T}) and (\ref{GGG}).
\\
Now, recalling the Taylor formula $e^{\alpha D}F(x)=F(x+\alpha)$
we may generalize it to
\begin{eqnarray}\label{gentaylor}
    :e^{\alpha(X) D}:\ F(x)=F(x+\alpha(x)).
\end{eqnarray}
It can be seen from the power series expansion of the exponential.
\\
This allows to write
\begin{eqnarray}\nonumber
e^{\lambda \left(q(X)D+v(X)\right)}F(x)&=& :G(\lambda,X)\cdot
e^{[T(\lambda,X)-X]D}:F(x)\ \\\nonumber &=& G(\lambda,X)\
:e^{[T(\lambda,X)-X]D}:F(x)\ \\\nonumber
&\stackrel{(\ref{gentaylor})}{=}& G(\lambda,X)\
F(x+[T(\lambda,x)-x])\\\nonumber &=& G(\lambda,x)\cdot
F(T(\lambda,x)).
\end{eqnarray}
This completes the proof.
\\
Here we list several examples illustrating
Eqs.(\ref{substitution})-(\ref{GGG}) for some choices of $q(x)$ and
$v(x)$. Since Eqs.(\ref{T}) and (\ref{GGG}) are first order linear
differential equations we shall simply write down their solutions
without dwelling on details. First we treat the case of $v(x) =
0$, which implies $g(\lambda,x)\equiv 1$:
\begin{itemize}
\item[{\bf Ex.1:}]{
\begin{eqnarray}\nonumber
&&q(x) = x,\ \ \ \ \ \ \ \ \ T(\lambda,x) = xe^{\lambda}\nonumber
\end{eqnarray}
which gives $\exp\left(\lambda x \frac{d}{{d}x}\right)F(x) =
F(xe^{\lambda})$, a well known illustration of the Euler dilation
operator $\exp
\left(\lambda x\frac{d}{{d}x}\right)$.}
\item[\bf{Ex.2:}]{
\begin{eqnarray}\nonumber
&&q(x) = x^{r},\ \ r>1,\ \ \ \ \ \ \ \ \ T(\lambda,x) =
\frac{\displaystyle x}{\displaystyle\left(1 - \lambda\left(r - 1\right)x^{r-1}\right)^{\frac{1}{r-1}}}\nonumber
\end{eqnarray}}
\end{itemize}
The above examples were considered in the literature
\cite{DattoliNuovoCim}\cite{Lang}\cite{BlasiakAoC2003}\cite{BlasiakPLA2003}.
\\
We shall go on to examples with $v(x)\neq 0$ leading to nontrivial
prefunctions:
\begin{itemize}
\item[{\bf Ex.3:}]{
\begin{eqnarray}\nonumber
&&q(x) = 1\ \ \ \text{and}\ \ \ v(x)\ -\
\text{arbitrary},\\\nonumber &&T(\lambda,x) = x +
\lambda,\nonumber
\\\nonumber
&&g(\lambda,x) = e^{\int\limits_{0}^{\lambda}{\rm d}u\ v(x+u)}.
\end{eqnarray}}
\item[{\bf Ex.4:}]{
\begin{eqnarray}\nonumber
&&q(x) = x,\ \ \ \text{and}\ \ \ v(x) = x^2\\\nonumber
&&T(\lambda,x) = xe^{\lambda},\\\nonumber &&g(\lambda,x) =
\exp{\left[\frac{x^{2}}{ 2}
\left(e^{2\lambda} - 1\right)\right]}.
\end{eqnarray}}
\item[{\bf Ex.5:}]{
\begin{eqnarray}\nonumber
&&q(x) = x^{r}\ \ r>1\ \ \ \text{and}\ \ \ v(x) = x^s\\\nonumber
&&T(\lambda,x) = \frac{\displaystyle x}{\displaystyle\left(1 -
\lambda\left(r - 1\right)x^{r-1}\right)^{\frac{1}{r-1}}}
\\\nonumber
&&g(\lambda,x) = \exp{\left[\frac{\displaystyle
x^{s-r+1}}{\displaystyle 1 - r}\left(\frac{\displaystyle
1}{\displaystyle\left(1 - \lambda\left(r -
1\right)x^{r-1}\right)^{\frac{s-r+1}{r-1}}} - 1\right)\right]}.
\end{eqnarray}}
\end{itemize}
Closer inspection of the  above examples (or at any other example
which the reader may easily construct) indicates that from the
analytical point of view the substitution theorem should be
supplemented by some additional assumptions, like  restrictions
on $\lambda$. In general we can say that it can be valid only
locally
\cite{Duchamp2004}. However, we note that a formulation in the
language of formal power series does not require such detailed
analysis and the resulting restrictions may be checked afterwords.
\\
We finally  note that the substitution theorem can be used to find
the normally ordered form of an exponential operator linear in $a$
or $a^\dag$ (as in Chapter \ref{Sheffer}). This idea has been
exploited in
\cite{BlasiakPLA2005}.

%%%%%%%%%%%%%%%%%%%%%%%%%%%%%%%%%%%%%%%%%%%%%%%%%%%%%%%%%%%%%%%%%%%%%%%%%%%%%%%%%%%%%

\chapter{Conclusions}
\label{Conclusions}

In this work we have considered the boson normal ordering problem
for powers and exponentials of two wide classes of operators. The
first one consists of boson strings and more generally homogenous
polynomials, while the second one treats operators linear in one
of the creation or annihilation operators. We have used the
methods of advanced combinatorial analysis to obtain  a thorough
understanding and efficient use of the proposed formalism. In all
cases we provided closed form expressions, generating functions,
recurrences etc. The analysis was based on the Dobi\'nski-type
relations and the umbral calculus methods. In general,  the
combinatorial analysis is shown to be  an effective and flexible
tool in this kind of problem.
\\
We also provided  a wealth of examples and pointed out possible
applications. The advantages may be neatly seen from the coherent
state perspective ({\it e.g.} we may use it for construction of
the phase space pictures of quantum
mechanics). We may also obtain  solutions of the moment problem,
enabling us to construct new families of generalized coherent
states. Moreover application to the operator calculus is noted 
and exemplified in the substitution theorem. 
We also observe  that the normal ordering
problems for deformed algebras may be handled within that setting.
\\
These few remarks on the possible applications and extensions
of the methods used in this work (others may be found in the
supplied references) yield a potentially wide field of possible
future research.
\\
We would like to point out some other interesting features. We
believe that similar combinatorial methods may be used to find the
normally ordered forms of other classes of operators, such as  the
exponential of a general boson polynomial. Moreover, the extension
to the multi-mode boson and fermion case should be susceptible to explicit
analysis.  This program would open a wide arena for immediate
application to variety of physical models.
\\
In conclusion, we would like to recall that the use of advanced
combinatorics provides a wealth of interpretative tools for the problems under
consideration. Combinatorial objects may be interpreted in terms
of graphs, rook polynomials, partitions, correlations etc. We
have not touched upon this aspect in this work but we consider it
as a very promising one.

%%%%%%%%%%%%%%%%%%%%%%%%%%%%%%%%%%%%%%%%%%%%%%%%%%%%%%%%%%%%%%%%%%%%%%%%%%%%%%%%%%

\appendix

\chapter{Coherent states}
\label{Coherent states}

Here we define and review some properties of coherent states
\cite{Klauder}\cite{Gilmore}.
\\
We define coherent states $|z\rangle$ as the eigenstates of the
annihilation operator
\begin{eqnarray}\label{az}
    a|z\rangle=z|z\rangle,
\end{eqnarray}
where $z\in\mathbb{C}$ is a complex number. They can be written
explicitly as
\begin{eqnarray}\label{zn}
    |z\rangle=e^{-\frac{1}{2}|z|^2}\ \sum_{n=0}^\infty\frac{z^n}{\sqrt{n!}}|n\rangle.
\end{eqnarray}
These states are normalized but not orthogonal. The coherent
states \emph{overlapping factor} is
\begin{eqnarray}\label{zz}
    \langle z|z'\rangle=e^{z^*z'-\frac{1}{2}|z^{'}|^2-\frac{1}{2}|z|^2},
\end{eqnarray}
The set of coherent states $\left\{|z\rangle:\
z\in\mathbb{C}\right\}$ constitute an overcomplete basis in the
Hilbert space $\cal{H}$. The following \emph{resolution of unity}
property holds
\begin{eqnarray}\label{unity}
    \frac{1}{\pi}\int_{\mathbb{C}}|z\rangle\langle z|\ d^2z=1.
\end{eqnarray}
Eq.(\ref{zn}) may be rewritten in the form (see Eqs.(\ref{n0}))
\begin{eqnarray}\label{z0}
    |z\rangle=e^{-\frac{|z|^2}{2}}e^{za^\dag}|0\rangle,
\end{eqnarray}
or with the use of Eq.(\ref{BCH}) as
\begin{eqnarray}
    |z\rangle=e^{za^\dag-z^*a}|0\rangle.
\end{eqnarray}
The operator $D(z)=e^{za^\dag-z^*a}$ is called a
\emph{displacement operator} and therefore coherent states are
sometimes called displaced vacuum. This group property may be
taken as the definition of the coherent states also for other
(than Heisenberg-Weyl) groups, see \cite{Perelomov}.
\\
Let us define two self-adjoint operators $Q$ and $P$ by
\begin{eqnarray}
\begin{array}{rcl}
    Q&=&\frac{\displaystyle a^\dag+a}{\displaystyle \sqrt{2}},\vspace{2mm}\\
    P&=&i\ \frac{\displaystyle a^\dag-a}{\displaystyle \sqrt{2}}.
\end{array}
\end{eqnarray}
They satisfy the Heisenberg-Weyl commutation relation $[Q,P]=i$.
These operators may be interpreted as the position and momentum
operators for the quantum particle in the harmonic oscillator
potential, in quantum optics they play the role of the field
quadratures and are also used in the phase space formulation of
Quantum Mechanics. It can be shown that the coherent states are
the \emph{minimum uncertainty states} for operators $Q$ and $P$,
{\it i.e.}
\begin{eqnarray}\label{uncertainty}
    \Delta_{|z\rangle}Q\cdot\Delta_{|z\rangle}P=\frac{1}{2},
\end{eqnarray}
where
$\Delta_{\psi}A=\sqrt{\langle\psi|(A-\langle\psi|A|\psi\rangle)^2|\psi\rangle}$
is the uncertainty of the operator $A$. Moreover they are the only
states if one additionally imposes the condition
$\Delta_{|z\rangle}Q=\Delta_{|z\rangle}P$ (otherwise the family of
\emph{squeezed states} is obtained) which serves as another
possible definition of coherent states.
\\
Finally we mention that the resolution of unity of
Eq.({\ref{unity}) together with the continuity of the mapping
$z\to|z\rangle$ are sometimes taken as the minimum requirements
for the coherent states. However, this definition is not unique
and leads to other families of so called \emph{generalized
coherent states}
\cite{KlauderJMP1}\cite{KlauderJMP2}\cite{Klauder} (see also
Section \ref{Generalized coherent states}).
\\
We note that, due to their special features, coherent states are widely used in quantum optics
\cite{Glauber}\cite{KlauderSudarshan} as well as in other areas of
physics \cite{Klauder}.
\\
In this text we especially exploit the property of Eq.(\ref{az}).
It is because for an operator $F(a,a^\dag)$ which is in the normal form,
$F(a,a^\dag)\equiv{\cal N}\left[F(a,a^\dag)\right]\equiv\
:F(a,a^\dag):$, its coherent state matrix elements may be readily
written as
\begin{eqnarray}
    \langle z|F(a,a^\dag)|z'\rangle=\langle z|z'\rangle\ F(z',z^*).
\end{eqnarray}
Also for the double dot operation it immediately yields
\begin{eqnarray}
    \langle z|:G(a,a^\dag):|z'\rangle=\langle z|z'\rangle\ G(z',z^*).
\end{eqnarray}
Unfortunately for the general operator none of these formulae
hold. Nevertheless, there is a very useful property which is true;
that is, if for an  arbitrary operator $F(a,a^{\dag})$ we have
\begin{eqnarray}\label{n}
\langle z|{F}(a,a^{\dag})|z'\rangle = \langle z|z'\rangle\ G(z^*,z')
\end{eqnarray}
\noindent then the normally ordered form of ${F}(a,a^{\dag})$ is given by
\begin{eqnarray}\label{N}
{\cal N}\left[{F}(a,a^{\dag})\right] =\ :G(a^{\dag},a):\, .
\end{eqnarray}
For other properties of coherent states we refer to
\cite{Klauder}\cite{Gilmore}.

%%%%%%%%%%%%%%%%%%%%%%%%%%%%%%%%%%%%%%%%%%%%%%%%%%%%%%%%%%%%%%%%%%

\chapter{Formal power series. Umbral calculus}
\label{Umbral}

Here we recall the basic definitions and theorems concerning
the formal power series calculus. As an illustration some topics
of the umbral calculus are reviewed. For a detailed discussion see
\cite{Niven}, \cite{Comtet}, \cite{Riordan}, \cite{Wilf},
\cite{Knuth}, \cite{Flajolet}, \cite{Roman}, \cite{Rainville},
\cite{Rota}, \cite{DiBucchianico}.
\newline

\section*{Formal power series}

Suppose we are given a series of numbers $(f_n)_{n=0}^\infty$.
We define a {\em formal power series} in indeterminate $x$ as
\begin{eqnarray}\label{formal}
F(x)=\sum_{n=0}^\infty f_n\frac{x^n}{n!}.
\end{eqnarray}
The set of formal power series constitutes a ring when the
following operations are imposed
\begin{itemize}
\item \underline{addition}:
\begin{eqnarray}
F(x)+G(x)=\sum_{n=0}^\infty f_n\frac{x^n}{n!}+\sum_{n=0}^\infty
g_n\frac{x^n}{n!}=\sum_{n=0}^\infty (f_n+g_n)\frac{x^n}{n!}
\end{eqnarray}
\item \underline{multiplication} (Cauchy product rule):
\begin{eqnarray}\label{Cauchy}
F(x)\cdot G(x)=\sum_{n=0}^\infty
f_n\frac{x^n}{n!}\cdot\sum_{n=0}^\infty g_n\frac{x^n}{n!}
=\sum_{n=0}^\infty
\sum_{k=0}^n\binom{n}{k}f_kg_{n-k}\frac{x^n}{n!}
\end{eqnarray}
\end{itemize}
Both operations have an inverses. The inverse of the series $F(x)$
with respect to addition is the series with negative coefficients
$-F(x)=\sum_{n=0}^\infty -f_n\frac{x^n}{n!}$. The (unique)
multiplicative inverse on the other hand is well defined only when
$f_0\neq 0$ and may be given recursively, see {\it e.g.} \cite{Wilf}.
\\
One can also define the \emph{substitution} of formal power series by
\begin{eqnarray}
F(G(x))=\sum_{n=0}^\infty f_n\frac{G(x)^n}{n!}
\end{eqnarray}
when $g_0=0$ in the expansion $G(x)=\sum_{n=0}^\infty
g_n\frac{x^n}{n!}$. Explicit expression for the coefficients
is given by the Fa\`a di Bruno formula
\cite{Comtet}\cite{Aldrovandi}. A series $F(x)$ may be shown  to
have a compositional inverse $F^{-1}(x)$ iff $f_0=0$ and
$f_1\neq0$.
\\
Other definitions and properties can be further given. Here is the
example of the \emph{derivative} and \emph{integral} of the formal power series
(note that these operators act like shift operators on the
sequence $(f_n)_{n=0}^\infty$)
\begin{eqnarray}\label{DF}
DF(x)=F'(x)=\sum_{n=1}^\infty
f_n\frac{x^{n-1}}{(n-1)!}=\sum_{n=0}^\infty f_{n+1}\frac{x^n}{n!}
\end{eqnarray}
\begin{eqnarray}
\int F(x)\ dx=\sum_{n=0}^\infty f_n\frac{x^{n+1}}{(n+1)!}=\sum_{n=1}^\infty f_{n-1}\frac{x^n}{n!}
\end{eqnarray}
Observe that these definitions mirror the corresponding operations
on  analytic functions. Nevertheless the limiting operations are
not needed in these definitions. The ring of formal power series
can be also be given the structure of a complete metric space (see
{\it e.g.} \cite{Flajolet}, \cite{Roman}).
\\
We note that $(f_n)_{n=0}^\infty$ may be also a sequence of
polynomials, functions or anything else.
\\
The use of power series is well suited in the context of
generating functions. The series as in Eq.(\ref{formal}) is called
the
\emph{exponential generating function}
(because of the factor $1/n!$ in the expansion). Functions
$G(x)=\sum_{n=0}^\infty g_n x^n$ are called
\emph{ordinary generating functions}. The use of generating functions is especially
useful in solving recurrences and enumerating combinatorial objects, see eg. \cite{Wilf}\cite{Knuth}\cite{Flajolet}.
\\
All these definitions and properties may be naturally extended to
the multivariable case.
\\
For a systematic description of the formal power series and
applications see
\cite{Niven}, \cite{Comtet}, \cite{Riordan}, \cite{Wilf}, \cite{Knuth}, \cite{Flajolet}, \cite{Roman}.

\section*{Umbral calculus. Sheffer-type polynomials}

Subject of the umbral calculus is the study of a Sheffer A-type
zero polynomials, called here briefly Sheffer-type \cite{Sheffer},
\cite{Rainville}. Without going into details we recall some basic
theorems and definitions \cite{Roman}, \cite{Rota},
\cite{DiBucchianico}.
\\
Suppose we have a polynomial sequence $s_n(x)$, $n=0,1,2,...$
($s_n(x)$ being the polynomial of degree $n$). It is called of a
{\em Sheffer-type} if it possesses an exponential generating
function of the form
\begin{eqnarray}\label{AB}
G(\lambda,x)=\sum_{n=0}^\infty
s_n(x)\frac{\lambda^n}{n!}=A(\lambda)e^{xB(\lambda)}.
\end{eqnarray}
for some (possibly formal) functions $A(\lambda)$ and $B(\lambda)$
such that $B(0)=0$, $B'(0)\neq0$ and $A(0)\neq0$. When
$B(\lambda)=1$ it is called {\em Apple sequence} for $A(\lambda)$.
For $A(\lambda)=1$ it is known as {\em associated sequence} for
$B(\lambda)$.
\\
Yet another definition of the Sheffer-type sequences can be given
through their lowering and rising operators, {\it i.e.} the
polynomial sequence $s_n(x)$ is of Sheffer-type iff there exist
some functions $f(x)$ and $g(x)$ (possibly formal) such that
$f(0)=0$, $f^{'}(0)\neq0$ and $g(0)\neq 0$ which satisfy
\begin{eqnarray}\label{fg}
&&f(D)s_n(x)=ns_{n-1}(x),\\
&&\left[X-\frac{g'(D)}{g(D)}\right]\frac{1}{f'(D)}s_n(x)=s_{n+1}(x).
\end{eqnarray}
These two definitions describe the Sheffer-type sequence uniquely
and the correspondence is given by
\begin{eqnarray}
&&A(x)=f^{-1}(x),\\
&&B(x)=\frac{1}{g(f^{-1}(x))}.
\end{eqnarray}
\\
Many curious properties of the Sheffer-type polynomials can be
worked out. We quote only one of them, called the \emph{Sheffer
identity}, to show the simplicity of formal manipulations. Using
definition Eq.(\ref{AB}) we have
\begin{eqnarray}\nonumber
\sum_{n=0}^\infty s_n(x+y)\frac{\lambda^n}{n!}&=&A(\lambda)e^{(x+y)B(\lambda)}=A(\lambda)e^{xB(\lambda)}e^{yB(\lambda)}\\\nonumber
&=&\sum_{n=0}^\infty
s_n(x)\frac{\lambda^n}{n!}\cdot\sum_{n=0}^\infty
p_n(y)\frac{\lambda^n}{n!}.
\end{eqnarray}
By the Cauchy product rule Eq.(\ref{Cauchy}) we obtain the Sheffer
identity:
\begin{eqnarray}
    s_n(x+y)=\sum_{k=0}^n\binom{n}{k}p_k(y)s_{n-k}(x)
\end{eqnarray}
where $p_n(x)$ is the associated sequence for $B(\lambda)$.
Observe that for associated sequences ($A(\lambda)=1$) this
property generalizes the binomial identity and therefore sometimes
they are called of {\em binomial type}.
\\
The Sheffer-type polynomials constitute a basis in the space of
polynomials and through that property any formal power series can
be developed in that basis also. On the other hand, formal power
series can be treated as  functionals on the space of polynomials.
Investigation of this connection is also the subject of umbral calculus
in which the Sheffer-type polynomials play a prominent role
\cite{Roman}.
\\
We emphasize that all the functions and series here can be formal
and then the operations are meant in the sense of the previous
Section.
\\
Thorough discussion of the umbral calculus, Sheffer-type
polynomials and applications can be found in
\cite{Roman}\cite{Rota}\cite{DiBucchianico}\cite{Rainville}.

%%%%%%%%%%%%%%%%%%%%%%%%%%%%%%%%%%%%%%%%%%%%%%%%%%%%%%%%%%%%%%%%%%%%%%%%%%%

\newpage

\bibliographystyle{alpha}
\bibliography{thesis}

\newcommand{\etalchar}[1]{$^{#1}$}
\begin{thebibliography}{CGH{\etalchar{+}}96}

\bibitem[Ald01]{Aldrovandi}
R.~Aldrovandi.
\newblock {\em Special Matrices of Mathematical Physics}.
\newblock World Scientific, Singapore, 2001.

\bibitem[AM77]{MehtaII}
G.~P. Agarawal and C.~L. Mehta.
\newblock Ordering of the exponential of a quadratic in boson operators. {II}.
  {M}ultimode case.
\newblock {\em J. Math. Phys.}, 18:408--409, 1977.

\bibitem[Bar61]{Bargmann}
V.~Bargmann.
\newblock On a {H}ilbert space of analytic functions and an associated integral
  transform.
\newblock {\em Commun. Pure and Appl. Math.}, 14:187--214, 1961.

\bibitem[BDHP05]{BlasiakJPA2005}
P.~Blasiak, G.~Dattoli, A.~Horzela, and K.A. Penson.
\newblock Representations of monomiality principle with {S}heffer-type
  polynomials and boson normal ordering.
\newblock 2005.
\newblock arXiv:quant-ph/0504009.

\bibitem[Ben91]{BenderHahn}
C.~M. Bender.
\newblock Solution of operator equations of motion.
\newblock In J.~Dittrich and P.~Exner, editors, {\em Rigorous Results in
  Quantum Dynamics}, pages 99--112, Singapore, 1991. World Scientific.

\bibitem[BHP{\etalchar{+}}05]{BlasiakPLA2005}
P.~Blasiak, A.~Horzela, K.A. Penson, G.H.E. Duchamp, and A.I. Solomon.
\newblock Boson normal ordering via substitutions and {S}heffer-type
  polynomials.
\newblock {\em Phys. Lett. A}, 338:108--116, 2005.
\newblock arXiv:quant-ph/0501155.

\bibitem[BHPS05]{BlasiakKerr}
P.~Blasiak, A.~Horzela, K.A. Penson, and A.~I. Solomon.
\newblock Combinatorics of generalized {K}err models.
\newblock 2005.

\bibitem[BN05]{Burdik}
C.~Burdik and O.~Navratil.
\newblock Normal ordering for the deformed {H}eisenberg algebra involving the
  reflection operator.
\newblock {\em J. Phys. A : Math. Gen.}, 38:2305--2310, 2005.

\bibitem[BPS03a]{BlasiakJPA2003}
P.~Blasiak, K.~A. Penson, and A.~I. Solomon.
\newblock Dobi\'nski-type relations and the log-normal distribution.
\newblock {\em J. Phys. A: Math. Gen.}, 36:L273--L278, 2003.
\newblock arXiv:quant-ph/0303030.

\bibitem[BPS03b]{BlasiakAoC2003}
P.~Blasiak, K.A. Penson, and A.I. Solomon.
\newblock The boson normal ordering problem and generalized {B}ell numbers.
\newblock {\em Ann. Combinat.}, 7:127--139, 2003.

\bibitem[BPS03c]{BlasiakPLA2003}
P.~Blasiak, K.A. Penson, and A.I. Solomon.
\newblock The general boson normal ordering problem.
\newblock {\em Phys. Lett. A}, 309:198--205, 2003.

\bibitem[BPS04]{BlasiakJPA2004}
P.~Blasiak, K.~A. Penson, and A.~I. Solomon.
\newblock Hierarchical {D}obi\'nski-type via substitution and the moment
  problem.
\newblock {\em J. Phys. A: Math. Gen.}, 37:3475--3487, 2004.
\newblock arXiv:quant-ph/0312202.

\bibitem[BPS{\etalchar{+}}05]{BlasiakJMP2005}
P.~Blasiak, K.~A. Penson, A.~I. Solomon, A.~Horzela, and G.~H.~E. Duchamp.
\newblock Some useful combinatorial formulas for bosonic operators.
\newblock {\em J. Math. Phys.}, 46:052110, 2005.
\newblock arXiv:quant-ph/0405103.

\bibitem[Buc98]{DiBucchianico}
A.~Di Bucchianico.
\newblock Introduction to umbral calculus.
\newblock http://www.win.tue.nl/{\textasciitilde}sandro/, 1998.

\bibitem[Ces00]{Cesarano2000}
C.~Cesarano.
\newblock Monomiality principle and {L}egendre polynomials.
\newblock In H.~M.~Srivastava G.~Dattoli and C.~Cesarano, editors, {\em
  Advanced Special Functions and Integration Methods}, page 147. Rome: Aracne
  Rditrice, 2000.

\bibitem[CG69]{Glauber1969}
K.~E. Cahill and R.J. Glauber.
\newblock Ordered expansions in boson amplitude operators.
\newblock {\em Phys.Rev.}, 177:1857--1881, 1969.

\bibitem[CGH{\etalchar{+}}96]{KnuthW}
R.~M. Corless, G.~H. Gonnet, D.~E.~G. Hare, D.~J. Jeffrey, and D.~E. Knuth.
\newblock On the {L}ambert {W} function.
\newblock {\em Adv. Comp. Math.}, 5:329--359, 1996.

\bibitem[Che03]{Cheikh}
Y.~Ben Cheikh.
\newblock Some results on quasi-monomiality.
\newblock {\em Appl. Math. Comput.}, 141:63--76, 2003.

\bibitem[Com74]{Comtet}
L.~Comtet.
\newblock {\em Advanced Combinatorics}.
\newblock Reidel, Dordrecht, 1974.

\bibitem[Das96]{DasGupta}
A.~DasGupta.
\newblock Disentanglement formulas: {A}n alternative derivation and some
  applications to squeezed coherent states.
\newblock {\em Am. J. Phys.}, 64:1422--1427, 1996.

\bibitem[Dat99]{Dattoli1999}
G.~Dattoli.
\newblock Hermite-{B}essel and {L}aguerre-{B}essel functions: a by-product of
  the monomiality principle.
\newblock In G.~Dattoli D.~Cocolicchio and H.~M. Srivastava, editors, {\em
  Advanced Special Functions and Applications}, page~83. Rome: Aracne Rditrice,
  1999.

\bibitem[DOTV97]{DattoliNuovoCim}
G.~Dattoli, P.L. Ottaviani, A.~Torre, and L.~Vasquez.
\newblock Evolution operator equations: {I}ntegration with algebraic and
  finite-difference methods. {A}pplications to physical problems in classical
  and quantum mechanics and quantum field theory.
\newblock {\em Riv. Nuovo Cim.}, 20:1--133, 1997.

\bibitem[DPS{\etalchar{+}}04]{Duchamp2004}
G.~Duchamp, K.~A. Penson, A.~I. Solomon, A.~Horzela, and P.~Blasiak.
\newblock One-parameter groups and combinatorial physics.
\newblock In M.~N.~Hounkonnou J.~Govaerts and A.~J. Msezane, editors, {\em
  Contemporary Problems in Mathematical Physics}, pages 436--449, Singapore,
  2004. World Scientific.
\newblock arXiv:quant-ph/0401126.

\bibitem[DSC01]{Dattoli2001}
G.~Dattoli, H.~M. Srivastava, and C.~Cesarano.
\newblock The {L}aguerre and {L}egendre polynomials from operational point of
  view.
\newblock {\em Appl. Math. Comput.}, 124:117, 2001.

\bibitem[FS05]{Flajolet}
P.~Flajolet and R.~Sedgewick.
\newblock Analytic {C}ombinatorics.
\newblock http://algo.inria.fr/flajolet/Publications/books.html, 2005.

\bibitem[GDM97]{Dattoli1997}
A.~Torre G.~Dattoli and G.~Mazzacurati.
\newblock Quasimonomials and isospectral problems.
\newblock {\em Nuov. Cim. B}, 112:133--138, 1997.

\bibitem[GKP94]{Knuth}
R.~L. Graham, D.~E. Knuth, and O.~Patashnik.
\newblock {\em Concrete Mathematics}.
\newblock Addison-Wesley, Massachusetts, 1994.

\bibitem[Gla63]{Glauber}
R.~J. Glauber.
\newblock The quantum theory of optical coherence.
\newblock {\em Phys. Rev.}, 130:2529--2539, 1963.

\bibitem[Gro78]{Grosswald}
E.~Grosswald.
\newblock {\em Bessel Polynomials}.
\newblock Springer, Berlin, 1978.

\bibitem[JPS01]{PensonJIntSeq2001}
J.M.Sixdeniers, K.A. Penson, and A.I. Solomon.
\newblock Extended {B}ell and {S}tirling numbers from hypergeometric
  exponentiation.
\newblock {\em J. Int. Seqs.}, Article 01.1.04:1--11, 2001.
\newblock http://www.research.att.com/{\textasciitilde}njas/sequences/JIS/.

\bibitem[Kat74]{Katriel}
J.~Katriel.
\newblock Combinatorial aspects of boson algebra.
\newblock {\em Lett. Nuovo Cimento}, 10:565--567, 1974.

\bibitem[Kat83]{Katriel1983}
J.~Katriel.
\newblock Normal ordering formulae for some boson operators.
\newblock {\em J. Phys. A : Math. Gen.}, 16:4171--4173, 1983.

\bibitem[Kat00]{Katriel2000}
J.~Katriel.
\newblock Bell numbers and coherent states.
\newblock {\em Phys. Lett. A}, 237:159--161, 2000.

\bibitem[Kat02]{Katriel2002}
J.~Katriel.
\newblock Coherent states and combinatorics.
\newblock {\em J. Opt. B: Quantum Semiclass. Opt.}, 4:S200--S203, 2002.

\bibitem[KD95]{Duchamp1995}
J.~Katriel and G.~Duchamp.
\newblock Ordering relations for $q$-boson operators, continued fraction
  techniques and the $q$-{B}{C}{H} enigma.
\newblock {\em J. Phys. A : Math. Gen.}, 28:7209--7225, 1995.

\bibitem[KK92]{KatrielKibler}
J.~Katriel and M.~Kibler.
\newblock Normal ordering for deformed boson operators and operator-valued
  deformed {S}tirling numbers.
\newblock {\em J. Phys. A: Math. Gen.}, 25:2683--2691, 1992.

\bibitem[Kla63a]{KlauderJMP1}
J.~R. Klauder.
\newblock Continuous-representation theory. {I}. {P}ostulates for
  continuous-representation theory.
\newblock {\em J. Math. Phys}, 4:1055--1058, 1963.

\bibitem[Kla63b]{KlauderJMP2}
J.~R. Klauder.
\newblock Continuous-representation theory. {II}. {G}eneralized relation
  between quantum and classical dynamics.
\newblock {\em J. Math. Phys}, 4:1058--1073, 1963.

\bibitem[KPS01]{KPS1}
J.~R. Klauder, K.A. Penson, and J.M. Sixdeniers.
\newblock Constructing coherent states through solutions of {S}tieltjes and
  {H}ausdorff moment problems.
\newblock {\em Phys. Rev. A}, 64:013817--013835, 2001.

\bibitem[KS68]{KlauderSudarshan}
J.~R. Klauder and E.~C.~G. Sudarshan.
\newblock {\em Fundamentals of Quantum Optics}.
\newblock Benjamin, New York, 1968.

\bibitem[KS85]{Klauder}
J.~R. Klauder and B-S. Skagerstam.
\newblock {\em Coherent States. Application in Physics and Mathematical
  Physics}.
\newblock World Scientific, Singapore, 1985.

\bibitem[KVD01]{Vogel2}
Z.~Kis, W.~Vogel, and L.~Davidovich.
\newblock Nonlinear coherent states of trapped-atom motion.
\newblock {\em Phys. Rev. A}, 64:033401--033411, 2001.

\bibitem[Lan00]{Lang}
W.~Lang.
\newblock On generalizations of the {S}tirling number triangles.
\newblock {\em J. Int. Seqs.}, Article 00.2.4:1--19, 2000.
\newblock http://www.research.att.com/~njas/sequences/JIS/.

\bibitem[Lou64]{Louisell}
W.~H. Louisell.
\newblock {\em Radiation and Noise in Quantum Electronics}.
\newblock McGrow-Hill Co., New York, 1964.

\bibitem[MBP05]{Mendez}
M.A. M\'endez, P.~Blasiak, and K.~A. Penson.
\newblock Combinatorial approach to generalized {B}ell and {S}tirling numbers
  and boson normal ordering problem.
\newblock {\em J. Math. Phys.}, 2005.
\newblock arXiv:quant-ph/0505180.

\bibitem[Meh77]{MehtaI}
C.~L. Mehta.
\newblock Ordering of the exponential of a quadratic in boson operators. {I}.
  {S}ingle mode case.
\newblock {\em J. Math. Phys.}, 18:404--407, 1977.

\bibitem[Mik83]{Mikhailov1983}
V.~V. Mikhailov.
\newblock Ordering of some boson operator functions.
\newblock {\em J. Phys. A : Math. Gen.}, 16:3817--3827, 1983.

\bibitem[Mik85]{Mikhailov1985}
V.~V. Mikhailov.
\newblock Normal ordering and generalized {S}tirling numbers.
\newblock {\em J. Phys. A : Math. Gen.}, 18:231--235, 1985.

\bibitem[MM98]{Manko2}
V.~I. Manko and V.I.~Vilela Mendes.
\newblock On the nonlinearity interpretation of $q$- and $f$-deformations and
  some applicationss.
\newblock {\em J. Phys. A}, 31:6037--6044, 1998.

\bibitem[MMSZ97]{Manko1}
V.~I. Manko, G.~Marmo, E.C.G. Sudarshan, and F.~Zaccaria.
\newblock f-{O}scilators and nonlinear coherent states.
\newblock {\em Phys.Scr.}, 55:528--541, 1997.

\bibitem[MSI93]{Mufti}
A.~Mufti, H.~A. Schmitt, and M.~Sargent III.
\newblock Finite-dimensional matrix representations as calculational tools in
  quantum optics.
\newblock {\em Am. J. Phys.}, 6!:729--733, 1993.

\bibitem[MV96]{Vogel1}
V.I.~Vilela Mendes and W.~Vogel.
\newblock Nonlinear coherent states.
\newblock {\em Phys. Rev. A}, 54:4560--4563, 1996.

\bibitem[MW95]{Mandel}
L.~Mandel and R.~Wolf.
\newblock {\em Optical Coherence and Quantum Optics}.
\newblock University Press, Cambridge, 1995.

\bibitem[Nav73]{Navon}
A.M. Navon.
\newblock Combinatorics and fermion algebra.
\newblock {\em Nuovo Cimento}, 16B:324--330, 1973.

\bibitem[Niv69]{Niven}
I.~Niven.
\newblock Formal power series.
\newblock {\em Am. Math. Monthly}, 76:871--889, 1969.

\bibitem[Per86]{Perelomov}
A.M. Perelomov.
\newblock {\em Generalized Coherent States and Their Application}.
\newblock Springer-Verlag, Berlin, 1986.

\bibitem[Pit99]{Pittman}
J.~Pittman.
\newblock A lattice path model for the {B}essel polynomials.
\newblock Technical Report 551, U.C. Berkeley, USA, 1999.

\bibitem[PS99]{PS}
K.~A. Penson and A.~I. Solomon.
\newblock Mittag-{L}effler coherent states.
\newblock {\em J. Phys. A: Math. Gen.}, 32:7543--7563, 1999.

\bibitem[PS01]{PSQTS}
K.~A. Penson and A.~I. Solomon.
\newblock Coherent states from combinatorial sequences.
\newblock In {\em 2nd International Symposium on Quantum Theory and
  Symmetries}, pages 527--530, Singapore, 2001. World Scientific.
\newblock arXiv:quant-ph/0111151.

\bibitem[Rai65]{Rainville}
E.D. Rainville.
\newblock {\em Special Functions}.
\newblock The Macmillan Company, New York, 1965.

\bibitem[Rio84]{Riordan}
J.~Riordan.
\newblock {\em An Introduction to Combinatorial Analysis}.
\newblock Wiley, New York, 1984.

\bibitem[Rom84]{Roman}
S.~Roman.
\newblock {\em The Umbral Calculus}.
\newblock Academic Press, Orlando, 1984.

\bibitem[RR78]{Rota}
S.~Roman and G.-C. Rota.
\newblock The umbral calculus.
\newblock {\em Adv. Math.}, 27:95--188, 1978.

\bibitem[Sch03]{Schork}
M.~Schork.
\newblock On the combinatorics of normal ordering bosonic operators and
  deformations of it.
\newblock {\em J. Phys. A: Math. Gen.}, 36:4651--4665, 2003.

\bibitem[SDB{\etalchar{+}}04]{QTS3}
A.I. Solomon, G.H.E. Duchamp, P.~Blasiak, A.~Horzela, and K.A. Penson.
\newblock Normal order: {C}ombinatorial graphs.
\newblock In {\em 3rd International Symposium on Quantum Theory and
  Symmetries}, pages 527--536, Singapore, 2004. World Scientific.
\newblock arXiv:quant-ph/0402082.

\bibitem[Seg63]{Segal}
I.~Segal.
\newblock {\em Mathematical Problems of Relativistic Physics}.
\newblock Providence, Rhode Island, 1963.

\bibitem[She39]{Sheffer}
I.M. Sheffer.
\newblock Some properties of polynomial sets of type zero.
\newblock {\em Duke Math. Journal}, 5:590--622, 1939.

\bibitem[Six01]{Sixdeniers}
J.~M. Sixdeniers.
\newblock {\em Constructions de nouveaux \'etats coh\'erentes \`a l'aide de
  solutions des probl\`emes des moments}.
\newblock PhD thesis, Universit\'e Paris 6, 2001.

\bibitem[Slo05]{Sloane}
N.J.A. Sloane.
\newblock Encyclopedia of integer sequences.
\newblock http://www.research.att.com/{\textasciitilde}njas/sequences, 2005.

\bibitem[Sol94]{SolomonPLA}
A.I. Solomon.
\newblock A characteristic functional for deformed photon phenomenology.
\newblock {\em Phys. Lett. A}, 196:29--34, 1994.

\bibitem[SP00]{PS2}
J.~M. Sixdeniers and K.~A. Penson.
\newblock On the completeness of coherent states generated by binomial
  distribution.
\newblock {\em J. Phys. A: Math. Gen.}, 33:2907--2916, 2000.

\bibitem[SPK01]{KPS2}
J.~M. Sixdeniers, K.~A. Penson, and J.~R. Klauder.
\newblock Tricomi coherent states.
\newblock {\em Int. J. Mod. Phys. B}, 15:4231--4243, 2001.

\bibitem[ST79]{Shalitin}
D.~Shalitin and Y.~Tikochinsky.
\newblock Transformation between the normal and antinormal expansions of boson
  operators.
\newblock {\em J. Math. Phys.}, 20:1676--1678, 1979.

\bibitem[Sta99]{Stanley}
R.P. Stanley.
\newblock {\em Enumerative Combinatorics}.
\newblock University Press, Cambridge, 1999.

\bibitem[TS95]{Turbiner}
A.~Turbiner and Y.~Smirnov.
\newblock Hidden {SL2}-algebra of finite difference equations.
\newblock {\em Mod. Phys. Lett. A}, 10:1795--1801, 1995.

\bibitem[Var04]{Varvak}
A.~Varvak.
\newblock Rook numbers and the normal ordering problem.
\newblock In {\em Formal Power Series and Algebraic Combinatorics}, pages
  259--268, University of British Columbia (Vancouver B.C., Canada), 2004.
\newblock arXiv:math.CO/0402376.

\bibitem[Wil67]{Wilcox}
R.M. Wilcox.
\newblock Exponential operators and parameter differentiation in quantum
  physics.
\newblock {\em J. Math. Phys.}, 8:962--982, 1967.

\bibitem[Wil94]{Wilf}
H.S. Wilf.
\newblock {\em Generatingfunctionology}.
\newblock Academic Press, New York, 1994.

\bibitem[Wit75]{Witschel1975}
W.~Witschel.
\newblock Ordered operator expansions by comparison.
\newblock {\em J. Phys. A : Math. Gen.}, 8:143--155, 1975.

\bibitem[Wit05]{Witschel2005}
W.~Witschel.
\newblock Ordering of boson operator functions by the {H}ausdorff similarity
  transform.
\newblock {\em Phys. Lett. A}, 334:140--143, 2005.

\bibitem[Yue76]{Yuen}
H.~P. Yuen.
\newblock Two-photon coherent states of the radiational field.
\newblock {\em Phys. Rev. A}, 13:2226--2243, 1976.

\bibitem[ZFR90]{Gilmore}
W.~M. Zhang, D.H. Feng, and R.Gilmore.
\newblock Coherent states: {T}heory and some applications.
\newblock {\em Rev. Mod. Phys.}, 62:867--927, 1990.

\end{thebibliography}

%%%%%%%%%%%%%%%%%%%%%%%%%%%%%%%%%%%%%%%%%%%%%%%%%%%%%%%%%%%%%%%%%%%%%%%%%%%

\chapter*{List of Publications}

\subsubsection{\large \underline{Journal publications}}\bigskip

\begin{itemize}

\item[\textbf{1.}]\textbf{P. B\l{}asiak, K.A. Penson and A.I. Solomon}\\
\textsc{The general boson normal ordering problem}\\
\textit{Phys. Lett. A} \textbf{309} 198-205 (2003)\\arXiv:quant-ph/0402027\bigskip

\item[\textbf{2.}]\textbf{P. B\l{}asiak, K.A. Penson and A.I. Solomon}\\
\textsc{Dobi\'nski-type relations and the log-normal distribution}\\
\textit{J. Phys. A: Math. Gen.} \textbf{36} L273-L278 (2003)\\arXiv:quant-ph/0303030\bigskip

\item[\textbf{3.}]\textbf{P. B\l{}asiak, K.A. Penson and A.I. Solomon}\\
\textsc{The boson normal ordering problem and generalized Bell numbers}\\
\textit{Ann. Combinat.} \textbf{7} 127-139 (2003)\\arXiv:quant-ph/0212072\bigskip

\item[\textbf{4.}]\textbf{P. B\l{}asiak, A. Horzela and E. Kapu\'scik}\\
\textsc{Alternative hamiltonians and Wigner quantization}\\
\textit{J. Opt. B: Quantum Semiclass. Opt.} \textbf{5} S245-S260 (2003)\bigskip

\item[\textbf{5.}]\textbf{P. B\l{}asiak and A. Horzela}\\
\textsc{Quantization of alternative Hamiltonians}\\
\textit{Czech. J. Phys.} \textbf{53} 985-991 (2003)\bigskip

\item[\textbf{6.}]\textbf{K.A. Penson, P. B\l{}asiak, G. Duchamp, A. Horzela}\\\textbf{and A.I. Solomon}\\
\textsc{Hierarchical Dobi\'nski-type relations via substitution and the moment problem}\\
\textit{J. Phys. A: Math. Gen.} \textbf{37} 3475-3487 (2004)\\arXiv:quant-ph/0312202\bigskip

\item[\textbf{7.}]\textbf{P. B\l{}asiak, K.A. Penson and A.I. Solomon}\\
\textsc{Combinatorial coherent states via normal ordering of bosons}\\
\textit{Lett. Math. Phys.} \textbf{67} 13-23 (2004)\\arXiv:quant-ph/0311033\bigskip

\item[\textbf{8.}]\textbf{P. B\l{}asiak, A. Horzela, K. A. Penson, A.I. Solomon}\\
\textsc{Deformed bosons: Combinatorics of the normal ordering}\\
\textit{Czech. J. Phys.} \textbf{54} 1179-1184 (2004)\\arXiv:quant-ph/0410226\bigskip

\item[\textbf{9.}]\textbf{P. B\l{}asiak, K.A. Penson, A.I. Solomon, A. Horzela}\\\textbf{and G. H. E. Duchamp}\\
\textsc{Some useful combinatorial formulas for bosonic operators}\\
\textit{J. Math. Phys.} \textbf{46} 052110 (2005)\\arXiv:quant-ph/0405103\bigskip

\item[\textbf{10.}]\textbf{P. B\l{}asiak, A. Horzela, K.A. Penson, G.H.E. Duchamp}\\\textbf{and A.I. Solomon}\\
\textsc{Boson Normal Ordering via Substitutions and Sheffer-type Polynomials}\\
\textit{Phys. Lett. A} \textbf{338} 108-116 (2005)\\arXiv:quant-ph/0501155\bigskip

\item[\textbf{11.}]\textbf{M.A. M\'{e}ndez, P. B\l{}asiak and K.A. Penson}\\
\textsc{Combinatorial approach to generalized Bell and Stirling numbers and boson normal ordering problem}\\
\textit{J. Math. Phys.}, in print (2005)\\arXiv:quant-ph/0505180\bigskip

\end{itemize}

\subsubsection{\large \underline{Conference proceedings}}\bigskip

\begin{itemize}

\item[\textbf{12.}]\textbf{A.I. Solomon, P. B\l{}asiak, G. Duchamp, A. Horzela}\\\textbf{and K.A. Penson}\\
\textsc{Combinatorial physics, normal order and model Feynman graphs}\\
\textit{Symmetries in Science XI}\\ B. Gruber, G. Marmo and N. Yoshinaga (eds.), p.527-536 (Kluwer Academic Publishers 2004)\\arXiv:quant-ph/0310174\bigskip

\item[\textbf{13.}]\textbf{A.I. Solomon, G. Duchamp, P. B\l{}asiak,
A. Horzela}\\\textbf{and K.A. Penson}\\
\textsc{Normal Order: Combinatorial Graphs}\\
\textit{Quantum Theory and Symmetries}\\ Proceedings of the 3rd International Symposium
\\ P.C. Argyres, T.J. Hodges, F. Mansouri, J.J. Scanio, P. Suranyi, L.C.R. Wijewardhana (eds.), p.398-406 (World Scientific Publishing 2004)\\arXiv:quant-ph/0402082\bigskip

\item[\textbf{14.}]\textbf{G. Duchamp, K.A. Penson, A.I. Solomon, A. Horzela }\\\textbf{and P. B\l{}asiak}\\
\textsc{One-Parameter Groups and Combinatorial Physics}\\
Proceedings of the
\textit{Third International Workshop on}\\\textit{ Contemporary Problems in
Mathematical Physics}\\ (Cotonu, B\'enin, November 1st-7th 2003)\\
J. Govaerts, M.N. Hounkonnou and A.Z. Msezane (eds.)\\ p. 436-449
(World Scientific Publishing,
2004)\\arXiv:quant-ph/0401126\bigskip

\item[\textbf{15.}]\textbf{A.I. Solomon, P. B\l{}asiak, G. Duchamp, A. Horzela}\\\textbf{and K.A. Penson}\\
\textsc{Partition functions and graphs: A combinatorial approach}\\
Proceedings of the \textit{XI International Conference on Symmetry
Methods in Physics (SYMPHYS-11)} (Prague, Czech Republic, June
2004), in press (JINR Publishers,
Dubna)\\arXiv:quant-ph/0409082\bigskip

\item[\textbf{16.}]\textbf{A. Horzela, P. B\l{}asiak, G.E.H. Duchamp, K.A. Penson}\\\textbf{and A.I. Solomon}\\
\textsc{A product formula and combinatorial field theory}\\
Proceedings of the \textit{XI International Conference on Symmetry
Methods in Physics (SYMPHYS-11)} (Prague, Czech Republic, June
2004), in press (JINR Publishers,
Dubna)\\arXiv:quant-ph/0409152\bigskip

\end{itemize}

\subsubsection{\large \underline{Preprints}}\bigskip

\begin{itemize}

\item[\textbf{17.}]\textbf{P. B\l{}asiak, G. Dattoli, A. Horzela and K.A. Penson}\\
\textsc{Representations of Monomiality Principle with Sheffer-type Polynomials
and Boson Normal Ordering}\\
\textit{J. Phys. A: Math. Gen.}, submitted (2005)\\arXiv:quant-ph/0504009\bigskip

\item[\textbf{18.}]\textbf{P. B\l{}asiak, A. Horzela, K.A. Penson and A.I. Solomon}\\
\textsc{Combinatorics of generalized Kerr models}\\
in preparation (2005)\bigskip

\end{itemize}

\end{document}